\documentclass[sn-mathphys-num]{sn-jnl}

\usepackage{graphicx}%
\usepackage{multirow}%
\usepackage{amsmath,amssymb,amsfonts}%
\usepackage{amsthm}%
\usepackage{mathrsfs}%
\usepackage[title]{appendix}%
\usepackage{xcolor}%
\usepackage{textcomp}%
\usepackage{manyfoot}%
\usepackage{booktabs}%
\usepackage{algorithm}%
\usepackage{algorithmicx}%
\usepackage{algpseudocode}%
\usepackage{listings}%
\usepackage[utf8]{inputenc}

\usepackage{tikz-cd}
\usepackage{enumerate}
\algnewcommand{\LeftComment}[1]{\Statex \(\triangleright\) #1}

\newcommand{\etal}{et~al.}
\newcommand{\RG}{\mathcal{RG}}
\newcommand{\RS}{\mathbb{W}}
\newcommand{\MDRG}{\mathrm{MDRG}}
\newcommand{\RSS}{\mathcal{RS}}

\newcommand{\JSet}{\mathbb{J}}
\newcommand{\JStruct}{\mathcal{J}}

\newcommand{\W}{\mathbb{W}}

\newcommand{\M}{\mathbb{M}}
\newcommand{\X}{\mathbb{X}}
\newcommand{\cP}{\mathcal{P}}

\newcommand{\cM}{\mathcal{M}}
\newcommand{\R}{\mathbb{R}}
\newcommand{\bc}{\mathbf{c}}
\newcommand{\bS}{\mathbb{S}}

\newcommand{\Z}{\mathbb{Z}}

\newcommand{\e}{\mathbf{e}}
\newcommand{\x}{\mathbf{x}}
\newcommand{\y}{\mathbf{y}}

\newcommand{\f}{\mathbf{f}}

\newcommand{\br}{\mathbf{r}}
\newcommand{\bu}{\mathbf{u}}
\newcommand{\bv}{\mathbf{v}}

\newcommand{\cO}{\mathcal{O}}

\newcommand{\Star}[1]{\text{St }#1}
\newcommand{\Link}[1]{\text{Lk }#1}

\newcommand{\LowerLink}[1]{\text{Lk}_{-}{#1}}
\newcommand{\UpperLink}[1]{\text{Lk}^{+}{#1}}

\newcommand{\ba}{\mathbf{a}}
\newcommand{\Net}{\mathbb{N}}

\algdef{SE}[DOWHILE]{Do}{doWhile}{\algorithmicdo}[1]{\algorithmicwhile\ #1}%

\newcommand{\secref}[1]{Section \ref{#1}}
\newcommand{\algoref}[1]{Algorithm \ref{#1}}
\newcommand{\figref}[1]{Fig. \ref{#1}}

\newcommand{\lemref}[1]{Lemma \ref{#1}}
\newcommand{\propref}[1]{Proposition \ref{#1}}
\newtheorem{lemma}{Lemma}[section]
\newtheorem{proposition}[lemma]{Proposition}
\newtheorem{remark}[lemma]{Remark}

\theoremstyle{thmstyleone}%

\theoremstyle{thmstyletwo}%

\theoremstyle{thmstylethree}%
\newtheorem{definition}{Definition}%

\begin{document}

\title[Article Title]{An Algorithm for Fast and Correct Computation of Reeb Spaces for PL Bivariate Fields}


\author*[1]{\fnm{Amit} \sur{Chattopadhyay}}\email{a.chattopadhyay@iiitb.ac.in}

\author[1]{\fnm{Yashwanth} \sur{Ramamurthi}}\email{yashwanth@iiitb.ac.in}

\author[2]{\fnm{Osamu} \sur{Saeki}}\email{saeki@imi.kyushu-u.ac.jp}

\affil[1]{\orgdiv{Computer Science}, \orgname{International Institute of Information Technology, Bangalore}, \orgaddress{\street{26/C Electronic City}, \state{Karnataka}, \postcode{560100}, \country{India}}}

\affil[2]{\orgdiv{Institute of Mathematics for Industry}, \orgname{Kyushu University}, \orgaddress{\street{Motooka 744}, \city{Nishi-ku},  \state{Fukuoka}, 
\postcode{819-0395},
\country{Japan}}}



\abstract{ 
The Reeb space is a fundamental data structure in computational topology that represents the fiber topology of
a multi-field (or multiple scalar fields), extending the level set topology of a scalar field.
For piecewise-linear (PL) bivariate fields, the Reeb spaces are $2$-dimensional polyhedrons while for PL scalar fields, the Reeb graphs (or Reeb spaces) are of dimension $1$. Efficient algorithms have been designed for computing Reeb graphs, however, computing correct Reeb spaces 
 for PL bivariate fields, is a challenging open problem. There are only a few implementable algorithms in the literature for computing Reeb space or its approximation, via range quantization or by computing a Jacobi fiber surface, which are computationally expensive or have correctness issues, i.e., the computed Reeb space may not be topologically equivalent or homeomorphic to the actual Reeb space.  In the current paper, we propose a novel algorithm for fast and correct computation of the  Reeb space corresponding to a generic PL bivariate field defined on a triangulation $\M$ of a $3$-manifold without boundary, leveraging the fast algorithms for computing Reeb graphs in the literature. 

Our algorithm is based on the computation of a Multi-Dimensional Reeb Graph (MDRG) which is first proved to be homeomorphic with the Reeb space. 
For the correct computation of the MDRG, we compute the Jacobi set of the PL bivariate field and its projection into the Reeb space, called the Jacobi structure.
Finally, the correct Reeb space is obtained by computing a net-like structure embedded in the Reeb space and then computing its $2$-sheets in the net-like structure. The time complexity of our algorithm is $\cO(n^2 + n\, c_{int}\, \log  n + nc_L^2)$, where $n$ is the total number of simplices in $\M$, $c_{int}$ is the number of intersection points of the projections of the non-adjacent Jacobi set edges on the range of the bivariate field and $c_L$ is the upper bound on the number of simplices in the link of an edge of $\M$. This complexity is comparable with the fastest algorithm available in the literature. Moreover, we claim to provide the first algorithm to compute the topologically correct Reeb space without using range quantization. }

\maketitle
\tableofcontents

\section{Introduction}
\label{sec:introduction}
Multi-field topology has become increasingly prominent due to its richness compared to scalar topology \cite{2012-Duke-Nuclear-Scission, 2015-Carr-Fiber, 2021-Ramamurthi-MRS}. Techniques for computing multi-field topology have been developed based on Jacobi sets \cite{2004-edelsbrunner-jacobi-set}, singular fibers \cite{2004-Saeki-topoology-of-singulat-fibers}, and Reeb spaces \cite{2008-edels-reebspace}.  
Tools in multi-field topology have proven effective in revealing features that cannot be detected using scalar topology tools \cite{2012-Duke-Nuclear-Scission, 2015-Carr-Fiber, 2021-Ramamurthi-MRS}. 
Carr \etal~\cite{2014-Carr-JCN, 2012-Duke-Nuclear-Scission} proposed a joint contour net (JCN), a quantized approximation of the Reeb space, and showcased its application in detecting nuclear scission of plutonium and fermium atom data. Towards this, the current paper addresses the correct computation of the Reeb space without quantization that captures the quotient topology of a piecewise-linear (PL) bivariate field generalizing the Reeb graph of a PL scalar field \cite{2008-edels-reebspace}. We note that for a PL bivariate field, the Reeb space is a $2$-dimensional CW-complex (or polyhedron) composed of $0$-, $1$-, and $2$-sheets (or cells), capturing the evolution of fiber topology in the domain, where each fiber corresponds to the intersection of level sets of the two component scalar fields. In contrast, for a PL scalar field, the Reeb graph (or Reeb space) is a one-dimensional complex composed of only $0$- and $1$-sheets, capturing the evolution of level set topology.

Efficient algorithms have been proposed for computing Reeb graphs. Shinagawa \etal \cite{1991-Shinagawa-Reeb-Graph} proposed an algorithm for computing the Reeb graph of a PL scalar field (function) defined on a triangulated surface, which takes $\cO(n_t^2)$ time, where $n_t$ is the number of triangles. Cole-McLaughlin \etal~\cite{2003-Cole-McLaughlin-Reeb-Graph} proposed a $\cO(n_e \log n_e)$ time algorithm for computing the Reeb graph of a PL Morse function defined on triangulation corresponding to a $2$-manifold, where $n_e$ is the number of edges in the triangulation. Tierny \etal~\cite{2009-Tierny-Reeb-Graph} computed the Reeb graph of a PL scalar field defined on a volumetric mesh by first transforming it to a loop-free mesh through a process called `loop surgery', which systematically removes loops from the domain. Then, the contour tree corresponding to the transformed mesh is computed, from which the Reeb graph of the original mesh is derived by reconstructing the removed loops. The time complexity of the algorithm is $\cO(n_v \log n_v + n \widetilde{\alpha}(n) + g n)$, where $\widetilde{\alpha}$ is the inverse Ackermann function, $g$ is the number of handles, $n_v$ is the number of vertices, and $n$ is the total number of simplices in the input mesh. Algorithms have been proposed for computing the Reeb graphs of PL functions defined on triangulations of $3$-manifolds, which take $\cO(n \log n )$ time, where $n$ is the number of simplices in the input triangulation (see \secref{subsec:Reeb-graph} for further details) \cite{2010-Harvey-Reeb-Graph, 2012-Parsa-Reeb-Graph}. However, computing fast and correct Reeb spaces for PL multi-fields, or even for PL bivariate fields, is a challenging open problem.

\subsection{Prior Works on Computing Reeb Spaces}
\label{subsec:prior-work}
There are a few algorithms in the literature for computing Reeb space or its approximations via quantization. The motivation for developing the current algorithm originated from the work by Edelsbrunner \etal \cite{2008-Edelsbrunner-Time-varying-Reeb-graphs} on time-varying Reeb graphs of a $1$-parameter family of smooth functions defined on a $3$-manifold without boundary (see \secref{subsec:time-varying-Reeb-graph} for more details). However, generalizing the results for PL bivariate fields is more challenging.  In another work, Edelsbrunner \etal~\cite{2008-edels-reebspace} studied the local and global structures of the Reeb space of generic PL multi-fields (or maps) on combinatorial manifolds for computing Reeb spaces. However, no practical algorithm has been developed based on this theory until now. For applications in topological data analysis (TDA) and visualization, range-based quantized approximations of the Reeb space have been proposed using Mapper \cite{2007-Mapper} and Joint Contour Net (JCN) \cite{2014-Carr-JCN}.
The challenging part of these quantization-based methods is the selection of appropriate quantization levels to capture the correct topology of the Reeb space. In other words, such quantized algorithms may miss the important critical features of the Reeb space which project to sub-pixel regions in the range \cite{2016-Julien-ReebSpace}. Moreover, such algorithms are computationally expensive. For a multi-field $\f$ with $r$ fields defined on a domain of dimension $d$, the complexity of the JCN algorithm is $\cO(r(2r+ d)n_\f + (2r + d)n_\f\widetilde{\alpha}((2r + d)n_\f))$, where $n_\f$ is the total number of fragments (a fragment is a part of a quantized contour in a simplex of the domain) and $\widetilde{\alpha}$ is the inverse Ackermann function. In general, the complexity is high depending on the number of resolutions of the quantization or the number of fragments.  

Similar to the multi-dimensional Reeb graph (MDRG) data-structure by Chattopadhyay \etal~\cite{2014-EuroVis-short}, Strodthoff \etal~\cite{2015-Strodthoff} introduced a layered Reeb graph for representing the Reeb space as a hierarchical collection of Reeb graphs. However, for the computation of the layered Reeb graph, the feasible functions are assumed to be very restricted with no critical points in the interior of the domain which is 3D solid (embedded three-dimensional manifold with boundary). Therefore, their algorithm computes the Jacobi set only on the boundary representation of the domain. Moreover, neither the Reeb space computation nor the relationship between the layered Reeb graph and the Reeb space has been addressed in \cite{2015-Strodthoff}. \textcolor{black}{Carr \etal~\cite{2015-Carr-Fiber} introduced the concept of a \emph{fiber surface} for a bivariate field as a generalization of an isosurface for a scalar field, and proposed an algorithm for computing a fiber surface corresponding to a path in the range of the bivariate field. Klacansky \etal~\cite{2016-Klacansky} subsequently extended this approach to enable the exact computation of fiber surfaces in tetrahedral meshes.} More recently, Tierny \etal~\cite{2016-Julien-ReebSpace} proposed an algorithm for computing the Reeb space of a PL bivariate field by computing the \emph{Jacobi fiber surface}, that is, the fiber surface passing through the Jacobi set edges, using the exact fiber surface algorithm of Klacansky \etal~\cite{2016-Klacansky}.  The time complexity of this Reeb space algorithm is $\cO(n_e n_T)$, where $n_e$ denotes the number of edges and $n_T$ the number of tetrahedra in the input mesh, which is comparable to the complexity of the algorithm proposed in the present paper. However, the Jacobi fiber surface–based approach to constructing the Reeb space introduced by Tierny \etal~\cite{2016-Julien-ReebSpace} suffers from the following two fundamental shortcomings:
\begin{figure}
    \centering
    \includegraphics[scale=0.26]{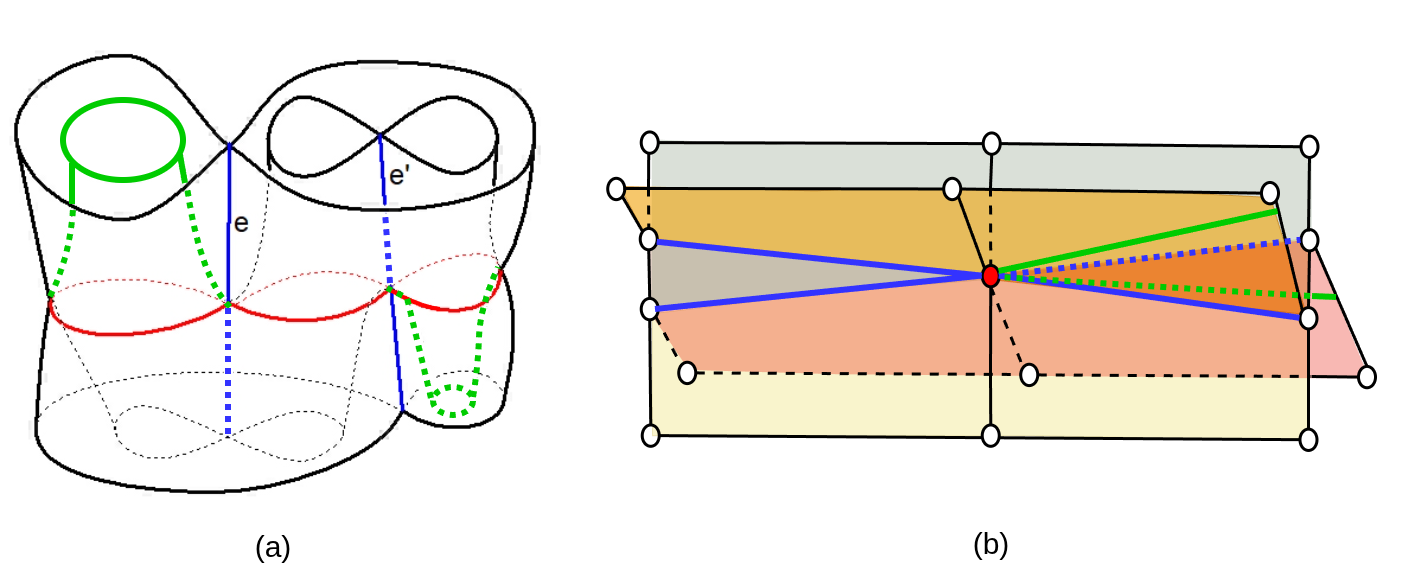}
    \caption{\textcolor{black}{
    \textbf{Jacobi fiber surface (in domain) and associated Reeb Space near a double point.}
(a) An example of a fiber surface (part of which is the Jacobi fiber surface) in the domain of a bivariate field, as illustrated in \figref{fig:violation-second-Morse-condition}. The fiber surface in (a) is obtained as the inverse image under the quotient map of the Jacobi structure edges (blue edges in (b)), together with two regular edges (green edges in (b)) in the Reeb space. The inverse images of the blue Jacobi structure edges form the \emph{Jacobi fiber surfaces} (shown in black in (a)), including the singular fiber component (red curve in (a)) and the Jacobi set edges (blue edges in (a)). The inverse images of the green regular edges in (b) form the \emph{regular fiber surfaces} (shown in green in (a)). The disjoint blue edges $\e$ and $\e'$ in the domain correspond to Jacobi set edges whose projections intersect as Jacobi structure edges in the Reeb space (b). The Jacobi fiber surfaces associated with these two disjoint Jacobi set edges intersect along a singular fiber (red curve in (a)).
(b) The Reeb space of the bivariate field near a double point, as shown in \figref{fig:violation-second-Morse-condition}. The Jacobi set edges project to Jacobi structure edges in the Reeb space, where they intersect at a double point (red point in (b)). The 2-sheets of the Reeb space (shown in different colors in (b)) meet along these intersecting Jacobi structure edges.
    }}
    \label{fig:jacobi-fiber-surface-double-pt}
\end{figure}

\begin{enumerate}
    \item While the terminology and treatment of $3$-sheets and $2$-sheets in the domain (corresponding to $2$-cells and $1$-cells in the Reeb space, respectively) by Tierny \etal's algorithm appear to be sound, the handling of $0$- and $1$-sheets is limited to the computation of the Jacobi set in the domain (see Section~3.1 in \cite{2016-Julien-ReebSpace}). However, as our analysis and the example in \figref{fig:jacobi-fiber-surface-double-pt} (and \figref{fig:violation-second-Morse-condition}) demonstrate, computing only the Jacobi set is insufficient for correctly capturing the Reeb space. One must also consider its image under the quotient map, that is, the \emph{Jacobi structure} in the Reeb space (see \secref{subsec:background-jacobi-structure} of the present paper). 
    
    More specifically, complications arise when two Jacobi set edges, say $\e$ and $\e'$, are mapped to intersecting arcs in the Reeb space, even though $\e$ and $\e'$ are disjoint in the domain (as shown by the blue lines in \figref{fig:jacobi-fiber-surface-double-pt}(a)). Such intersections may occur at a point in the Reeb space (as shown by the red point as the intersection of blue lines in \figref{fig:jacobi-fiber-surface-double-pt}(b)) for which the singular fiber does not contain a vertex of the domain mesh; instead, the singular fiber intersects $\e$ and $\e'$ in their interiors. In this case, when Tierny \etal's algorithm traces the Jacobi fiber surface starting from $\e$, it eventually intersects $\e'$, but the algorithm does not clarify how such situations are handled. These intersections correspond to complex singular fibers (red singular fiber in \figref{fig:jacobi-fiber-surface-double-pt}(a))---potentially involving multiple sheets---that do not originate from mesh vertices but arise due to the geometry of the Jacobi structure.

    Furthermore, the construction of the Reeb space by Tierny \etal~ introduces $0$-sheets only at non-manifold vertices of the Jacobi set (i.e., vertices of the Jacobi set which are not adjacent to exactly two of its edges), but our example in \figref{fig:jacobi-fiber-surface-double-pt}  demonstrates that additional ``new'' vertices or $0$-cells are necessary to represent crossing points in the Jacobi structure (referred to as \textbf{double points}). This omission can lead to an incorrect or incomplete representation of the Reeb space. We thus identify this as a fundamental limitation---or possibly a flaw---in Tierny \etal's algorithm: \emph{it fails to handle Jacobi structure crossings in the Reeb space, which are essential for constructing correct Jacobi fiber surfaces and accurate Reeb space topology}. Our algorithm resolves this issue, as detailed in \secref{subsec:computing-Jacobi-structure}.

    \item Tierny \etal~\cite{2016-Julien-ReebSpace} state that each $3$-sheet corresponds to a $2$-cell in the Reeb space. However, their algorithm does not make clear how these $2$-cells are attached to one another. While attachment via Jacobi fiber surfaces may work in simple configurations, it becomes problematic in the presence of more intricate singular fibers---as shown in the example \figref{fig:jacobi-fiber-surface-double-pt}. In such cases, without a precise treatment of the Jacobi structure, it is unclear how the $2$-cells should be consistently glued near these singularities. In other words, the algorithm does not describe how the $0$-, $1$-, and $2$-cells are assembled to form a coherent Reeb space. The term ``adjacent'' is used informally, without any rigorous definition of how adjacency is determined or enforced in \cite{2016-Julien-ReebSpace}. This lack of detail leaves the combinatorial structure of their Reeb space construction ambiguous and, in complex cases, potentially incorrect. Our algorithm addresses this issue explicitly, as detailed in \secref{subsec:Compute-Reeb-Space}.
\end{enumerate}

\subsection{Problem Statement}
\label{subsec:problem-stmt}
In this paper, we address the problem of fast and correct computation of the Reeb space $\RS_\f$ associated with a generic PL bivariate field $\f = (f_1, f_2): \M \rightarrow \mathbb{R}^2$, where $\M$ is a triangulation of a compact, orientable $3$-manifold without boundary (or an orientable combinatorial $3$-manifold without boundary). Generically, the Reeb space of a PL bivariate field is a $2$-dimensional polyhedron consisting of $2$-dimensional sheets joined along $1$-dimensional components of the \emph{Jacobi structure}, which is the projection of the Jacobi set into the Reeb space. These attachments may occur in complex ways, reflecting the fiber topology (or the topology of intersections of the level sets). Combinatorially, the Reeb space $\RS_\f$ can be described as a $2$-dimensional CW-complex (or polyhedral complex) composed of $0$-, $1$-, and $2$-sheets (or cells), derived from the connectivity of fiber components in the domain. The goal of this paper is to design an algorithm that correctly computes the combinatorial structure of the Reeb space for a PL bivariate field $\f$ under a set of genericity assumptions. 

Our algorithm assumes the following genericity conditions on the input field $\f$:  
(i) $\f$ is a simple PL bivariate field;  
(ii) $f_1$ is PL Morse; and  
(iii) the restrictions of $f_2$ to the contours (where a contour is a connected component of a level set) of $f_1$ are PL Morse, except at a finite number of contours.
A PL bivariate field $\f$ is said to be \emph{simple} if it is generic, and every $1$-simplex in its Jacobi set is a simple critical edge \cite{book-herbert-computational-topology}. The field $\f$ is \emph{generic} if the image of each $i$-simplex under $\f$ is an $i$-simplex for $i = 0, 1, 2$. The PL Morse conditions in (ii) and (iii) are essential for constructing the Reeb graphs of $f_1$, and of $f_2$ restricted to the contours of $f_1$, respectively. Additionally, we assume that there is at most one violation of the PL Morse condition in (iii) for any given contour of $f_1$. The assumption in (iii)—that only a finite number of such violations may occur—is inspired by \emph{Cerf theory} in differential topology and singularity theory, which studies families of smooth real-valued functions on smooth manifolds~\cite{cerf-1968-diffeomorphismes}. A key challenge addressed in this paper is characterizing these violations of the PL Morse condition in the family of functions $f_2$ restricted to the contours of $f_1$, as they play a central role in correctly constructing the Reeb space.
The above assumptions also imply that the Jacobi set of $\f$ forms an embedded PL $1$-manifold (or a $1$-dimensional PL submanifold) in $\M$~\cite{2004-edelsbrunner-jacobi-set}. Our current algorithm does not handle degenerate cases where multiple genericity violations occur along a single contour of $f_1$. However, standard perturbation techniques, such as the \emph{simulation of simplicity} framework by Edelsbrunner~\etal~\cite{edelsbrunner-simulation-of-simplicity}, can be employed to address such cases.

Our approach to computing the Reeb space is based on constructing the \emph{Multi-Dimensional Reeb Graph (MDRG)}, a hierarchical decomposition of the Reeb space into lower-dimensional Reeb graphs. To this end, we first address the theoretical problem of showing that the MDRG is topologically equivalent (homeomorphic) to the Reeb space.
Next, we investigate the problem of correctly identifying points on the first-dimensional Reeb graph at which the topology of the second-dimensional Reeb graphs changes. This leads to four core algorithmic problems necessary for computing the Reeb space:
\begin{enumerate}
    \item Computing the correct \emph{Jacobi structure} by projecting the Jacobi set edges and identifying its intersections in the Reeb space, which form the $0$-sheets and $1$-sheets in the Reeb space;
    \item Computing the correct \emph{MDRG}, where the second-dimensional Reeb graphs are embedded within the Reeb space;
    \item Constructing a \emph{net-like structure} by connecting the second-dimensional Reeb graphs using the  Jacobi structure embedded in the Reeb space; and
    \item Computing the $2$-sheets (or, $2$-cells) of the Reeb space within this net-like structure.
\end{enumerate}
Finally, we consider providing a formal \emph{proof of correctness} and a \emph{complexity analysis} of our algorithm.

\subsection{Contributions}
\label{subsec:contributions}
The core contribution of this paper is a complete and provably correct algorithm for computing the Reeb space of a generic PL bivariate field on a triangulated compact, orientable $3$-manifold $\M$ without boundary. Central to this is the introduction and theoretical validation of the Multi-Dimensional Reeb Graph (MDRG) framework, which decomposes the Reeb space hierarchically and enables its correct and efficient computation without field quantization. Our specific contributions are as follows:

\begin{itemize}
    \item We provide a mathematical proof that the MDRG of a bivariate field is homeomorphic to its corresponding Reeb space. This foundational result ensures that the MDRG accurately captures the topology of the Reeb space (\secref{subsec:homeomorphism-proof}).

    \item We characterize the discrete set of points on the first-dimensional Reeb graph where the topology of the second-dimensional Reeb graphs changes in the MDRG hierarchy. This characterization is critical for ensuring the correctness of the MDRG construction (\secref{subsec:mdrg-topological-changes}).

    \item We design an algorithm to compute the \emph{Jacobi structure} by projecting the Jacobi set of the PL bivariate field and identifying its intersections in the Reeb space (\secref{subsec:computing-Jacobi-structure}).

    \item We present an algorithm for the correct computation of the MDRG of a PL bivariate field using the computed Jacobi structure, without requiring any field quantization (\secref{subsec:algorithm-computing-MDRG}).

    \item We propose an algorithm for constructing a \emph{net-like structure} in the Reeb space by connecting the second-dimensional Reeb graphs along the Jacobi structure (\secref{subsec:Compute-Net-Structure}).

    \item Based on this net-like structure, we develop an algorithm to reconstruct the full Reeb space by computing its $2$-sheets (\secref{subsec:Compute-Reeb-Space}). We also provide a formal proof of the correctness of this construction (\secref{subsec:proof-correctness}).

    \item Finally, we analyze the computational complexity of the entire Reeb space computation pipeline (\secref{sec:complexity-analysis}).
\end{itemize}

\paragraph*{Overview}
\secref{sec:background} offers the essential background for understanding the proposed algorithm. This section outlines computing critical points and the Reeb graph of a PL scalar field. Then it provides a background of the Jacobi set and Reeb space as generalizations to PL multi-fields.  Next, it introduces multi-dimensional Reeb graph, Jacobi structure, and time-varying Reeb graphs which are important to understand the rest of our paper.
\secref{sec:theory-contributions} provides two important theoretical contributions of the paper. First,
a mathematical proof of homeomorphism between the Reeb space and the MDRG for a generic PL bivariate field is given in \secref{subsec:homeomorphism-proof}. Then \secref{subsec:mdrg-topological-changes} provides characterizations of the topological change points on the first-dimensional Reeb graph of the MDRG.
\secref{sec:MDRG} provides our main algorithm for computing the correct Reeb space of a generic PL bivariate field and a proof of topological correctness of the computed Reeb space. 
 In \secref{sec:complexity-analysis}, we provide the complexity analysis of our algorithm by analyzing each of the sub-parts for computing the Reeb space of a PL bivariate field. 
Finally, in \secref{sec:conclusion}, we conclude by discussing the main contributions and future works of the current paper.

\section{Background}
\label{sec:background}
In this section, we describe the necessary background of  scalar and multi-field topology defined on a smooth, compact, orientable $d$-dimensional manifold $\cM$ without boundary. For the current paper, we need to consider $d=3$ and $d=2$. Since most of the real data comes as a discrete set of real numbers at the grid
points (vertices) of a mesh, we consider a simplicial complex approximation of $\cM$.

\subsection{Simplicial Complex}
\label{subsec:simplicial-complex}
An \emph{i-simplex} $\sigma$ is the convex hull of a set $S$ of $i + 1$ affinely independent points, and its dimension is $i$ \cite{book-herbert-computational-topology}. A \emph{face} of $\sigma$ is the convex hull of a non-empty subset of $S$. A \emph{simplicial complex} $K$ is a finite collection of simplices, where the faces of a simplex in $K$ also belong to $K$, and the intersection of any two simplices in $K$ is either empty or a face of both the simplices. For a simplex $\sigma \in K$, its \emph{star} is denoted by $\Star{\sigma}$, and is defined as the set of simplices which contain $\sigma$ as a face. The \emph{closed star} of $\sigma$ is obtained by adding all the faces of the simplices in $\Star{\sigma}$. The \emph{link} of $\sigma$, denoted as $\Link{\sigma}$, is the set of simplices belonging to the \emph{closed star} of $\sigma$ that do not intersect $\sigma$. 
Let $|K|$ be the underlying space described by $K$.
If there exists a homeomorphism $h: |K|\rightarrow\cM$, then we say $\M=(|K|,h)$ is a triangulation or mesh of $\cM$.
Further, $\M$ is a \emph{combinatorial} $d$-manifold if the link of every $i$-simplex in $\M$  triangulates a combinatorial $(d - i - 1)$-sphere \cite{2008-edels-reebspace}.

\subsection{PL Scalar Field}
\label{subsec:scalar-field}
Scalar data is usually presented as a discrete set of real values at the vertices of a triangulation $\M$ corresponding to the $d$-manifold $\cM$. \textcolor{black}{Note, $\M$ is a combinatorial $d$-manifold, where} the vertex set of $\M$ is represented as $V(\M) = \{\bv_0, \bv_1, \ldots, \bv_{n_v-1}\}$, where $n_v$ is the number of vertices in $\M$. The discrete scalar data can be mathematically represented by a function $\hat{f}: V(\M) \rightarrow \R$. From this discrete map $\hat{f}$, a piecewise-linear (PL) scalar field $f: \M \rightarrow \R$ can be obtained as follows. At the vertices of $\M$, $f$ takes the values of $\hat{f}$, and the values in higher dimensional simplices are determined through linear interpolation. The PL scalar field $f$ is said to be \emph{generic} if no two adjacent vertices of $\M$ have the same $f$-value.

\subsubsection{PL Critical Point}
\label{subsubsec:pl-critical}
Consider a generic PL scalar field $f: \M \rightarrow \R$. Then, if $\bv$ and $\bv'$ are the endpoints of an edge in $\M$, it follows that $f(\bv) \neq f(\bv')$. The \emph{lower link} of a vertex $\bv$, denoted by $\LowerLink{\bv}$, is the collection of simplices in $\Link{\bv}$ whose vertices have smaller $f$-values than $f(\bv)$. The \emph{upper link} $\UpperLink{\bv}$ is defined, similarly. To determine the type of vertices we compute the reduced Betti numbers of their lower links.

Following the usual convention, the $i$-th Betti number $\beta_i$ is the rank of the $i$-th homology group in $\Z_2$ coefficients. The reduced Betti number, denoted by $\widetilde{\beta_i}$, is obtained as follows. If $i \geq 1$, then $\tilde{\beta_i} = \beta_i$. For $i = 0$ or $-1$, there are two possibilities.
If the lower link is non-empty, then $\tilde{\beta_0} = \beta_0 - 1$ and $\tilde{\beta}_{-1} = 0$. Otherwise, $\tilde{\beta}_0 = \beta_0 = 0$ and $\tilde{\beta}_{-1} = 1$. We note, the reduced Betti numbers $\tilde{\beta_i}$ are non-negative integers. If all reduced Betti numbers of the lower link corresponding to a vertex $\bv$ vanish, then $\bv$ is called a \emph{PL regular} point (vertex) of $f$. Otherwise, $\bv$ is a \emph{PL critical} point (vertex),  and the corresponding function value $f(\bv)$ is a \emph{critical value}. Further,  if the reduced Betti numbers of $\LowerLink{\bv}$  in all dimensions sum up to $1$, then $\bv$ is called a \emph{simple critical} point,
otherwise, $\bv$ is called a \emph{degenerate critical} point. The \emph{index} of a simple critical point $\bv$ is $i$ if $\tilde{\beta}_{i-1} = 1$. 
A simple critical vertex of index $0$ is called a 
 minimum and a simple critical vertex of index $d$ is called a maximum. Any other critical point of index $i$ is called an $i$-saddle when $i$ is an integer that varies from $1$ to $d-1$.
In particular, for $d = 3$ the simple critical vertices of indices $0, 1, 2$ and $3$ are referred to as \emph{minima}, \emph{$1$-saddles}, \emph{$2$-saddles}, and \emph{maxima}, respectively. 
The pre-image $f^{-1}(a)$ corresponding to a level value $a \in \R$ is called the \emph{level set} of $f$, and each connected component of the level set is called a \emph{contour}. A value $a \in \R$ is a \emph{regular value} of $f$ if its level set $f^{-1}(a)$ does not pass through a PL critical point. We note, a generic PL function $f$ is said to be \emph{PL Morse} if: 
\begin{enumerate}[I.]
    \item every critical point of $f$ on $\M$ is simple, and 
    \item no two critical vertices of $f$ on $\M$ lie on  the same level set of $f$.
\end{enumerate}
Next, we discuss the Reeb graph that captures the level-set topology of a PL Morse function.

\subsubsection{Reeb Graph}
\label{subsec:Reeb-graph}
\textbf{Quotient space.} Let $\X$ be a topological space and $\cP$ be a partition of $\X$ corresponding to an equivalence relation $\sim$. A new space $\W$ is called a \emph{quotient space} if (i) each point of $\W$ corresponds to a member of $\cP$ by a mapping, say $q:\X\rightarrow \W$ and (ii) the topology of $\W$ is the largest such that $q$ is continuous. The map $q$ is called the \emph{quotient map}.

For the PL scalar  field $f:\M \rightarrow \R$, a partition of the triangulation $\M$ can be obtained naturally by the equivalence relation: $\x \sim \y$ if and only if $f(\x) = f(\y) = c$, and both $\x$ and $\y$ belong to the same contour of $f^{-1}(c)$. The corresponding quotient space and quotient map are denoted as $\W_f$ and $q_f$, respectively. Thus we obtain a factorization of $f$ as $f=\overline{f}\circ q_f$, where $\overline{f}:\W_f\to\R$. 
In particular, if $f: \M \rightarrow \R$ is a PL Morse function, the quotient space $\W_f$ has a graph structure which is known as \emph{Reeb graph} and is denoted by $\RG_{f}$. If $\M$ is a triangulation corresponding to a simply connected domain, then $\RG_{f}$ has no loop and is called a \emph{contour tree}. A Reeb graph consists of a set of nodes, and arcs connecting the nodes. A point in the Reeb graph is referred to as a \emph{node} if the corresponding contour passes through a critical point of $f$. A point on an arc of the Reeb graph is called a \emph{regular point} if the corresponding contour of $f$ does not contain any critical point of $f$. The degree of a node is defined as the number of arcs incident to it.  The number of such arcs joining adjacent nodes with lesser $\overline{f}$-values is called the \emph{down-degree} of the node and the number of such arcs joining adjacent nodes with higher $\overline{f}$-values is called the \emph{up-degree} of the node. Each node of $\RG_{f}$ is one of the following types \cite{book-herbert-computational-topology}:
\begin{enumerate}[(i)]

\item \emph{minimum} (down-degree: $0$, up-degree: $1$) - corresponding to a minimum of $f$ where a contour \textcolor{black}{starts or is born},
\item  \emph{maximum} (down-degree: $1$, up-degree: $0$) - corresponding to a maximum of $f$ where a contour dies,
\item \emph{down-fork} (down-degree: $2$, up-degree: $1$) - corresponding to a $1$-saddle of $f$ which merges two contours of $f$ into a single contour,   
\item \emph{up-fork} (down-degree: $1$, up-degree: $2$) - corresponding to an index $d-1$ saddle of $f$ (here, $d$ is the dimension of the PL manifold $\M$) which splits a contour of $f$ into two contours, and
\item \emph{degree-2 critical node} (up-degree: $1$, down-degree: $1$) - corresponding to other critical points of indices between $1$ and $d-1$ which correspond to a change in the genus and not in the number of contours.
\end{enumerate}
A Reeb graph with degree-$2$ critical nodes is also known as an \emph{augmented Reeb graph}. Since $f$ is PL Morse, there is a one-to-one correspondence between critical points of $f$ and nodes of augmented $\RG_{f}$.  We denote the collection of nodes and arcs of an augmented $\RG_{f}$ by $V(\RG_{f})$ and $Arcs(\RG_{f})$, respectively. 
The evolution of the level set topology of $f$, for increasing values of $f$, can be traced by its Reeb graph. In particular, for $d=3$, a minimum node of $\RG_{f}$ corresponds to a minimum point where a contour \textcolor{black}{is born}. Similarly, a maximum node corresponds to a maximum point where a contour dies. A down-fork corresponds to a $1$-saddle where two contours merge into a single contour. Similarly, an up-fork corresponds to a $2$-saddle where a contour splits into two contours. A degree-$2$ node indicates a change in the genus of the contour, and the corresponding critical points are also known as genus-change critical points \cite{2003-Chiang-Simplification-Tetrahedral-Meshes}.

\noindent
\textbf{Computing Reeb Graphs.} Numerous algorithms for computing Reeb graphs are available in the literature. Here, we spotlight a few of them. 
Harvey \etal~\cite{2010-Harvey-Reeb-Graph} presented a randomized algorithm to compute the Reeb graph of a PL Morse function $f$ defined on a \textcolor{black}{combinatorial $2$-manifold $\M$} by collapsing the contours of $f$ in random order.  The expected time complexity of the algorithm is $\cO(n \log n)$, where $n$ is the number of simplices in $\M$. Parsa~\etal~\cite{2012-Parsa-Reeb-Graph} introduced a method that involves sweeping the vertices in $\M$ (the input simplicial complex) with increasing values of $f$ and monitoring the connected components of the level sets of $f$. The changes in level set correspond to the merge, split, creation, or removal of components in the Reeb graph. The time complexity of the algorithm is $\cO(n \log n)$, where $n$ is the number of simplices in the $2$-skeleton of $\M$ (i.e. union of simplices of $\M$ of dimensions $\leq 2$). Doraiswamy \etal \cite{2012-Doraiswamy-Reeb-Graph} devised a Reeb graph computation algorithm by first partitioning the input domain into interval volumes, each having Reeb graphs without loops. Then, the contour trees corresponding to each of the subdivided volumes are constructed,  and these are interconnected to obtain the Reeb graph.  The algorithm has a time complexity of $\cO(n_v \log(n_v) + s n_t)$, where $n_v$ and $n_t$ represent the numbers of vertices and triangles in the input triangle mesh, respectively, and $s$ is the number of saddles.

In the current paper, we need to encode the genus-change critical points (degree-$2$ critical nodes) in the Reeb graph as they are essential for computing the correct multi-dimensional Reeb graph and the Reeb space (see \secref{sec:MDRG} for more details). Therefore, we construct the augmented Reeb graph, by projecting these genus-change saddle points on $\RG_{f}$ as discussed by Chiang \etal \cite{2003-Chiang-Simplification-Tetrahedral-Meshes}. For the identification of genus-change saddle points, we test the criticality of each vertex in $\M$, and identify the saddle points that map to the interior of an arc in $\RG_{f}$ by the quotient map $q_{f}$. The augmented Reeb graph is obtained by subdividing arcs of $\RG_{f}$ based on the insertion of degree-$2$ nodes corresponding to these saddle points. In our algorithm in \secref{sec:MDRG},
the procedure {\sc ConstructReebGraph} computes the ordinary Reeb graph without augmentation and \textsc{AugmentReebGraph} procedure computes an augmented Reeb graph with additional points of topological changes, including the genus change critical points.

\subsection{PL Multi-Field}
\label{subsec:PL-Multi-Field}
Analogous to the definition for PL scalar field, a PL multi-field $\f = (f_1, \ldots, f_r): \M \rightarrow \R^r$  on the triangulation $\M$ corresponding to the $d$-manifold $\mathcal{M}$ (with $d$ $\geq r \geq 1$) is defined at the vertices of $\M$  and linearly interpolated within each simplex of $\M$.  
The preimage of the map $\f$ associated with a value $\bc \in \R^r$, denoted as $\f^{-1}(\bc)$, is known as a \emph{fiber}, and each connected component of a fiber is referred to as a \emph{fiber-component} \cite{2004-Saeki-topoology-of-singulat-fibers,2014-saeki-visualizing-multivariate-data}. 
Specifically, in the case of a scalar field, these are called \emph{level sets} and \emph{contours}, respectively (see \secref{subsec:Reeb-graph} for more details). We assume that $\f$ is a \emph{generic PL mapping}: i.e., the image of every $i$-simplex $\sigma$ of dimension at most $r$ is an $i$-simplex. Specifically, for $r = 1$ and $r = 2$, $\f$ is called a generic PL scalar and a generic PL bivariate field, respectively.

Next, we briefly introduce the Jacobi set which is the generalization of the notion of critical points for the multi-fields.

\subsubsection{Jacobi Set}
\label{subsubsec:jacobiset}
The \emph{Jacobi set} is an extension of the notion of critical points for multi-fields \cite{2004-edelsbrunner-jacobi-set}. Intuitively, the Jacobi set of the multi-field, comprising $r$ functions, is the collection of critical points of one function restricted to the intersection of the level sets of the remaining $r-1$ functions. For a generic PL multi-field $\f: \M \rightarrow \R^r$, its Jacobi set consists of $(r-1)$-simplices of $\M$ which are critical. We briefly describe the determination of these critical simplices here and refer the readers to \cite{2008-edels-reebspace} for more details.

Let $\sigma$ be an $(r-1)$-simplex of $\M$. Consider a unit vector $\bu$ in the $(r-1)$-sphere $\bS^{r-1}$, and let $h_{\bu} : \M \rightarrow \R$ be the PL function defined as $h_{\bu}(\x) = \langle \f(\x), \bu \rangle$, which is the height of the image of $\x$ in the direction $\bu$. If the value of $h_{\bu}$ is constant on the simplex $\sigma$ in $\M$, the lower (upper) link of $\sigma$ consists of simplices in the link of $\sigma$ having $h_{\bu}$-values 
strictly less (greater) than the values at the vertices of $\sigma$. From the genericity condition,  the upper and lower links of $\sigma$ cover all vertices of  $\Link{\sigma}$ \cite{2016-Julien-ReebSpace}. Then by applying reduced homology of the lower link, as discussed in \secref{subsubsec:pl-critical}, we determine whether the simplex $\sigma$ is regular or critical for $h_{\bu}$. Furthermore, it can be determined whether a critical simplex is simple critical or not. 

If $\sigma$ is an $(r - 1)$-simplex, then precisely two unit vectors exist for which their height functions remain constant on $\sigma$. Specifically, these vectors are the unit normals $\bu$ and $-\bu$ corresponding to the image of $\sigma$ in $\R^r$. The lower link of $\sigma$ for the height function $h_{\bu}$ is its upper link for the other height function $h_{-\bu}$. We note, $\sigma$ has essentially only a single chance to be critical, as it is critical for $h_{\bu}$ if and only if it is critical for $h_{-\bu}$. We say that an $(r-1)$-simplex $\sigma$ is \emph{critical} if it is critical for some $h_\bu$, otherwise it is \emph{regular}. The \emph{Jacobi set} of $\f$, denoted by $\JSet_{\f}$, consists of the set of critical $(r-1)$-simplices in $\M$, along with their faces. A point $\x \in \M$ is a \emph{singular (critical) point} of $\f$ if $\x \in \JSet_{\f}$ and $\f(\x)$ is a \emph{singular (critical) value}. Otherwise, $\x$ is said to be a \emph{regular point}. A point $\y \in \R^r$ is said to be a \emph{regular value} if $\f^{-1}(\y)$ does not contain a singular point. We note, the preimage of a singular value is termed as a \emph{singular fiber}, while the preimage of a regular value is known as a \emph{regular fiber}. A fiber-component is categorized as a \emph{singular fiber-component} if it traverses a singular point. Otherwise, it is called a \emph{regular fiber-component}. It should be noted that a singular fiber may include one or more regular fiber-components.

A generic PL multi-field $\f$ is said to be \emph{simple} if every ($r-1$)-simplex of $\JSet_{\f}$ is simple critical. If $\f$ is a simple PL multi-field, then for sufficiently small values of $r$, $\JSet_{\f}$ is a PL $(r-1)$-dimensional manifold \cite{2008-edels-reebspace, 1973-Golubitsky-Stable-Maps}. This paper deals with simple PL bivariate fields and assumes that the Jacobi set is a PL $1$-manifold. The procedure \textsc{ComputeJacobiSet} provides the pseudo-code for computing the Jacobi set $\JSet_{\f}$ of a bivariate field $\f$ defined on $\M$ which will be used in \secref{sec:MDRG}.

\begin{algorithmic}[1]
\Procedure{ComputeJacobiSet}{$\M, \f$}
\label{proc:Jacobi-Set}
\State $\JSet_{\f} \gets \emptyset$
\For{each edge $\e$ of $\M$}
	\State  Compute the unit normal $\mathbf{n}$ corresponding to $\f(\e)$
    \State $\LowerLink{\mathbf{n}} \gets \textsc{ComputeLowerLink}(\mathbf{n})$
	\If{$\exists i \geq 0$ such that the reduced Betti number $\widetilde{\beta_i}$ of $\LowerLink{\mathbf{n}}$ is non-zero }
	
		\State Add $\e$ to $\JSet_{\f}$
	\EndIf
\EndFor
\State\Return{$\JSet_{\f}$}
\EndProcedure
\end{algorithmic}

Next, we briefly describe the Reeb space which captures the topology of a multi-field.

\subsubsection{Reeb Space}
\label{subsec:background-ReebSpace}
For a generic PL multi-field $\f:\M \rightarrow \R^r$, and a point $\bc \in \R^r$, the inverse image $\f^{-1}(\bc)$ is called a \emph{fiber}, and each connected component of $\f^{-1}(\bc)$ is called a \emph{fiber-component} \cite{2004-Saeki-topoology-of-singulat-fibers, 2014-saeki-visualizing-multivariate-data}. 
We note, a fiber-component of $\f$ can be considered as an equivalence class determined by an equivalence relation $\sim$ on $\M$. Here, two points $\x, \y \in \M$ are considered equivalent (or $\x \sim \y$) if and only if $\f(\x) = \f(\y) = \bc$, and both $\x$ and $\y$ belong to the same fiber-component of $\f^{-1}(\bc)$.  
The Reeb space of $\f$ is the quotient space $\RS_{\f}$, determined by the quotient  map $q_{\f}: \M \rightarrow \RS_{\f}$, which contracts each fiber-component in $\M$ to a unique point in $\RS_{\f}$ \cite{2008-edels-reebspace}. The Stein factorization of $\f$ is the representation of $\f$ as the composition of $q_{\f}$ and the unique continuous map $\overline{\f} : \RS_{\f} \rightarrow \R^r$. The following commutative diagram depicts the relationship between the maps $\f, q_{\f}$ and $\overline{\f}$.

\begin{figure}
    \centering
    \includegraphics[width = \textwidth]{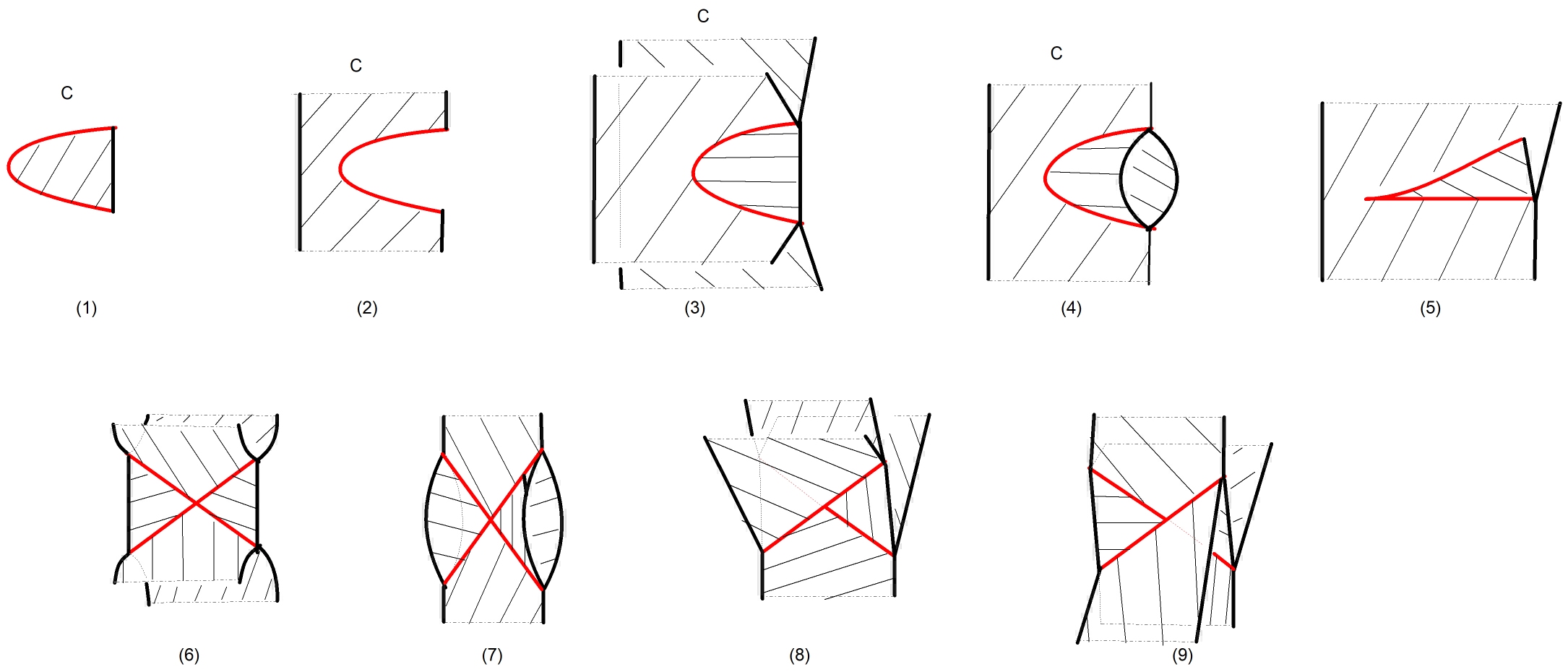}
    \caption{A classification list of local structures of the Reeb space for the smooth stable map case. The horizontal direction corresponds to $pr_1\circ\f$ and the vertical direction to $pr_2\circ\f$, where $pr_i$ projects the range of $\f$ onto the range of $f_i$, for $i=1,2$. 
    Red curves depict the Jacobi structure and the thick graphs on the left and the right-hand sides depict the corresponding Reeb graphs of $f_2$, restricted to contours of $f_1$. Each figure with the letter ``C'' contains the image of exactly one critical point of $f_1$. There are also up-side down or right-left reversed versions (see \cite{Kushner-1984}, \cite{Levine-2006} for more details). }
    \label{fig:RS-classification-smooth}
\end{figure}

\begin{center}
\begin{tikzcd}[column sep=normal]
\M \arrow{dr}[swap]{q_\f}\arrow{rr}{\f} & & \mathbb{R}^r\\
& \RS_\f \arrow{ur}[swap]{\overline{\f}} &
\end{tikzcd}
\end{center}
In particular, 
 the combinatorial structure of the Reeb space $\RS_{\f}$ of a generic PL bivariate field $\f : \M \to \mathbb{R}^2$ can be described as a $2$-dimensional CW-complex (or polyhedral complex), composed of $0$-, $1$-, and $2$-sheets (or cells) \cite{Osamu-2013-Triangulating}. These sheets are derived from the connectivity of the fiber-components of $\f$ in the domain $\M$. Formally, $0$-, $1$-, and $2$-sheets of the $\RS_\f$ can be described as follows:
\begin{itemize}
    \item \textbf{$0$-sheets (or, $0$-cells)}: These correspond to: (i) Intersections of the $q_\f$-images of Jacobi set edges (i.e., double points of the Jacobi structure; see Section~\ref{subsec:background-jacobi-structure}), and
    (ii) $q_\f$-images of vertices of the (PL $1$-manifold) Jacobi set that correspond to critical points of $f_1$ (or $f_2$) restricted to the Jacobi set.

    \item \textbf{$1$-sheets  (or, $1$-cells)}: 
    $1$-sheets are the connected components obtained by partitioning the $q_\f$-image of the Jacobi set in the Reeb space, by excluding the $0$-sheets. 
    
    \item \textbf{$2$-sheets  (or, $2$-cells)}: $2$-sheets are the connected components of the image \( q_\f(\M) \) obtained by removing the $0$- and $1$-sheets. Each $2$-sheet corresponds to a connected component of a regular region in the Reeb space, within which the fiber topology remains invariant—that is, the corresponding region in the domain contains no critical points of \( \f \).
\end{itemize}
The current paper presents an algorithm for computing the $0$-, $1$-, and $2$-sheets of the Reeb space of a simple PL bivariate field on a $3$-manifold without boundary.
\figref{fig:RS-classification-smooth} is a list of possible local structures of the Reeb space of a smooth stable bivariate map $\f$, defined on a smooth closed orientable $3$-manifold without boundary. The horizontal direction corresponds to $pr_1\circ\f$ and the vertical direction to $pr_2\circ\f$, where $pr_i$ projects the range of $\f$ onto the range of $f_i$ for $i=1,2$ (as shown in the commutative diagram in \secref{subsec:homeomorphism-proof}). There are also up-side down and left-right reversed versions (see \cite{Kushner-1984}, \cite{Levine-2006} for more details). 

Next, we describe a multi-dimensional Reeb graph representation of the Reeb space which is used to compute the correct Reeb space in the current paper.

\subsubsection{Multi-Dimensional Reeb Graph}
A Multi-Dimensional Reeb Graph (MDRG) is a hierarchical decomposition of the Reeb space into a collection of lower-dimensional quotient spaces (in particular, Reeb graphs) \cite{2016-Chattopadhyay-CGTA-simplification}. For a Reeb space $\RS_\f$ of a generic PL bivariate field $\f=(f_1,f_2):\M\rightarrow \mathbb{R}^2$, we consider the decomposition as follows. First, we consider the quotient space $\W_{f_1}$ of $f_1$. Then for each  $p\in\W_{f_1}$, we consider the restricted field $\widetilde{f_2^p}\equiv f_2|_{C_p}: C_p\rightarrow \mathbb{R}$, where $C_p:=q_{f_1}^{-1}(p)$, and its corresponding quotient space $\W_{\widetilde{f_2^p}}$. These quotient spaces are shown by the following commutative diagrams: 
\begin{center}
$
\begin{array}{cc}
\begin{tikzcd}[column sep=normal]
\M \arrow{dr}[swap]{q_{f_1}}\arrow{rr}{f_1} & & \R\\
& \W_{f_1} \arrow{ur}[swap]{\overline{f_1}} &
\end{tikzcd} \qquad
\begin{tikzcd}[column sep=normal]
C_p \arrow{dr}[swap]{q_{\widetilde{f_2^p}}}\arrow{rr}{\widetilde{f_2^p}} & & \mathbb{R}\\
& \W_{\widetilde{f_2^p}} \arrow{ur}[swap]{\overline{\widetilde{f_2^p}}}
\end{tikzcd}
\end{array}
$
\end{center}

The hierarchical decomposition of the Reeb Space $\RS_\f$ into the quotient spaces  $\W_{f_1}$ and  $\W_{\widetilde{f_2^p}}$ for each $p\in\W_{f_1}$ is called the Multi-Dimensional Reeb Graph (MDRG) and is denoted by $\MDRG_\f$. Thus the decomposition of the Reeb Space of $\f=(f_1,f_2)$ into an MDRG can be defined as: 
\begin{align}
\MDRG_\f &=\left\{(p_1, p_2): p_1\in \W_{f_1}, p_2\in \W_{\widetilde{f_2^{p_1}}}\right\}.
\end{align}

\noindent
Similarly, for a generic PL multi-field $\f=(f_1, f_2,\ldots,f_r):\M \rightarrow \mathbb{R}^r$ the definition can be generalized as:
\begin{align}
\MDRG_\f &=\left\{(p_1, p_2, \ldots, p_r): p_1\in \W_{f_1}, p_2\in \W_{\widetilde{f_2^{p_1}}},\, \ldots,\, p_r\in \W_{\widetilde{f_r^{p_{r-1}}}} \right\}.
\end{align}

In the current paper, we develop an algorithm for computing the  MDRG (see \secref{subsec:algorithm-computing-MDRG}) for a generic PL bivariate field $(f_1, f_2)$ where we assume $f_1$ is PL Morse and $\widetilde{f_2^{p}}$ is PL Morse except for a discrete set of points $p\in \W_{f_1}$. Under such assumption, the corresponding quotient spaces $\W_{f_1}$ and $\W_{\widetilde{f_2^{p}}}$ are the Reeb graphs, denoted as  $\RG_{f_1}$ and $\RG_{\widetilde{f_2^{p}}}$, respectively. The MDRG is then utilized in computing the correct Reeb space (see \secref{subsec:Compute-Reeb-Space}).

Next, we provide a brief description of the Jacobi structure, which is the projection of Jacobi set into the Reeb space and has a significant role in the correct computation of the Reeb space.

\subsubsection{Jacobi Structure}
\label{subsec:background-jacobi-structure}
The Jacobi structure of the Reeb space $\RS_{\f}$ of a generic PL multi-field $\f: \M \rightarrow \mathbb{R}^r$ is denoted by $\JStruct_{\f}$, and is defined as the projection of $\JSet_{\f}$ to $\RS_{\f}$ by the quotient map $q_{\f}$ \cite{2016-Chattopadhyay-CGTA-simplification}. A point in $\RS_{\f}$ represents a singular fiber-component only if it belongs to $\JStruct_{\f};$ otherwise, it represents a regular fiber-component. Therefore, $\JStruct_{\f}$ partitions the Reeb space into regular and singular components, and thereby plays an important role in capturing the Reeb space topology. As described in \secref{subsec:background-ReebSpace}, generically the Jacobi structure of a bivariate field $\f$ is composed of $0$- and $1$-sheets in the Reeb space. We note, with suitable PL Morse assumptions on the component functions,  each point of the Jacobi structure is guaranteed to appear as a critical node of the lowest level Reeb graphs of an MDRG.
In particular, for a generic PL bivariate field $\f=(f_1, f_2)$ (with suitable PL Morse assumptions on the component functions) the Jacobi structure of $\f$ is captured by the critical nodes of the second dimensional Reeb graphs $\RG_{\widetilde{f_2^{p}}}$ for $p\in \RG_{f_1}$. In the current paper, we assume that the functions $\widetilde{f_2^{p}}$ are PL Morse except at a discrete set of points $p$ on $\RG_{f_1}$. In Section \ref{subsec:mdrg-topological-changes}, 
 we detect these points (where one of the PL Morse conditions is violated) by analyzing the Jacobi structure to track the topological changes in the second-dimensional Reeb graphs of the MDRG.

Next, we briefly outline the topological changes in a time-varying Reeb graph which is a special case of the MDRG.

\subsection{Time-Varying Reeb Graph}
\label{subsec:time-varying-Reeb-graph}
Edelsbrunner~\etal \cite{2008-Edelsbrunner-Time-varying-Reeb-graphs} studied the topological changes in a time-varying Reeb graph of a $1$-parameter family $f: \cM \times \R \rightarrow \R$ of smooth scalar fields based on the Jacobi set of the corresponding bivariate field $(t, f(\x, t)):\cM \times \R \rightarrow \R^2$ where $\cM$ is a $3$-manifold without boundary. The restriction of $f$ to a level set of the first field is denoted by $f_t: \cM \times \{t\} \rightarrow \{t\}\times \R$ and the corresponding time-varying Reeb graph is denoted as $\RG_{f_t}$. 
The nodes of $\RG_{f_t}$ correspond to the critical points of $f_t$ which trace out the segments of the Jacobi structure as $t$ varies. The function 
$f_t$ is assumed to be a Morse function except at a finite set of values of $t$ where one of the Morse conditions may be violated. The topological changes in
 $\RG_{f_t}$, when $t$ varies, are classified into two categories: (i) birth-death of a node - this happens when the Morse condition I is violated in $f_t$ and (ii) swapping of nodes in the Reeb graphs - this happens when the Morse condition II is violated in $f_t$. The birth-death points correspond to points where the Jacobi set and the level sets of the component scalar fields (of the bivariate field) have a common normal. The Jacobi set is decomposed into segments by disconnecting at the birth-death points. It is shown that the indices of critical points remain the same on a segment and the indices of two critical points created or destroyed at a birth-death point differ by one. This is stated as \emph{index lemma} as follows.

\begin{lemma}
\textbf{Index Lemma} \cite{2008-Edelsbrunner-Time-varying-Reeb-graphs}: If $f: \cM \times \R \rightarrow \R$ is a $1$-parameter family of Morse functions, then at a birth-death point, the indices of the two critical points which are created or destroyed differ by exactly one.
\label{lem:index-lemma}
\end{lemma}
We utilize these observations in constructing the MDRG of a generic PL bivariate field $\f = (f_1, f_2)$. Specifically, we compute the points in the Reeb graph of $f_1$, where there is a change in the topology of the second-dimensional Reeb graphs. We observe that these points are associated with critical points of $f_1$ restricted to the Jacobi set, and the double points of the Jacobi structure $\JStruct_{\f}$ as stated in Lemma~\ref{lem:topological-changes}.
In the next section, we provide two important theoretical contributions which are key to develop
our Reeb space algorithm. 

\section{Theoretical Contributions}
\label{sec:theory-contributions}
The current paper introduces an algorithm for computing the correct Reeb space of a generic PL bivariate field based on the  MDRG. This stems from two theoretical or mathematical contributions - (1) homeomorphism between the Reeb space and the MDRG and (2) characterization of topological change points on the first-dimensional Reeb graph of the  MDRG, which we discuss in the next two subsections.

\subsection{Homeomorphism between the Reeb Space and the MDRG}
\label{subsec:homeomorphism-proof}
In this subsection, we prove that the MDRG corresponding to a bivariate field is homeomorphic to its Reeb space. Consider a continuous map $\f = (f_1, f_2) : \cM \to \R^2$.
Note, $\cM$ is a $d$-dimensional manifold and
$d \geq 2$. (However, in the 
statements and proofs of this section, $\cM$ can be any
topological space.) Let us define $\omega_i : \RS_\f \to \W_{f_i}$, $i = 1, 2$, as
follows. Take $p \in \RS_\f$.
Set $\br = \overline{\f}(p) \in \R^2$ and $\br = (r_1, r_2)$.
The point $p$ corresponds to a connected component of 
$$\f^{-1}(\br) = f_1^{-1}(r_1) \cap f_2^{-1}(r_2).$$
This is a nonempty connected subset of $f_i^{-1}(r_i)$: therefore,
it is contained in a unique connected component of $f_i^{-1}(r_i)$ for $i=1, 2$.
This corresponds to a point in $\W_{f_i}$, which we define to be
$\omega_i(p)$. By this description, we see easily that
$\omega_i$ is well defined.

By definition, it is clear that $\omega_i \circ q_\f = q_{f_i}$.
As $\RS_\f$ and $\W_{f_i}$ are endowed with the quotient topologies, we
see immediately that $\omega_i$ is continuous. Thus we have the following commutative diagram of continuous maps:

\begin{center}
\begin{tikzpicture}[commutative diagrams/every diagram]
\node (P0) at (0+175:0.3cm) {$\cM$};
\node (P1) at (0+60:3cm) {$\R^2$} ;
\node (P2) at (0+60*2:3cm) {$\R$};
\node (P3) at (0+3*60:3cm) {$\W_{f_2}$};
\node (P4) at (0+4*60:3cm) {$\RS_\f$};
\node (P5) at (0+5*60:3cm) {$\W_{f_1}$};
\node (P6) at (0+6*60:3cm) {$\R$};
\path[commutative diagrams/.cd, every arrow, every label]
(P0) edge[color=black] node {$\f$} (P1)
(P0) edge node[swap] {$f_2$} (P2)
(P0) edge node[swap] {$q_{f_2}$} (P3)
(P0) edge[color=black] node[swap] {$q_\f$} (P4)
(P0) edge node[swap] {$q_{f_1}$} (P5)
(P0) edge node {$f_1$} (P6)
(P1) edge node[swap] {$pr_2$} (P2)
(P1) edge node {$pr_1$} (P6)
(P3) edge node {$\overline{f_2}$} (P2)
(P4) edge node {$\omega_2$} (P3)
(P4) edge node[swap] {$\omega_1$} (P5)
(P5) edge node[swap] {$\overline{f_1}$} (P6)
(P4) edge[bend right, dashed, color=black]
node[swap] {$\overline{\f}$} (P1);
\end{tikzpicture}
\end{center}

Note that $pr_i$ projects the range of the map $\f$ onto the range of $f_i$, for $i=1,2$.
Next, we provide the proof of homeomorphism between $\RS_\f$ and $\MDRG_\f$.

\begin{lemma}
For $p_1 \in \W_{f_1}$, the space
$\W_{\widetilde{f_2^{p_1}}}$ can be identified
with the subspace $\omega_1^{-1}(p_1)$ of $\RS_\f$
in a canonical way.
\label{lem:homeomorphism-embedding}
\end{lemma}

\begin{proof}
Recall that $\widetilde{f_2^{p_1}} = f_2|q_{f_1}^{-1}(p_1)$.
Let us first observe that $\W_{\widetilde{f_2^{p_1}}}$
can be regarded as a subspace of $\RS_\f$. First,
a point in $\W_{\widetilde{f_2^{p_1}}}$ corresponds
to a connected component of $(f_2|q_{f_1}^{-1}(p_1))^{-1}(r_2)
= q_{f_1}^{-1}(p_1) \cap f_2^{-1}(r_2)$
for some $r_2 \in \R$. This component coincides with a unique
connected component of $\f^{-1}(r_1, r_2)
= f_1^{-1}(r_1) \cap f_2^{-1}(r_2)$, where $r_1 = \overline{f}_1(p_1)$,
since $q_{f_1}^{-1}(p_1)$ is a connected component
of $f_1^{-1}(r_1)$. This corresponds to a unique point
of $\RS_\f$. Furthermore, the mapping $\varphi : \W_{\widetilde{f_2^{p_1}}}
\to \RS_\f$ thus obtained is obviously injective, since
a point in $\W_{\widetilde{f_2^{p_1}}}$ and its associated
point in $\RS_\f$ both correspond to the same 
connected component of an $\f$-fiber. Furthermore,
the identification is canonical in this sense. In the following, we canonically identify  $\W_{\widetilde{f_2^{p_1}}}$ with its image by $\varphi$ as a set.

Then, by definition, we see that $\omega_1(x) = p_1$
for every $x \in \W_{\widetilde{f_2^{p_1}}}$.
Therefore, we have
$$\W_{\widetilde{f_2^{p_1}}} \subset \omega_1^{-1}(p_1).$$

On the other hand, for a point $y \in \RS_\f$,
suppose $\omega_1(y) = p_1$. Set $\overline{\f}(y) = (r_1, r_2)
\in \R^2$. Then, $y$ corresponds to a connected
component of $\f^{-1}(r_1, r_2) = f_1^{-1}(r_1) \cap
f_2^{-1}(r_2)$. As $\omega_1(y) = p_1$,
this is a connected component of $q_{f_1}^{-1}(p_1) \cap
f_2^{-1}(r_2)$. This can be regarded
as a point of $\W_{\widetilde{f_2^{p_1}}}$.
Thus, we have $\W_{\widetilde{f_2^{p_1}}} = \omega_1^{-1}(p_1)$
as sets.

Let us now prove that their topologies coincide.
For this, we need to show that the canonical injection
$\varphi : \W_{\widetilde{f_2^{p_1}}}
\to \RS_\f$ is actually an embedding.
Since $\varphi \circ q_{\widetilde{f_2^{p_1}}}
= q_\f|q_{f_1}^{-1}(p_1)$, we see that
$\varphi$ is continuous.

Let us take a closed subset $C$ of $\W_{\widetilde{f_2^{p_1}}}$.
By definition, $q_{\widetilde{f_2^{p_1}}}^{-1}(C)$
is a closed subset of $q_{f_1}^{-1}(p_1)$.
As $q_{f_1}^{-1}(p_1)$ is a closed subset of $\cM$,
this means that $q_{\widetilde{f_2^{p_1}}}^{-1}(C)$
is a closed subset of $\cM$.
Note that $q_\f^{-1}(\varphi(C)) = q_{\widetilde{f_2^{p_1}}}^{-1}(C)$.
This implies that $\varphi(C)$ is a closed
subset of $\RS_\f$. Thus, this is also a closed subset
of the image of $\varphi$. Hence, $\varphi$ is a closed map.

Consequently, $\varphi$ is a homeomorphism onto its image, i.e. an embedding.
This completes the proof. 
\end{proof}

Then, by the definition of the multi-dimensional
Reeb graph together with the above lemma, we have
\begin{align}
    \mathrm{MDRG}_\f = \{(p_1, p_2)\,|\,
p_1 \in \W_{f_1}, \, p_2 \in \omega_1^{-1}(p_1)\}.
\end{align}
As $p_1 = \omega_1(p_2)$ for $p_2 \in \omega_1^{-1}(p_1)$,
and $p_2$ sweeps out all the points of $\RS_\f$
as $p_1$ ranges over all the points of $\W_{f_1}$,
we see that this space coincides with
$$\Gamma = \{(\omega_1(p_2), p_2) \,|\, p_2 \in \RS_\f\}
\subset \W_{f_1} \times \RS_\f,$$
which is endowed with the product topology.

\begin{remark}
In fact, $\mathrm{MDRG}_\f$ is
topologized through the above identification with $\Gamma$.
\end{remark}

Let us define the map $h : \RS_\f \to \Gamma$
by $h(p) = (\omega_1(p), p)$ for $p \in \RS_\f$.
This is obviously continuous and bijective.
Furthermore, the inverse map of $h$
is given by the restriction to $\Gamma$ of the projection
$\W_{f_1} \times \RS_\f \to \RS_\f$ to the second factor,
and is therefore continuous.
This implies that $h$ is a homeomorphism.
Thus, we get the following proposition.

\begin{proposition}
\label{prop:homeo}
$\mathrm{MDRG}_\f
= \{(p_1, p_2) \, | \, p_1 \in \W_{f_1}, \,
p_2 \in \W_{\widetilde{f_2^{p_1}}}\}$
is homeomorphic to $\RS_\f$.
\end{proposition}

\textbf{Genericity Conditions:} In the current paper, we develop an algorithm for computing the correct Reeb space by computing the corresponding correct MDRG of a PL bivariate field $\f = (f_1,f_2): \M\rightarrow \R^2$ where $\M$ is a triangulation of a compact, orientable  $3$-manifold $\cM$ without boundary. To develop our algorithm, we assume $\f$ satisfies the following genericity conditions.
\begin{enumerate}[(i)]
    \item $\f = (f_1,f_2)$ is a simple PL multi-field,
    \item $f_1$ is PL Morse. Under such assumption, the corresponding quotient space $\W_{f_1}$ is the Reeb graph, denoted as  $\RG_{f_1}$.
    \item The functions $\widetilde{f_2^{p}}$ are PL Morse except at a finite set of points $p$ on $\RG_{f_1}$. Under such assumption, the corresponding quotient spaces $\W_{\widetilde{f_2^{p}}}$ are the Reeb graphs,  denoted as $\RG_{\widetilde{f_2^{p}}}$. Moreover, we assume that at most one of the PL Morse conditions of $\widetilde{f_2^{p}}$ is violated at each point.
\end{enumerate}
Now for the correct computation of the MDRG, we need to detect all the points on the first-dimensional Reeb graph $\RG_{f_1}$ where the topology of the family of second-dimensional Reeb graphs $\RG_{\widetilde{f_2^{p}}}$ changes when $p$ varies over $\RG_{f_1}$. The next section provides the theoretical results for characterizing such topological change points on the first-dimensional Reeb graph of MDRG.

\subsection{Characterizing the Points of Topological Change on $\RG_{f_1}$}
\label{subsec:mdrg-topological-changes}
 In this subsection, we provide a method for characterizing (or detecting) the set $P$ of points in $\RG_{f_1}$ where the topology of $\RG_{\widetilde{f_2^p}}$ changes as $p$ varies in $\RG_{f_1}$. 
 First, we provide the following definition of topological equivalence between two Reeb graphs $\RG_{\widetilde{f_2^{p_1}}}$ and $\RG_{\widetilde{f_2^{p_2}}}$ for $p_1, p_2\in \RG_{f_1}$.

\begin{definition}
\label{def:topo-equivalent}
 Two Reeb graphs $\RG_{\widetilde{f_2^{p_1}}}$ and
 $\RG_{\widetilde{f_2^{p_2}}}$ are \emph{topologically
 equivalent} if there exists a homeomorphism
 $\Phi : \RG_{\widetilde{f_2^{p_1}}} \to
 \RG_{\widetilde{f_2^{p_2}}}$ such that for each
 point $x$ of $\RG_{\widetilde{f_2^{p_1}}}$,
 there exists an orientation (or direction) preserving
 homeomorphism $\Psi_x$ between small open neighborhoods of $\overline{\widetilde{f_2^{p_1}}}(x)$ and $\overline{\widetilde{f_2^{p_2}}}(\Phi(x))$, respectively, in $\R$ such that
 $\Psi_x \circ \overline{\widetilde{f_2^{p_1}}} = 
 \overline{\widetilde{f_2^{p_2}}} \circ
 \Phi$ holds on a small open neighborhood of $x$ in 
 $\RG_{\widetilde{f_2^{p_1}}}$.
\end{definition}

\begin{figure}
    \begin{center}
    \includegraphics[width=0.8\textwidth]{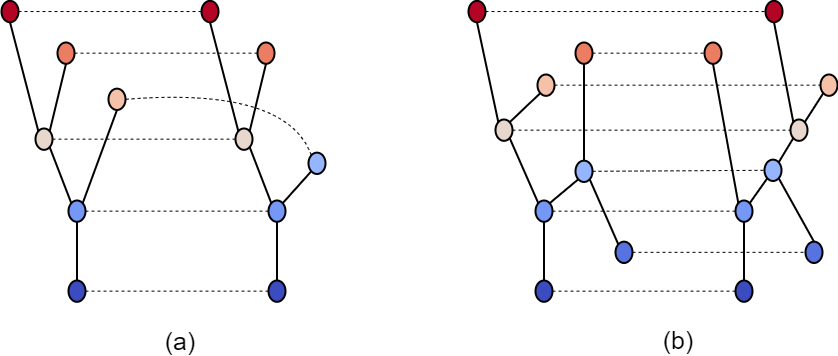}
    \caption{(a) Each function corresponding to two topologically equivalent Reeb graphs has the same set of indices corresponding to its critical points, (b) Converse is not true: each function corresponding to two Reeb graphs has the same set of indices corresponding to its critical points, but the Reeb graphs are not topologically equivalent.}
    \label{fig:topo-equiv-rgs}
    \end{center}
\end{figure}

Since $\Phi$ is a homeomorphism between
the Reeb graphs $\RG_{\widetilde{f_2^{p_1}}}$ and $\RG_{\widetilde{f_2^{p_2}}}$, the degrees of the
corresponding nodes are equal.
Furthermore, the homeomorphism $\Phi$ locally respects
the behaviors of functions $\overline{\widetilde{f_2^{p_1}}}$ and
$\overline{\widetilde{f_2^{p_2}}}$, including the direction of $\R$. Therefore, around each node,
a portion of the domain graph of map $\Phi$
looks locally the same as that of the corresponding
node of the range graph. Thus, around each node,
the indices of the corresponding
critical points are the same.
In other words, if $\RG_{\widetilde{f_2^{p_1}}}$ and  $\RG_{\widetilde{f_2^{p_2}}}$ are topologically equivalent, then the sets of indices corresponding to the sets of critical points of the underlying functions $\widetilde{f_2^{p_1}}$ and  $\widetilde{f_2^{p_2}}$, respectively, are the same; however, the converse may not be true (see \figref{fig:topo-equiv-rgs}).

\begin{figure}
    \centering
    \includegraphics[width = \textwidth]{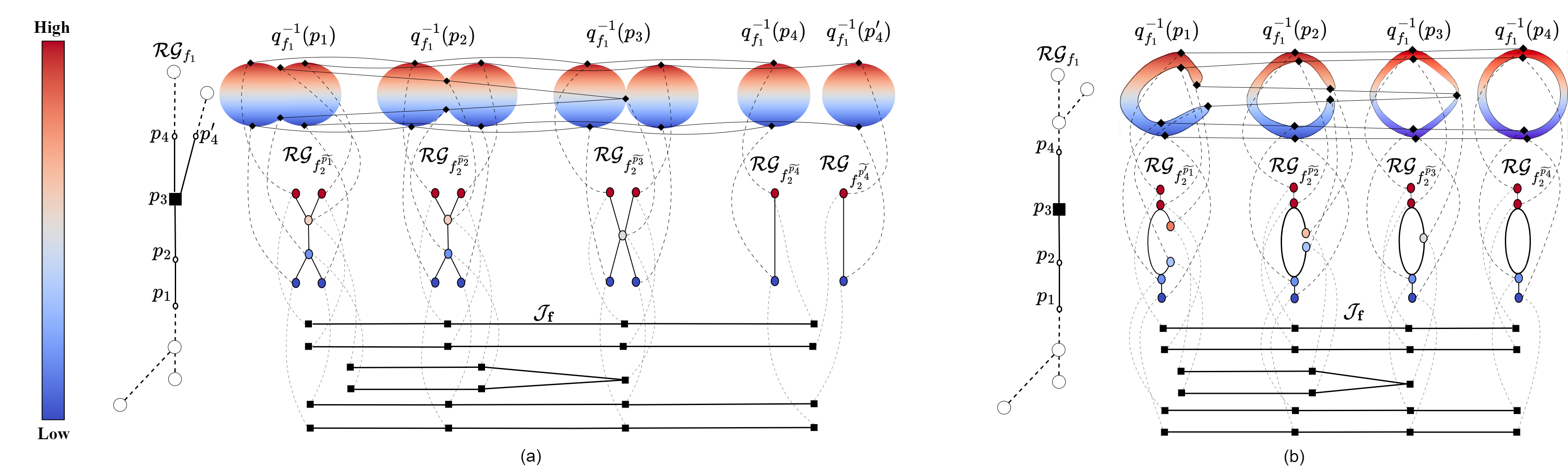}
    \caption{Topological changes in the second-dimensional Reeb graphs of the MDRG for a bivariate field $\f = (f_1, f_2)$ due to saddle critical points of $f_1$. In (a) and (b), points along arcs of $\RG_{f_1}$ are shown on the left. On the right, the top row shows contours of $f_1$ colored based on the values of $f_2$, critical points of $f_2$ restricted to the contours of $f_1$, and the connectivity between the critical points based on the segments of the Jacobi set $\JSet_{\f}$. The middle row displays the corresponding second-dimensional Reeb graphs, while the Jacobi structure $\JStruct_{\f}$ is presented in the bottom row. Dotted lines illustrate the relationship between the critical points, Reeb graph nodes, and the points in $\JStruct_{\f}$. In both cases, a topological change in the second-dimensional Reeb graphs occurs at the node $p_3$ of $\RG_{f_1}$ due to a saddle critical point of $f_1$. In (a), this critical point causes a split of a contour into two, thereby making $p_3$ an up-fork. In (b), the critical point causes a change in the genus of the contour of $f_1$ due to the addition of a handle, making $p_3$ a degree-$2$ node of $\RG_{f_1}$.}
    \label{fig:loop-formation}
\end{figure}

We observe that the detection of a point $p \in \RG_{f_1}$ as a point of topological change is attributed to either by (i) a change in the topology of the domain on which the function $\widetilde{f_2^p}$ is defined, i.e. $q_{f_1}^{-1}(p)$ or by (ii) $\widetilde{f_2^p}$ violating one of the two genericity conditions of Morse function (in Section \ref{subsubsec:pl-critical}). 
The first case occurs when $q_{f_1}^{-1}(p)$ contains a critical point of $f_1$, say $\x$. This critical point can induce the following topological changes in the contours of $f_1$: (a) birth or death of a contour, (b) split or merge of contours, and (c) genus change of a contour. If $\x$ belongs to the first two categories, then $p$ will be either a minimum, a maximum, an up-fork, or a down-fork (as described in \secref{subsec:Reeb-graph}). \figref{fig:loop-formation}(a) shows a scenario where a contour of $f_1$ splits into two. In the third case of genus change, either a handle is added to $q_{f_1}^{-1}(p)$, or a handle is deleted from $q_{f_1}^{-1}(p)$. This results in a change in the topology of the contours of $\widetilde{f_2^p}$ and, consequently, a change in the topology of $\RG_{\widetilde{f_2^p}}$ (Figure~\ref{fig:loop-formation}(b)). In all three cases, $p$ is detected as a node of the augmented Reeb graph $\RG_{f_1}$. The following lemma gives a characterization of the critical points (including genus change critical points) of $f_1$ using Jacobi set of $\f$.
\begin{lemma}
\label{lem:genus-change-cp}
 Every critical point of $f_1$ can be captured as a critical point of $f_1$ restricted to the Jacobi set $\JSet_\f$.
\end{lemma}
\begin{proof}
Let $\x$ be a point of $\M$.
If $\x$ is not a point of the Jacobi set,
then near $\x$, the map $\f$ is like a usual
projection, so it cannot be a critical
point of $f_1$. Thus all critical points of $f_1$ must lie on $\JSet_\f$.  If $\x$ is a critical point of $f_1$, and if $\x$ is not a critical
point of $f_1$ restricted to the Jacobi set,
then a small neighborhood of $\x$ 
on the Jacobi set is mapped PL homeomorphically
into $\R$ by $f_1$, so $\x$ cannot be a critical
point of $f_1$.
\end{proof}

\begin{figure}[t]
    \centering
    \includegraphics[width=\textwidth]{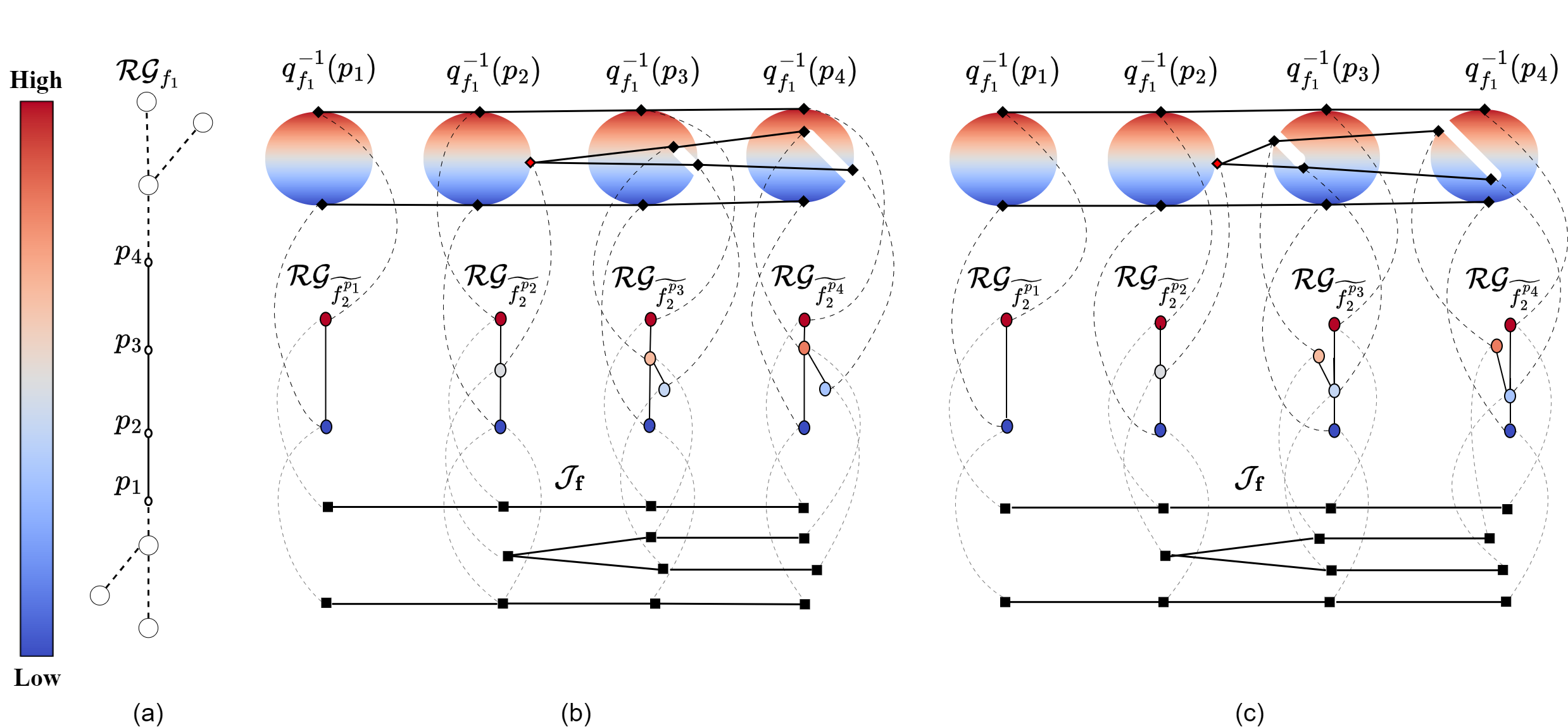}
        \caption{Topological changes in the second-dimensional Reeb graphs of the MDRG for a bivariate field $\f = (f_1, f_2)$ due to the violation of the first Morse condition. (a) Points along an arc of $\RG_{f_1}$. (b) and (c) depict the birth of an arc in the second-dimensional Reeb graphs:  (b) involving a minimum and down-fork, and (c) involving an up-fork and maximum. In both (b) and (c), the top row shows contours of $f_1$ colored based on the values of $f_2$, critical points of $f_2$ restricted to the contours of $f_1$, and the connectivity between the critical points based on the segments of the Jacobi set $\JSet_{\f}$. The middle row displays the corresponding second-dimensional Reeb graphs, while the Jacobi structure $\JStruct_{\f}$ is presented in the bottom row. Dotted lines illustrate the relationship between the critical points, Reeb graph nodes, and the points in $\JStruct_{\f}$. In both cases, the birth event is captured by a minimum of $f_1$ restricted to $\JSet_{\f}$.}
    \label{fig:Birth-arc}
\end{figure}

However, the converse of the above lemma is not true. For example, in \figref{fig:Birth-arc}(b)-(c) the critical points of $f_1$ restricted to $\JSet_\f$ do not correspond to critical points of $f_1$.
Next, we discuss the topological changes arising from the violation of Morse criteria. Note that we assume there can be a violation of exactly one of the Morse criteria at a time.

\begin{figure}[t]
    \centering
    \includegraphics[width=\textwidth]{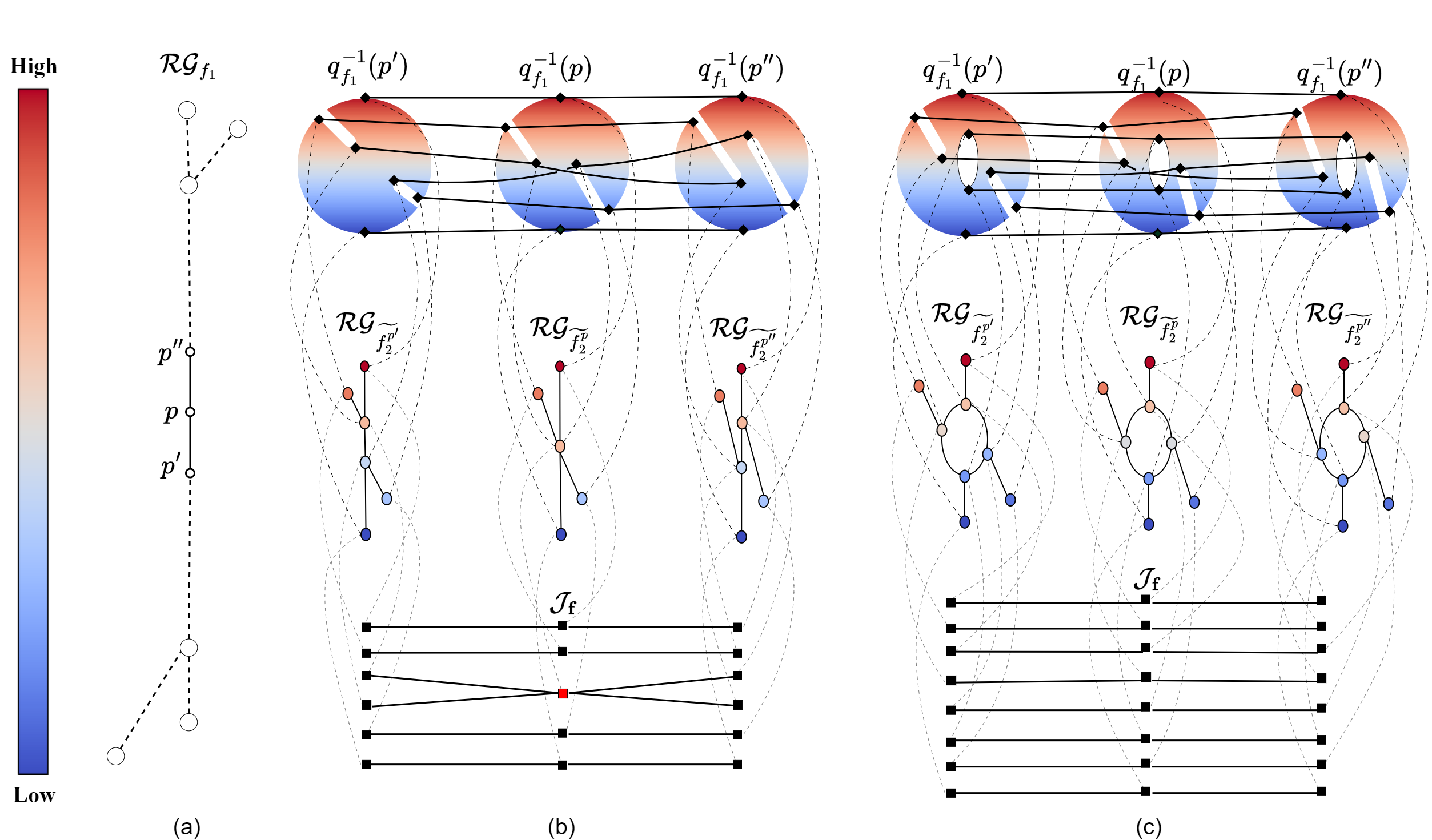}
    \caption{Topological change or not in the second-dimensional Reeb graphs of the MDRG corresponding to a bivariate field $\f = (f_1, f_2)$ due to the violation of the second Morse condition. (a) Points $p,p',p''$ along an arc of $\RG_{f_1}$. (b) and (c) depict two configurations of the second-dimensional Reeb graphs. In both (b) and (c), the top row shows the contours of $f_1$ colored based on the values of $f_2$, critical points of $f_2$ restricted to the contours, and the connectivity between these critical points based on segments of the Jacobi set $\JSet_{\f}$. The middle row displays the corresponding second-dimensional Reeb graphs, while the bottom row shows the Jacobi structure $\JStruct_{\f}$.
    In (b), two critical points of $\widetilde{f_2^p}$, corresponding to the middle Reeb graph, are part of the same contour and the Reeb graph undergoes a topological change, which is captured by a self-intersection point of $\JStruct_{\f}$ (shown in red).
    In (c), two critical points of $\widetilde{f_2^p}$, corresponding to the middle Reeb graph, share the same critical value but belong to different contours and the Reeb graph does not correspond to a topological change.}
    \label{fig:violation-second-Morse-condition}
\end{figure}

Generically, the function $\widetilde{f_2^p}$ is PL Morse. However, there are discrete points $p$ on the arcs of $\RG_{f_1}$ at which $\widetilde{f_2^p}$ violates one of the Morse conditions. We detect topological changes in the second-dimensional Reeb graphs $\RG_{\widetilde{f_2^p}}$ as point $p$ varies on the arcs of  $\RG_{f_1}$, by examining the violation of one of the Morse criteria of the functions $\widetilde{f_2^p}$. 
We note, the nodes of the second-dimensional Reeb graphs correspond to points in Jacobi structure $\JStruct_{\f}$. 
As $p$ varies along an arc of $\RG_{f_1}$, the nodes of $\RG_{\widetilde{f_2^p}}$ are traced out by $\JStruct_{\f}$. Figures \ref{fig:loop-formation}-\ref{fig:violation-second-Morse-condition} show the relationship between the nodes of the second-dimensional Reeb graphs with the Jacobi structure and Jacobi set. Thus, we detect the points of topological change by examining $\JSet_{\f}$ and $\JStruct_{\f}$. The following lemma characterizes the points of topological changes on $\RG_{f_1}$.

\begin{lemma}
\label{lem:topological-changes}
\textbf{Topology-Change-Characterization Lemma.} The topology of $\RG_{\widetilde{f_2^p}}$ changes at a point $p \in \RG_{f_1}$  if and only if one of the following criteria is satisfied:

\begin{enumerate}[(C1)]
    \item $q_{f_1}^{-1}(p)$ contains a critical point of $f_1$.
    
    \item $q_{f_1}^{-1}(p)$ does not contain a critical point of $f_1$ and $\widetilde{f_2^p}$ violates the first Morse condition. Corresponding $q_{f_1}^{-1}(p)$ contains a critical point of $f_1$  restricted to the Jacobi set $\JSet_{\f}$ (which is not a critical point of $f_1$).    

    \item  $q_{f_1}^{-1}(p)$ does not contain a critical point of $f_1$ and $\widetilde{f_2^p}$ violates the second Morse condition, such that there are two critical points of $\widetilde{f_2^p}$ belonging to the same contour of $\widetilde{f_2^p}$. In other words, $q_{f_1}^{-1}(p)$ contains a point $\x$ such that $q_{\f}(\x)$ is a double point on the Jacobi structure $\JStruct_{\f}$.   
\end{enumerate}
\end{lemma}
\noindent

\begin{proof}
Let us first show if one of (C1)--(C3) occurs, then a topological change happens at $p$.\\
\textbf{(C1):} 
Let $\x_p \in \M$ be a critical point of $f_1$ and  $q_{f_1}(\x_p) = p$. Then $p$ can indicate a change in the number of contours of $f_1$, or a change in the genus of a contour 
\cite{2003-Chiang-Simplification-Tetrahedral-Meshes}. If $p$ belongs to the first category, then it is a minimum, a maximum, an up-fork, or a down-fork (as described in \secref{subsec:Reeb-graph}). Therefore, $p$ is a node of $\RG_{f_1}$. \figref{fig:loop-formation}(a) shows an example of this scenario.

However, in the second case, the contour $q_{f_1}^{-1}(p)$  corresponds to a genus change. 
This event affects the topology of the domain on which $\widetilde{f_2^p}$ is defined,
leading to a consequential change in the topology of $\RG_{\widetilde{f_2^p}}$. In \figref{fig:loop-formation}(b), the addition of a handle in the level set of $f_1$ results in the formation of a loop in the second-dimensional Reeb graph. 
 We note, a change in the level set topology of $f_1$ by removal of a handle results in the deletion of a loop in the second-dimensional Reeb graph.\\

\noindent
\textbf{(C2):}
Note that  $q_{f_1}^{-1}(p)$ does not contain a critical point of $f_1$. In other words, the topology of the contour should not change near $p$. In fact, there is a possibility that $f_1$ restricted to $\JSet_\f$ has a critical point $\x$ in $q_{f_1}^{-1}(p)$, and at the same time $\x$ is a critical point of $f_1$. This case is covered by (C1).

If $\widetilde{f_2^p}$ violates the first Morse condition, then $\widetilde{f_2^p}$ has a degenerate critical point, say $\x_p$. 
This corresponds to birth-death of a pair of nodes in $\RG_{\widetilde{f_2^{p}}}$ similar to that discussed in Section~\ref{subsec:time-varying-Reeb-graph}.
 Let, $N(p)$ be a neighborhood of $p$ in  $\RG_{f_1}$ which does not contain any node of $\RG_{f_1}$ or any point $t$ (other than $p$) where  $\widetilde{f_2^t}$ violates one of the Morse conditions. Let $p', p'' \in N(p)$ such that $\overline{f_1}(p') < \overline{f_1}(p) < \overline{f_1}(p'')$. In the case of a birth event, $\widetilde{f_2^{p''}}$ has a pair of critical points that are not present in $\widetilde{f_2^{p'}}$.   Further, each of the two critical points of $\widetilde{f_2^{p''}}$ corresponds to a node in $\RG_{\widetilde{f_2^{p''}}}$, and these nodes are connected by an arc. Hence, a birth event signifies the birth of an arc in the second-dimensional Reeb graphs. 
According to the Index Lemma (see Lemma 1 of the present paper), the indices of two critical points created or destroyed at a birth-death point differ by an index of $1$. Since the function $\widetilde{f_2^{p''}}$ is defined on $q_{f_1}^{-1}(p'')$, which is a PL $2$-manifold, critical points of $\widetilde{f_2^{p''}}$ can have indices $0, 1$, or $2$. So there are two possibilities of indices: $0-1$ or $1-2$. If the two critical points have indices $0$ and $1$, then an arc connecting a minimum and a down-fork is born, as illustrated in \figref{fig:Birth-arc}(b). Otherwise, if the indices are $1$ and $2$, then an arc connecting an up-fork and a maximum is born, as depicted in \figref{fig:Birth-arc}(c).

The point $\x_p$ corresponds to a birth-death point of the Jacobi set $\JSet_{\f}$. 
Specifically, two segments of $\JSet_{\f}$ diverge from or converge to $\x_p$ referred to as birth or death events, respectively. 
In other words, locally, $f_1$ restricted to $\JSet_{\f}$, is monotonic along each of the Jacobi set segments meeting at $\x_p$. 
In the case of a birth event, $\x_p$ is a minimum of $f_1$ restricted to $\JSet_{\f}$, and in the case of a death event, it is a maximum. Thus a birth-death point is a critical point of $f_1$ restricted to  $\JSet_{\f}$.\\

\noindent
\textbf{(C3):} 
If $\widetilde{f_2^p}$ does not satisfy the second Morse condition, then $\widetilde{f_2^p}$ has two critical points $\x_p$ and $\y_p$ such that $\widetilde{f_2^{p}}(\x_p) = \widetilde{f_2^{p}}(\y_p)$. Let, $N(p)$ be a neighborhood of $p$ in $\RG_{f_1}$ which does not contain any critical node of $\RG_{f_1}$ or any point $t$ (other than $p$) such that  $\widetilde{f_2^t}$ violates one of the genericity conditions. Consider $p', p'' \in N(p)$ such that $\overline{f_1}(p') < \overline{f_1}(p) < \overline{f_1}(p'')$. Then, $\widetilde{f_2^{p'}}$ and $\widetilde{f_2^{p''}}$ are PL Morse functions. Let $\x_{p'}$ and $\y_{p'}$ be the critical points of $\widetilde{f_2^{p'}}$ traced from $\x_p$ and $\y_p$, respectively, each  along a segment of $\JSet_{\f}$. Similarly, let $\x_{p''}$ and $\y_{p''}$ be the critical points of $\widetilde{f_2^{p''}}$ traced from $\x_p$ and $\y_p$, respectively. Since $\x_{p'}$ and $\y_{p'}$ are critical points of the PL Morse function $\widetilde{f_2^{p'}}$, it follows that $f_2(\x_{p'}) \neq f_2(\y_{p'})$. Thus, $\x_{p'}$ and $\y_{p'}$ lie on different contours of $\widetilde{f_2^{p'}}$, and therefore, $q_{\widetilde{f_2^{p'}}}(\x_{p'})$ and $q_{\widetilde{f_2^{p'}}}(\y_{p'})$ are two different nodes of the Reeb graph $\RG_{\widetilde{f_2^{p'}}}$. Similarly, we have $f_2(\x_{p''}) \neq f_2(\y_{p''})$, and $q_{\widetilde{f_2^{p''}}}(\x_{p''}) \neq q_{\widetilde{f_2^{p''}}}(\y_{p''})$. However, to identify whether $p$ is a point of topological change, we need to check whether or not $\x_p$ and $\y_p$ belong to the same contour of $\widetilde{f_2^p}$. If $\x_p$ and $\y_p$ belong to the same contour of $\widetilde{f_2^p}$, then they correspond to the same node of the Reeb graph $\RG_{\widetilde{f_2^p}}$, i.e. $q_{\widetilde{f_2^p}}(\x_p) = q_{\widetilde{f_2^p}}(\y_p)$. Thus, the nodes $q_{\widetilde{f_2^{p'}}}(\x_{p'})$ and $q_{\widetilde{f_2^{p'}}}(\y_{p'})$ of $\RG_{\widetilde{f_2^{p'}}}$ merge into a single node $q_{\widetilde{f_2^p}}(\x_p) = q_{\widetilde{f_2^p}}(\y_p)$ of $\RG_{\widetilde{f_2^p}}$, which later splits into two nodes $q_{\widetilde{f_2^{p''}}}(\x_{p''})$ and $q_{\widetilde{f_2^{p''}}}(\y_{p''})$ of $\RG_{\widetilde{f_2^{p''}}}$. Thus, $p$ is a point of topological change in the second-dimensional Reeb graphs. Further, since each node in a second-dimensional Reeb graph of $\MDRG_{\f}$ corresponds to a singular fiber component, this event signifies two singular fiber components merging into a single singular fiber component and later splitting into two singular fiber components. The Jacobi structure $\JStruct_{\f}$, which captures the connectivity of singular fiber components, encodes this event as a self-intersection or double point. \figref{fig:violation-second-Morse-condition}(b) shows an illustration of this case. However, if $\x_{p}$ and $\y_{p}$ belong to different contours of $\widetilde{f_2^p}$, then they correspond to different nodes of $\RG_{\widetilde{f_2^p}}$, i.e. $q_{\widetilde{f_2^p}}(\x_p) \neq q_{\widetilde{f_2^p}}(\y_p)$. Thus, even though $\x_p$ and $\y_p$ share the same $f_2$-value, they do not induce merge or split of the contours of $\widetilde{f_2^t}$ for $t\in N(p)$. As a result, there is no change in the topology of the second-dimensional Reeb graphs. \figref{fig:violation-second-Morse-condition}(c) illustrates an example of this scenario.

Conversely, suppose that none of (C1)--(C3) occurs. Then we see that no topological change happens. This completes the proof of Lemma \ref{lem:topological-changes}.
\end{proof}

\noindent
\textbf{Notes:} 
1. In this paper, we call the points $p\in\RG_{f_1}$ satisfying (C1) as \emph{Type~I} points,   $p\in\RG_{f_1}$ satisfying (C2) as \emph{Type~II} points and  $p\in\RG_{f_1}$ satisfying (C3) as \emph{Type~III} points of topological changes. The augmented Reeb graph based on genus change Type~I saddle points will be denoted by $\RG_{f_1}^{AugI}$, the augmented Reeb graph based on Type~I and Type~II points will be denoted by $\RG_{f_1}^{AugII}$ and the augmented Reeb graph based on Type~I, Type~II and Type~III points will be denoted by $\RG_{f_1}^{AugIII}$.

2. The second dimensional Reeb graph $\RG_{\widetilde{f_2^p}}$, corresponding to a Type~I, Type~II or Type~III point $p\in\RG_{f_1}$, will be called a \emph{critical Reeb graph}. Since there is a violation of exactly one genericity condition at a time, the critical Reeb graph $\RG_{\widetilde{f_2^p}}$ has a unique point corresponding to this topological change for the family of the second dimensional Reeb graphs around $p$. This point in the critical Reeb graph $\RG_{\widetilde{f_2^p}}$ will be called a \emph{topological change point}.\\

Based on the above theoretical results, in the next section, we provide an algorithm for computing a topologically correct Reeb space $\RS_\f$ by computing a correct $\mathrm{MDRG}_\f$.

\section{Algorithmic Contributions}
\label{sec:MDRG}
This section describes our algorithm for computing the correct Reeb space $\RS_{\f}$ of $\f$ based on $\MDRG_\f$. The outline of our algorithm is as follows:

\begin{enumerate}
    \item \textbf{Computing the Jacobi Structure:}  To compute the correct MDRG we first compute the Jacobi structure by computing the projections of the Jacobi set edges and their intersections in the Reeb space. This step corresponds to the computation of the $0$-sheets and $1$-sheets of the Reeb space, which form the combinatorial backbone of the structure. The algorithm for this computation is detailed in \secref{subsec:computing-Jacobi-structure}.
    
    \item  \textbf{Computing the Correct MDRG:} The MDRG is computed in three steps:
    First, we build the Reeb graph of the first field, i.e. $\RG_{f_1}$, using the procedure {\sc ConstructReebGraph}($\M$, $f_1$) as discussed in \secref{subsec:Reeb-graph}. 
    In the second step, we identify the discrete points $p$ on $\RG_{f_1}$ where the second-dimensional Reeb graph $\RG_{\widetilde{f_2^{p}}}$ experiences a topological change. 
    These include (i) the nodes of $\RG_{f_1}$ corresponding to the critical points   (including the genus change critical points) of $f_1$ and 
    (ii) the points of $\RG_{f_1}$  at which $\widetilde{f_2^{p}}$ violates one of the two Morse conditions as discussed in \lemref{lem:topological-changes}.
    Thus, we introduce a minimal set of points in $\RG_{f_1}$, denoted by $P$, such that if
    $\RG_{f_1}$ is augmented based on the points in $P$, then each arc $\alpha$ of the augmented Reeb graph $\RG_{f_1}^{Aug III}$ fulfills the following two conditions: (i) $\overline{f_1}$ is monotonic along $\alpha$, and (ii) for two distinct points $p_1,p_2 \in \alpha$, the Reeb graphs $\RG_{\widetilde{f_2^{p_1}}}$ and $\RG_{\widetilde{f_2^{p_2}}}$ are topologically equivalent. We denote the set of arcs obtained by the augmentation of $P$ as $Arcs(\RG_{f_1}^{Aug III})$.
    The detailed procedure for determining the points in $P$ is given in \secref{subsec:algorithm-computing-MDRG}.
    Finally, corresponding to each arc $\alpha$ in $Arcs(\RG_{f_1}^{Aug III})$ we select a representative point $p$. We denote the set of representative points by $P_R$. For each point $p$ in $P_R$, we compute the second dimensional Reeb graph $\RG_{\widetilde{f_2^{p}}}$, using {\sc ConstructReebGraph}($q_{f_1}^{-1}$, $\widetilde{f_2^{p}}$). These Reeb graphs, along with $\RG_{f_1}^{Aug III}$, effectively capture the topology of $\MDRG_{\f}$.

\item \textbf{Computing the Net-Like Structure:} From the computed Jacobi Structure and MDRG of $\f$, we first compute a net-like structure $\Net_\f$ by connecting the nodes of the second-dimensional Reeb graphs of the MDRG using the Jacobi Structure. We note, the nodes of $\RG_{\widetilde{f_2^{p}}}$ correspond to the critical points of $\widetilde{f_2^p}$, and as we vary $p$, they trace out the segments of the Jacobi structure in the Reeb space. $\Net_\f$ gives a topological skeleton embedded in the Reeb space. The algorithm for computing the net-like structure is discussed in detail in \secref{subsec:Compute-Net-Structure}.

\item \textbf{Computing the Reeb space with $2$-sheets:} Finally, from the net-like structure we compute the complete $2$-sheets of the Reeb space $\W_f$. A complete $2$-sheet consists of one or more simple $2$-sheets. Two simple $2$-sheets belong to the same complete $2$-sheet if two regular points, in the domain, corresponding to the simple sheets can be connected by a path without crossing any singular fiber.  The algorithm for computing the complete $2$-sheets and Reeb space is detailed in \secref{subsec:Compute-Reeb-Space}.  
\end{enumerate}

From \lemref{lem:topological-changes}, it is evident that determining the points of topological change on $\RG_{f_1}$ requires the following computations: (i) critical points of $f_1$ associated with genus changes, (ii) critical points of $f_1$ restricted to $\JSet_{\f}$, and (iii) double points of $\JStruct_{\f}$. Using \lemref{lem:genus-change-cp}, the first two requirements are fulfilled by examining the criticality of $f_1$ at the vertices of $\JSet_{\f}$. However, to fulfill the third requirement, we need to compute the Jacobi structure $\JStruct_{\f}$ which is discussed next.

\subsection{Algorithm 1: Computing Jacobi Structure}
\label{subsec:computing-Jacobi-structure}
Consider a PL bivariate field $\f=(f_1,f_2)$ satisfying the genericity conditions (i)-(iii) in \secref{sec:theory-contributions}. 
\begin{algorithm}
\caption{\textsc{ComputeJacobiStructure}}
\label{alg:Jacobi-Structure}
\raggedright\textbf{Input:} $\M,\,\f,\,\JSet_{\f}$\\
\textbf{Output:} $\JStruct_{\f}$
\begin{algorithmic}[1]
\State Initialize: $\JStruct_{\f}=\emptyset$
\State \% \textit{Augmenting with Type~I and Type~II topological change points}
\State $\RG_{f_1} \gets $ {\sc ConstructReebGraph}($\M, f_1$)
\State $J_{min} \gets $ {\sc ComputeJacobiMinima}($\JSet_{\f}, f_1$)
\State $J_{max} \gets $ {\sc ComputeJacobiMaxima}($\JSet_{\f}, f_1$)
\State $P' \gets$  $J_{min}\cup J_{max}$
\State $\RG_{f_1}^{Aug II} \gets $ {\sc AugmentReebGraph}($\RG_{f_1}$, $P'$)
    \For{each edge $e(\bu,\bv)$ in $\JSet_{\f}$}
        \State \%\textit{Compute vertices for  $q_{\f}(e(\bu, \bv))$}
        \If{$q_{\f}(\bu)$ is   not defined}
            \State Add a vertex $u$ in $\JStruct_{\f}$
            \State Set $q_{\f}(\bu) \gets u$ and $\overline{\f}(u) \gets \f(\bu)$
        \Else
            \State $u \gets q_{\f}(\bu)$
        \EndIf
        \If{$q_{\f}(\bv)$ is not defined}
            \State Add a vertex $v$ in $\JStruct_{\f}$
            \State Set $q_{\f}(\bv) \gets v$ and $\overline{\f}(v) \gets \f(\bv)$            
        \Else
            \State $v \gets q_{\f}(\bv)$
        \EndIf
        \State Add the edge $e(u, v)$ in $\JStruct_{\f}$

        \For{each previously processed edge $e(\bu',\bv')$ of $\JSet_{\f}$ non-adjacent to $e(u, v)$}
        \State \%\textit{Compute the intersection of $q_{\f}(e(\bu,\bv))$ with $q_{\f}(e(\bu',\bv'))$ for non-adjacent Jacobi edges $e(\bu,\bv)$ and $e(\bu',\bv')$}
            \State \Call{Intersection}{$e(\bu,\bv), e(\bu',\bv'), \f, \, \RG_{f_1}^{AugII}$}            
        \EndFor
        \State Mark $e(\bu,\bv)$ as processed
    \EndFor
\State \Return $\JStruct_{\f}$
\end{algorithmic}
\end{algorithm}
The Jacobi set $\JSet_{\f}$ of $\f$ is first computed, using the procedure {\sc ComputeJacobiSet}, as described in Section \ref{subsubsec:jacobiset}. In this subsection, we describe the computation of the Jacobi structure $\JStruct_{\f}$, which is obtained as the projection of the Jacobi set $\JSet_{\f}$ into the Reeb space $\RS_{\f}$. Each point in $\JStruct_{\f}$ represents a singular fiber-component of $\f$. Thus, $\JStruct_{\f}$ is vital in determining the topology of $\RS_{\f}$. To compute the Jacobi structure $\JStruct_{\f}$, we leverage the observation that the functions $f_1$ and $f_2$ are monotonic along the edges of $\JSet_{\f}$. This follows from the genericity conditions of $f_1$ and $f_2$.

\begin{figure}[t]
    \centering
    \includegraphics[width=\textwidth]{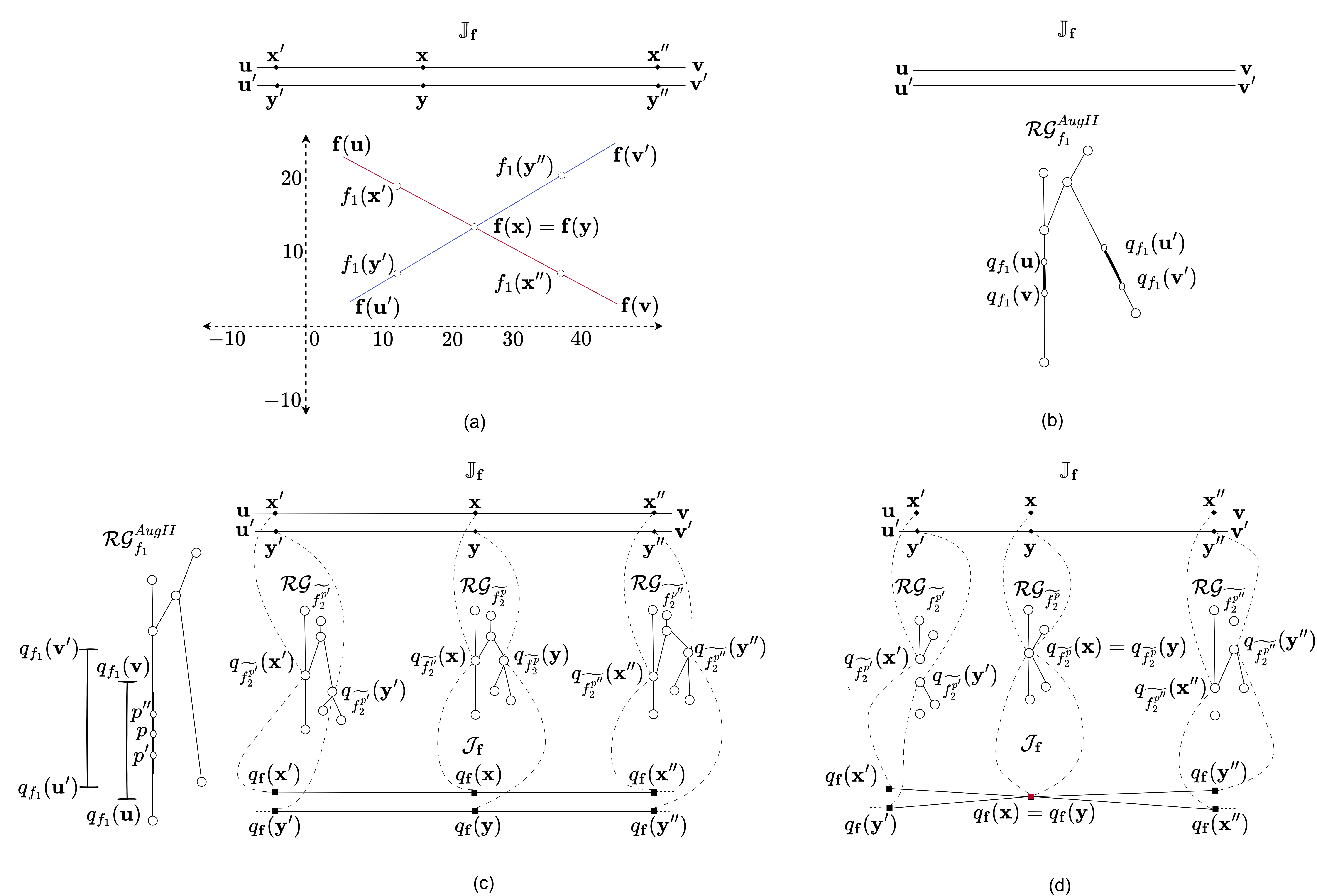}
        \caption{\textbf{Self-intersection (double) points on the Jacobi structure.} For a bivariate field $\f = (f_1,f_2)$, (a) shows two edges $e(\mathbf{u},\mathbf{v})$, $e(\mathbf{u'},\mathbf{v'})$ of the Jacobi set $\JSet_{\f}$ with intersecting projections to the range of $\f$. If $\exists$ $\x \in e(\mathbf{u},\mathbf{v}) \text{ and } \exists\, \y \in e(\mathbf{u'},\mathbf{v'})$ such that $\f(\x) = \f(\y)$, then consider points $\x', \x'' \in e(\mathbf{u},\mathbf{v})$ and $\y', \y'' \in e(\mathbf{u'},\mathbf{v'})$ with $f_1(\x') = f_1(\y') < f_1(\x) = f_1(\y) < f_1(\x'') = f_1(\y'')$. Three configurations of their projections to the second-dimensional Reeb graphs of $\MDRG_{\f}$ and the Jacobi structure $\JStruct_{\f}$ are shown: (b) $\x$ and $\y$ lie in different contours of $f_1$, (c) $\x$ and $\y$ belong to the same contour of $f_1$ but different contours of $\widetilde{f_2^p}$ (here, $p=q_{f_1}(\x)=q_{f_1}(\y)$), (d) $\x$ and $\y$ are in the same fiber-component of $\f$, and consequently $q_{\f}(\x) = q_{\f}(\y)$ is a double point of $\JStruct_{\f}$ (shown in red). The dotted lines illustrate the correspondence between points in the Jacobi set, Jacobi structure, and nodes in the Reeb graphs.}
    \label{fig:Jacobi-structure-Reeb-graphs}
\end{figure}

Generically, $\JSet_{\f}$ is a PL $1$-manifold \cite{2004-edelsbrunner-jacobi-set}. However, the restriction of $q_\f$ to $\JSet_{\f}$ may have a crossing, so the image may not be a $1$-manifold (as shown in Figure~\ref{fig:violation-second-Morse-condition}(b)). The procedure for computing $\JStruct_{\f}$ is outlined in \algoref{alg:Jacobi-Structure}. For each edge $e(\bu, \bv)$ of $\JSet_{\f}$, an edge $q_{\f}(e(\bu, \bv))$ is added to $\JStruct_{\f}$ (lines $5$-$18$, \algoref{alg:Jacobi-Structure}). However,  $q_{\f}(e(\bu, \bv))$ may intersect with a previously added edge $q_{\f}(e(\bu', \bv'))$ in $\JStruct_{\f}$, for two non-adjacent Jacobi edges $e(\bu,\bv)$ and $e(\bu',\bv')$,  as illustrated in \figref{fig:Jacobi-structure-Reeb-graphs}. Such an intersection occurs when two non-adjacent edges $e(\bu, \bv)$ and $e(\bu', \bv')$ intersect the same singular fiber-component of $\f$.
As shown in \figref{fig:Jacobi-structure-Reeb-graphs}(d), the points $\x \in e(\bu, \bv)$ and $\y \in e(\bu', \bv')$ lie on the same fiber-component of $\f$. Thus  $q_{\f}(e(\bu, \bv))$ and $q_{\f}(e(\bu', \bv'))$ intersect at $q_{\f}(\x) = q_{\f}(\y)$. However, the determination of such intersections requires the computation of the augmented Reeb graph $\RG_{f_1}^{AugII}$ which is obtained by augmenting with \emph{Type I} and \emph{Type II} points of topological changes in  $\RG_{f_1}$ (lines $3$-$7$, \algoref{alg:Jacobi-Structure}). Determining the set of points $P'$ of \emph{Type I}, and \emph{Type II} topological change requires the computation of (i) the minima of $f_1$ restricted to the Jacobi set $\JSet_{\f}$ and (ii) the maxima of $f_1$ restricted to $\JSet_{\f}$. The procedure $\textsc{ComputeJacobiMinima}$ provides the pseudo-code for determining the minima of $f_1$ on $\JSet_{\f}$ (line $4$, \algoref{alg:Jacobi-Structure}). A vertex $\bv \in \JSet_{\f}$ is identified as a minimum if $f_1(\bv)$ is not greater than the $f_1$-values of its adjacent vertices in $\JSet_{\f}$. The procedure for determining the maxima of $f_1$ in $\JSet_{\f}$ follows a similar approach (line $5$, \algoref{alg:Jacobi-Structure}).

\paragraph*{Procedure: Computing Jacobi Minima.}
We note, $\JSet_\f$ is a collection of PL $1$-manifold components consisting of vertices and edges. The procedure \textsc{ComputeJacobiMinima} iterates over the vertices of $\JSet_{\f}$, denoted by $V(\JSet_\f)$, to identify those that are minima for $f_1$ restricted to $\JSet_{\f}$. A vertex $\bv$ is a minimum of $f_1$ restricted to $\JSet_{\f}$ only if there is no adjacent vertex $\bv'$ of $\JSet_{\f}$ for which $f_1(\bv') < f_1(\bv)$. 
\begin{algorithmic}[1]
\Procedure{ComputeJacobiMinima}{$\JSet_{\f}, f_1$}
\label{proc:compute-Jacobi-Minima}
\State Initialize: $J_{min} \gets \emptyset$
\For{$\bv \in V(\JSet_{\f})$}
        \State $N_{\bv} \gets $ $\JSet_{\f}$.GetNeighbours($\bv$)
        \State isMinimum $\gets$ True
        \For{$\bv' \in N_{\bv}$}
            \If{$f_1(\bv') < f_1(\bv)$}
                \State isMinimum $\gets$ False
            \EndIf
        \EndFor
        \If{isMinimum $\gets$ True}
            \State Add $\bv$ to $J_{min}$
        \EndIf
\EndFor
\State \Return $J_{min}$
\EndProcedure
\end{algorithmic}

The procedure $\textsc{Intersection}$ (called in line $24$, \algoref{alg:Jacobi-Structure}) checks if projections of two non-adjacent Jabobi edges have an intersection on the Reeb space which we detail next.

\paragraph*{Procedure: Computing Intersection Points.} 
To check whether $q_{\f}(e(\bu, \bv))$ has an intersection with $q_{\f}(e(\bu', \bv'))$ for two non-adjacent Jacobi edges $e(\bu,\bv)$ and $e(\bu',\bv')$, we proceed as follows. We compute the projections of $e(\bu, \bv)$ and $e(\bu', \bv')$ on the range of $\f$, i.e. $\R^2$. If the line segments $\f(e(\bu, \bv))$ and $\f(e(\bu', \bv'))$ do not intersect, then for any $\x \in e(\bu, \bv)$ and $\y \in e(\bu', \bv')$, we have $\f(\x) \neq \f(\y)$, indicating that $\x$ and $\y$ do not lie on the same fiber, therefore, they cannot lie on the same fiber-component. On the other hand, if the line segments $\f(e(\bu, \bv))$ and $\f(e(\bu', \bv'))$ intersect, then there exist points $\x \in e(\bu, \bv)$ and $\y \in e(\bu', \bv')$ such that $\x$ and $\y$ lie on the same fiber of $\f$. We then check if $\x$ and $\y$ also belong to the same fiber-component of $\f$. 

We observe, if $q_{f_1}(\x) = q_{f_1}(\y) = p$ and $q_{\widetilde{f_2^{p}}}(\x) = q_{\widetilde{f_2^{p}}}(\y)$ then $\x$ and $\y$ lie on the same fiber-component of $\f$, i.e. $q_{\f}(\x) = q_{\f}(\y)$. In other words, if $\x$ and $\y$ are mapped to the same point in the first and corresponding second-dimensional Reeb graphs of  $\MDRG_{\f}$, then they lie on the same fiber-component. However, determining this requires exact computation of the intersection point of the line segments $\f(e(\bu, \bv))$ and $\f(e(\bu', \bv'))$, and checking if $q_{f_1}(\x) = q_{f_1}(\y) = p$ and $q_{\widetilde{f_2^{p}}}(\x) = q_{\widetilde{f_2^{p}}}(\y)$ hold,  overcoming floating-point errors, which are computationally challenging. Hence, we adopt the following strategy of analyzing the corresponding Reeb graphs in $\MDRG_{\f}$ to decide  
if $q_{\f}(\x)=q_{\f}(\y)$. 

We note, if $\f(e(\bu, \bv))$ and $\f(e(\bu', \bv'))$ intersect, then there are three different possibilities, as illustrated in \figref{fig:Jacobi-structure-Reeb-graphs}(b)-(d). First, we check how $q_{f_1}$  maps $e(\bu, \bv)$ and $e(\bu', \bv')$ in $\RG_{f_1}^{Aug II}$. If $q_{f_1}(e(\bu, \bv))$ and $q_{f_1}(e(\bu', \bv'))$ have no intersection in $\RG_{f_1}^{Aug II}$, then $q_{f_1}(\x) \neq q_{f_1}(\y)$ for any $\x\in e(\bu, \bv)$ and $\y \in e(\bu', \bv')$ with $\f(\x)=\f(\y)$ (see \figref{fig:Jacobi-structure-Reeb-graphs}(b)). Therefore, $q_{\f}(\x)\neq q_{\f}(\y)$. However, if $q_{f_1}(e(\bu, \bv))$ and $q_{f_1}(e(\bu', \bv'))$ intersect, for $p \in q_{f_1}(e(\bu, \bv))\cap q_{f_1}(e(\bu', \bv'))$,  let $q_{f_1}^{-1}(p)$ intersect $e(\bu, \bv)$ and $e(\bu', \bv')$ at $\x$ and $\y$, respectively. Therefore, $q_{f_1}(\x) = q_{f_1}(\y)=p$. We assume, $q_{f_1}(e(\bu, \bv))\cap q_{f_1}(e(\bu', \bv'))$ contains no Type I or Type II topological change points and can have at most one Type III topological change point.  In this case, there are two possibilities. If $\x$ and $\y$ belong to different contours of $\widetilde{f_2^{p}}$ (i.e. $q_{\widetilde{f_2^{p}}}(\x) \neq  q_{\widetilde{f_2^{p}}}(\y)$), then $q_{\f}(\x) \neq q_{\f}(\y)$ (see \figref{fig:Jacobi-structure-Reeb-graphs}(c)). Otherwise, $\x$ and $\y$ are in the same fiber-component, i.e. $q_{\f}(\x) = q_{\f}(\y)$, resulting in the intersection of $q_{\f}(e(\bu, \bv))$ and $q_{\f}(e(\bu', \bv'))$ (see \figref{fig:Jacobi-structure-Reeb-graphs}(d)). 
We note, this intersection point corresponds to the critical points of $\widetilde{f_2^{p}}$ where the second Morse condition is violated (as in Lemma~\ref{lem:topological-changes}-(C3)). In other words, this corresponds to the swapping of nodes in the second-dimensional Reeb graphs, as observed in Figure~\ref{fig:violation-second-Morse-condition}(b). This event can be detected uniquely by analyzing a second-dimensional Reeb graph corresponding to a point $p \in q_{f_1}(e(\bu, \bv))\cap q_{f_1}(e(\bu', \bv'))$. 

More precisely, for any $p \in q_{f_1}(e(\bu, \bv))\cap q_{f_1}(e(\bu', \bv'))$,  let $q_{f_1}^{-1}(p)$ intersect $e(\bu, \bv)$ and $e(\bu', \bv')$ at $\x$ and $\y$, respectively. Then, if the nodes $q_{\widetilde{f_2^{p}}}(\x)$ and $q_{\widetilde{f_2^{p}}}(\y)$ are not connected by an edge in $\RG_{\widetilde{f_2^{p}}}$ (case \figref{fig:Jacobi-structure-Reeb-graphs}(c)), $q_{\f}(e(\bu, \bv))$ and $q_{\f}(e(\bu', \bv'))$ do not intersect. Otherwise, if the nodes $q_{\widetilde{f_2^{p}}}(\x)$ and $q_{\widetilde{f_2^{p}}}(\y)$ are connected by an edge (or coincide) in $\RG_{\widetilde{f_2^{p}}}$ (case \figref{fig:Jacobi-structure-Reeb-graphs}(d)), $q_{\f}(e(\bu, \bv))$ and $q_{\f}(e(\bu', \bv'))$ intersect. At the point of intersection the nodes $q_{\widetilde{f_2^{p}}}(\x)$ and $q_{\widetilde{f_2^{p}}}(\y)$ coincide.

\begin{algorithmic}[1]
\Procedure{Intersection}{$e(\bu,\bv),e(\bu',\bv'), \f, \RG_{f_1}^{Aug II}$}
        \State \% \textit{Check for the intersection of $\f(e(\bu,\bv))$ and $\f(e(\bu',\bv'))$ for two non-adjacent Jacobi edges $e(\bu,\bv)$ and $e(\bu',\bv')$}    
        \If{$\f(e(\bu,\bv))$ and $\f(e(\bu',\bv'))$ intersect}
            \State Compute: $\ba \gets \f(e(\bu, \bv)) \cap \f(e(\bu',\bv'))$

            \State \% \textit{Check for the intersection of $q_{f_1}(e(\bu,\bv))$ and $q_{f_1}(e(\bu',\bv'))$}
            
            \If{$q_{f_1}(e(\bu,\bv))$ and $q_{f_1}(e(\bu',\bv'))$ intersect}
            
                \State $p \gets q_{f_1}(e(\bu,\bv)) \cap q_{f_1}(e(\bu',\bv'))$
                \State $\x \gets q_{f_1}^{-1}(p)\cap e(\bu,\bv)$
                \State $\y \gets q_{f_1}^{-1}(p)\cap e(\bu',\bv')$
                
                \State \% \textit{Construct the Reeb graph of $\widetilde{f_2^{p}}$}
                
                \State $\RG_{\widetilde{f_2^{p}}} \gets $ {\sc ConstructReebGraph}($q_{f_1}^{-1}(p), \widetilde{f_2^{p}}$)
                
                \If{$q_{\widetilde{f_2^{p}}}(\x)$ and $q_{\widetilde{f_2^{p}}}(\y)$ are adjacent nodes of an arc in $\RG_{\widetilde{f_2^{p}}}$}
                    \State \% \textit{$q_{\f}(e(\bu, \bv))$ and $q_{\f}(e(\bu',\bv'))$ have an intersection}
      
                    \State Add a vertex $w$ in $\JStruct_{\f}$
                
                    \State Subdivide $e(q_{\f}(\bu), q_{\f}(\bv))$ into edges $e(q_{\f}(\bu), w)$ and $e(w, q_{\f}(\bv))$
                    \State Subdivide $e(q_{\f}(\bu'),q_{\f}(\bv'))$ into edges $e(q_{\f}(\bu'), w)$ and $e(w, q_{\f}(\bv'))$

                    \State Set $\overline{\f}(w) \gets \ba$
                    \State $\x_0 \gets \f^{-1}(\ba)\cap e(\bu,\bv)$
                    \State $\y_0 \gets \f^{-1}(\ba)\cap e(\bu',\bv')$
                    \State Set $q_{\f}(\x_0) \gets w$ and $q_{\f}(\y_0) \gets w$ 
                    \State Mark edges $e(q_{\f}(\bu), w), e(w, q_{\f}(\bv)), e(q_{\f}(\bu'), w), e(w, q_{\f}(\bv'))$ as processed
                \Else
                    \State $q_{\f}(e(\bu, \bv))$ and $q_{\f}(e(\bu',\bv'))$ do not intersect
                \EndIf
            \Else
                \State $q_{\f}(e(\bu, \bv))$ and $q_{\f}(e(\bu',\bv'))$ do not intersect
            \EndIf
        \Else
            \State $q_{\f}(e(\bu, \bv))$ and $q_{\f}(e(\bu',\bv'))$ do not intersect
        \EndIf
\EndProcedure
\end{algorithmic}

\algoref{alg:Jacobi-Structure} constructs the $0$-sheets of the Reeb space by computing:  
(i) the critical points of $f_1$ restricted to the Jacobi set and projecting them to the Reeb space (lines: $4$-$5$ and $9$-$21$), and  
(ii) the double points of the Jacobi structure (line $25$).  
The remaining portions of the Jacobi structure, excluding these $0$-sheets, constitute the $1$-sheets of the Reeb space.
Next, we discuss our algorithm for computing MDRG based on the computed Jacobi structure in more detail.

\subsection{Algorithm 2: Computing the MDRG}
\label{subsec:algorithm-computing-MDRG}
More specifically, the computation of MDRG consists of the following four steps:
\begin{enumerate}
    \item Computing the Reeb graph $\RG_{f_1}$, 
    
    \item Determining the \emph{Type~I}, \emph{Type~II} and \emph{Type~III} points of topological change along the arcs of $\RG_{f_1}$ and augmenting the Reeb graph $\RG_{f_1}$ based on these points,
    
    \item Selecting a representative point $p$ from each subdivided arc of $\RG_{f_1}^{Aug III}$, 
    
    \item Computing the Reeb graph $\RG_{\widetilde{f_2^p}}$ corresponding to each representative point $p$ and building the MDRG.
\end{enumerate}
We note, the nodes in second dimensional Reeb graphs $\RG_{\widetilde{f_2^p}}$ correspond to critical points of $\widetilde{f_2^p}$, and consequently represent points in the Jacobi structure $\JStruct_{\f}$ (see \secref{subsec:computing-Jacobi-structure}). Therefore, these nodes are crucial in capturing the topology of the Reeb space. Note that since we have assumed the domain $\M$ is a triangulation of an orientable $3$-manifold without boundary, the (regular) level surfaces of $f_1$ are orientable and also have no boundary, and the regular level sets of $\widetilde{f_2^p}$ are finite disjoint unions of circles. Therefore, $\widetilde{f_2^p}$ has no genus change critical points and consequently, $\RG_{\widetilde{f_2^p}}$ will not have any degree-$2$ critical node.

\algoref{alg:compute-MDRG} provides the pseudo-code for computing $\MDRG_{\f}$. The first step for constructing $\MDRG_{\f}$ involves the computation of the Reeb graph $\RG_{f_1}$ (line $3$, \algoref{alg:compute-MDRG}) using an algorithm discussed in \secref{subsec:Reeb-graph}.
Next, we augment this Reeb graph further by determining the set of points $P$ of \emph{Type I}, \emph{Type II} and \emph{Type III} topological change which requires the computation of: (i) the minima of $f_1$ restricted to the Jacobi set $\JSet_{\f}$, (ii) the maxima of $f_1$ restricted to $\JSet_{\f}$, and (iii) double points of $\JStruct_{\f}$ (see \lemref{lem:topological-changes}).
\begin{algorithm}
\caption{\sc{ComputeMDRG}}
\label{alg:compute-MDRG}
\raggedright\textbf{Input:} $\M, \f, \JSet_{\f}, \JStruct_{\f}$\\
\textbf{Output:} $\MDRG_{\f}$
\begin{algorithmic}[1]
\State $\MDRG_{\f} \gets \emptyset$
\State \% \textit{Augment First-Dimensional Reeb Graph with Points of Topological Changes}
\State $\RG_{f_1} \gets $ {\sc ConstructReebGraph}($\M, f_1$)
\State $J_{min} \gets $ {\sc ComputeJacobiMinima}($\JSet_{\f}, f_1$)
\State $J_{max} \gets $ {\sc ComputeJacobiMaxima}($\JSet_{\f}, f_1$)
\State $DP \gets $ {\sc DoublePoints}($\JStruct_{\f}$)
\State $P \gets  J_{min} \cup J_{max} \cup DP$
\State $\RG_{f_1}^{AugIII} \gets $ {\sc AugmentReebGraph}($\RG_{f_1}$, $P$)
\State $\MDRG_{\f}$.{\sc Add}($\RG_{f_1}^{AugIII}$)
\State \% \textit{Computing Second-Dimensional Reeb Graphs}
\For{ arc $\alpha \in Arcs(\RG_{f_1}^{AugIII})$}
    \State $p \gets $ {\sc GetRepresentativePoint}($\RG_{f_1}^{AugIII}$, $\alpha$)
    \State $\RG_{\widetilde{f_2^p}} \gets $ {\sc ConstructReebGraph}($q_{f_1}^{-1}(p), f_2$)
    \State $\MDRG_{\f}$.{\sc Add}($\RG_{\widetilde{f_2^p}}$)
\EndFor
\State \Return $\MDRG_{\f}$
\end{algorithmic}
\end{algorithm}
The procedures $\textsc{ComputeJacobiMinima}$ and $\textsc{ComputeJacobiMaxima}$ compute the minima and maxima of $f_1$ on $\JSet_{\f}$, respectively (line $4$-$5$, \algoref{alg:compute-MDRG}), as discussed in \secref{subsec:computing-Jacobi-structure}. The \textsc{DoublePoints} procedure computes points in $\M$ mapped (by the quotient map $q_{\f}$) to double points in the Jacobi structure $\JStruct_{\f}$ (line $6$, \algoref{alg:compute-MDRG}). The collective outcomes of these procedures constitute the points of topological change, denoted as $P$ (line $7$, \algoref{alg:compute-MDRG}).

After determining $P$, the Reeb graph $\RG_{f_1}$ is augmented by creating degree $2$-nodes corresponding to the points in $P$. This is performed by the procedure \textsc{AugmentReebGraph} (line $8$, \algoref{alg:compute-MDRG}). For each arc in the augmented Reeb graph $\RG_{f_1}^{Aug III}$, a representative $p$ is selected by the procedure \textsc{GetRepresentativePoint}. Then, the Reeb graph $\RG_{\widetilde{f_2^p}}$ is computed by the procedure \textsc{ConstructReebGraph} (lines $12$-$13$, \algoref{alg:compute-MDRG}). The resulting Reeb graphs $\RG_{\widetilde{f_2^p}}$ (with $p$ as the representative point of an arc), along with the Reeb graph $\RG_{f_1}^{Aug III}$, collectively represent $\MDRG_{\f}$. These Reeb graphs are added to the  $\MDRG_{\f}$ structure by the \textsc{ADD} procedure (lines $9, 14$, \algoref{alg:compute-MDRG}). The obtained MDRG is then utilized in the construction of the Reeb space.

\paragraph*{Procedure: Computing Double Points.}
This procedure identifies the vertices of $\JStruct_{\f}$ that are double (or self-intersection) points. When projected onto the Reeb graph $\RG_{f_1}$, these points represent topological changes in the second-dimensional Reeb graphs (see \lemref{lem:topological-changes}). A vertex $v \in \JStruct_{\f}$ is identified as a double point if it is adjacent to four vertices of $\JStruct_{\f}$. \figref{fig:violation-second-Morse-condition}(b) and \figref{fig:Jacobi-structure-Reeb-graphs}(d) illustrate scenarios where the Jacobi structure has a double point.

\begin{algorithmic}[1]
\Procedure{DoublePoints}{$\JStruct_{\f}$}
\label{proc:double-points}
\State Initialize: $DP \gets \emptyset$
\For{$v \in \JStruct_{\f}$}
    \If{$v$ is adjacent to four vertices}
        \State Get an arbitrary vertex $\bv$  from  $q_{\f}^{-1}(v)$
        \State Add $\bv$ to $DP$
    \EndIf
\EndFor
\State \Return $DP$
\EndProcedure
\end{algorithmic}

In the next subsection, we discuss computing a net-like structure using the computed Jacobi structure and MDRG.
The net-like structure is embedded in the Reeb space and provides a topological skeleton for visualizing the correct Reeb space.

\subsection{Algorithm 3: Computing the Net-Like Structure}
\label{subsec:Compute-Net-Structure}
\begin{figure}[t]
    \centering
    \includegraphics[width=\textwidth]{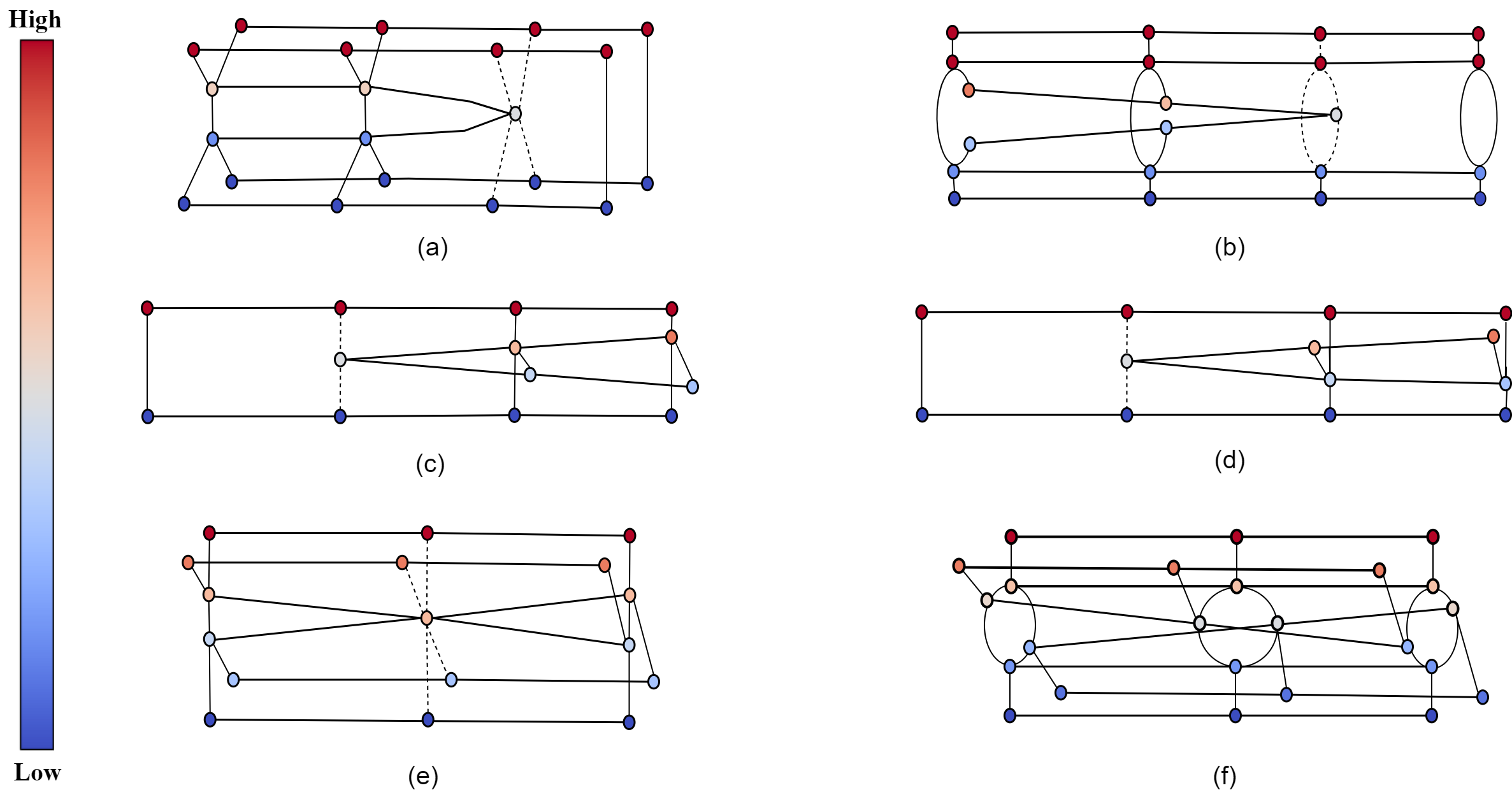}
        \caption{\textbf{Net-like structures:} (a)-(f) show the (local) net-like structures corresponding to the bivariate fields in Figures \ref{fig:loop-formation}(a), \ref{fig:loop-formation}(b), \ref{fig:Birth-arc}(b), \ref{fig:Birth-arc}(c), \ref{fig:violation-second-Morse-condition}(b), and \ref{fig:violation-second-Morse-condition}(c), respectively. The dotted lines constitute the edges of the second-dimensional Reeb graphs corresponding to the points of topological change. The edges corresponding to the other Reeb graphs are depicted as thin solid lines, and the thick solid lines constitute the Jacobi structure. The coloring of the nodes in the net-like stucture is based on the coloring of the corresponding nodes in the second-dimensional Reeb graphs.}
    \label{fig:net-like-structure}
\end{figure}

In this subsection, we provide the algorithm for computing the net-like structure corresponding to the Reeb space. From \lemref{lem:homeomorphism-embedding}, we note, the second-dimensional Reeb graphs in $\MDRG_{\f}$ have an embedding in the Reeb space $\RS_{\f}$. Therefore, to compute the net-like structure $\Net_{\f}$ corresponding to the Reeb space $\RS_{\f}$, we compute a topologically correct embedding of the second-dimensional Reeb graphs in $\MDRG_{\f}$ by connecting them based on the computed Jacobi structure $\JStruct_\f$. Thus, we obtain a net-like structure or a skeleton corresponding to the Reeb space, as shown in Figure~\ref{fig:net-like-structure}. 

\algoref{alg:net-like-structure} provides the pseudo-code for computing $\Net_{\f}$. The net-like structure $\Net_{\f}$ is first initialized to $\JStruct_{\f}$ (line $1$, \algoref{alg:net-like-structure}). After this step, the first-dimensional augmented Reeb graph $\RG_{f_1}^{AugIII}$ is retreived from $\MDRG_{\f}$ by the procedure \textsc{GetFirstDimensionalReebGraph} (line $2$, \algoref{alg:net-like-structure}). Then, for each arc $\alpha$ of $\RG_{f_1}^{AugIII}$, a representative point $p$ is obtained by the procedure \textsc{GetRepresentativePoint} (line $4$, \algoref{alg:net-like-structure}). For each representative point $p$, the  Reeb graph $\RG_{\widetilde{f_2^p}}$ is retreived from $\MDRG_{\f}$, by the procedure \textsc{GetSecondDimensionalReebGraph} (line $5$, \algoref{alg:net-like-structure}). Then, $\RG_{\widetilde{f_2^p}}$ is embedded in a net-like structure corresponding to $\RS_{\f}$ (line $6$, \algoref{alg:net-like-structure}). The procedure \textsc{EmbedReebGraph} provides the pseudo-code for embedding a Reeb graph $\RG_{\widetilde{f_2^p}}$ in a net-like structure $\Net_{\f}$ corresponding to $\RS_{\f}$ which is detailed next.

\begin{algorithm}
\caption{\sc{ComputeNetLikeStructure}}
\label{alg:net-like-structure}
\raggedright\textbf{Input:} $\JStruct_{\f}, \MDRG_{\f}$\\
\textbf{Output:} $\Net_{\f}$
\begin{algorithmic}[1]
\State Initialize: $\Net_{\f} \gets \JStruct_{\f}$
\State $\RG_{f_1}^{AugIII} \gets $ \textsc{GetFirstDimensionalReebGraph}($\MDRG_{\f}$)
\For{ arc $\alpha \in Arcs(\RG_{f_1}^{AugIII})$}
\State $p \gets $ {\sc GetRepresentativePoint}($\RG_{f_1}^{AugIII}$, $\alpha$)
    \State $\RG_{\widetilde{f_2^p}} \gets \textsc{GetSecondDimensionalReebGraph}(\MDRG_{\f}, p)$
    \State \textsc{EmbedReebGraph}($\RG_{\widetilde{f_2^p}}, \Net_{\f}$)
\EndFor
\State \Return $\Net_{\f}$
\end{algorithmic}
\end{algorithm}

\paragraph*{Procedure: Embed Reeb Graph.}
For an arc of $\RG_{\widetilde{f_2^p}}$, the start and end nodes are extracted by the procedures \textsc{GetStartNode} and \textsc{GetEndNode}, respectively (lines $3$ and $5$, procedure \textsc{EmbedReebGraph}). For each arc $\beta^p$ between two nodes $p_1$ and $p_2$ of $\RG_{\widetilde{f_2^p}}$, an edge is introduced between the corresponding vertices in $\JStruct_{\f}$, as follows. Since $p$ is a point on an arc of the augmented Reeb graph $\RG_{f_1}^{AugIII}$, the function $\widetilde{f_2^p}$ is Morse (see \secref{subsec:mdrg-topological-changes} for more details). Therefore, the fiber-component of $\f$ corresponding to $p_1$ contains exactly one critical point of $\widetilde{f_2^p}$, denoted as $\x_1$. This point is computed by the procedure \textsc{GetJacobiSetPoint} (line $4$, procedure \textsc{EmbedReebGraph}). As $\x_1$ is on the Jacobi set $\JSet_{\f}$, its projection $q_{\f}(\x_1)$ into $\RS_{\f}$ lies on $\JStruct_{\f}$. Similarly, let $\x_2$ be the unique critical point of $\widetilde{f_2^p}$ corresponding to $p_2$, and $q_{\f}(\x_2)$ denote its projection in $\JStruct_{\f}$ (line $6$, procedure \textsc{EmbedReebGraph}). Then an edge between $q_{\f}(\x_1)$ and $q_{\f}(\x_2)$ is added to build the net-like structure $\Net_{\f}$ corresponding to  $\RS_{\f}$ (line $7$, procedure \textsc{EmbedReebGraph}).

\begin{algorithmic}[1]
\Procedure{EmbedReebGraph}{$\RG_{\widetilde{f_2^p}}, \Net_{\f}$}
\label{proc:Embed-Reeb-Graph}
\For{$\beta^p \in Arcs(\RG_{\widetilde{f_2^p}})$}
    \State $p_1 \gets \textsc{GetStartNode}(\beta^p)$
    \State $\x_1 \gets \textsc{GetJacobiSetPoint}(p_1)$
    \State $p_2 \gets \textsc{GetEndNode}(\beta^p)$    
    \State $\x_2 \gets \textsc{GetJacobiSetPoint}(p_2)$    
    \State Add edge e($q_{\f}(\x_1), q_{\f}(\x_2)$) in $\Net_{\f}$
\EndFor
\EndProcedure
\end{algorithmic}

We note, the net-like structure is the Jacobi structure connecting the second-dimensional Reeb graphs (corresponding to the representative points on the arcs of the first-dimensional Reeb graphs) embedded in the Reeb space. However, at this stage, the second-dimensional Reeb graphs corresponding to the points of topological changes of the first-dimensional Reeb graph are not part of the net-like structure (since computing such Reeb graphs is challenging). In the next subsection, we discuss the main algorithm for computing the Reeb space by computing its $2$-sheets of the Reeb space in the computed net-like structure.

\subsection{Algorithm 4: Computing the Reeb Space with $2$-Sheets}
\label{subsec:Compute-Reeb-Space}
\begin{figure}[t]
    \centering
    \includegraphics[width=\textwidth]{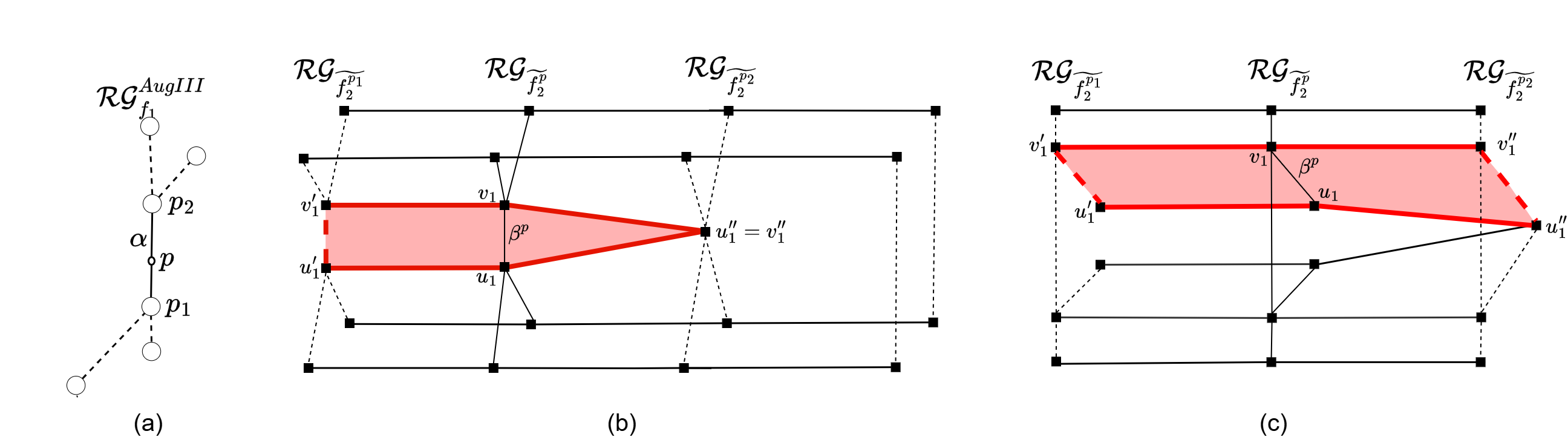}
    \caption{\textbf{Computing the simple Reeb sheet $ReebSheet(\alpha, \beta^p)$:} (a) The Reeb graph $\RG_{f_1}^{Aug III}$ showing the arc $\alpha$ (between nodes $p_1$ and $p_2$) and  the representative point $p$. (b) and (c) show two cases of the Reeb sheet boundary obtained by tracing the arc $\beta^{p}$ of the second-dimensional Reeb graph $\RG_{\widetilde{f_2^p}}$. In both (b) and (c), corresponding to $\beta^p$ an edge $e(u_1,v_1)$ is added in $\Net_{\f}$. The points $u_1', v_1'$ (and $u_1'', v_1''$) on $\Net_{\f}$ are obtained by tracing $u_1\; (\text{similarly}, v_1)$ along the monotonically decreasing (increasing) direction of $f_1$ until reaching the value of $\overline{f_1}(p_1)$  (and $\overline{f_1}(p_2)$), respectively.
    (b) A dummy edge $e(u_1',v_1')$ is added, (c) dummy edges $e(u_1',v_1')$ and $e(u_1'',v_1'')$ are added. The Reeb sheet is shown in red color, the edges in the Jacobi structure belonging to its boundary are depicted as thick red lines, and the dummy edges as dotted red lines. Figure 7(b) corresponds to Figure 1(3) (right-left reversed). Figure 7(c) corresponds to Figure 1(2) (right-left reversed).}
    \label{fig:trace-boundary}
\end{figure}

The net-like structure computed in \secref{subsec:Compute-Net-Structure} provides a topologically correct skeleton embedded in the Reeb space. However, the $2$-sheets of the Reeb space and their connectivities are still missing. In this subsection, we provide the final algorithm for computing the $2$-sheets of the Reeb space $\RS_{\f}$ which are connected along the Jacobi structure components in the computed $\Net_\f$, as shown in Figure~\ref{fig:net-like-structure}.
Note that an arc $\alpha$ of the augmented Reeb graph $\RG_{f_1}^{AugIII}$ corresponds to a set of second-dimensional Reeb graphs $\{\RG_{\widetilde{f_2^p}} \mid p \in \alpha\}$ which are topologically equivalent.
As the second-dimensional Reeb graphs are topologically equivalent, once we choose an arc $\beta^p$ of $\RG_{\widetilde{f_2^p}}$ for some $p \in \alpha$, then an arc $\beta^{p'}$ of  $\RG_{\widetilde{f_2^{p'}}}$ for any other $p' \in \alpha$ is naturally determined uniquely.
Thus, each arc $\beta^p$ of the second-dimensional Reeb graph $\RG_{\widetilde{f_2^p}}$, while $p$ varies in $\alpha$, traces out a unique $2$-sheet component which is called \emph{simple} Reeb sheet and is denoted by $ReebSheet(\alpha, \beta^p)$ where $\alpha$ and $\beta^p$ are called the first and second representative arcs, respectively. Note that depending on its representative arcs, a simple Reeb sheet may be 
\emph{complete} or \emph{incomplete}. If the boundary of a simple Reeb sheet consists only of the Jacobi structure components, then the Reeb sheet is called a complete Reeb sheet. Otherwise, the Reeb sheet is called an incomplete Reeb sheet. We note, in our construction, the boundary of an incomplete Reeb sheet will have one or more dummy edges  (as will be discussed in \textsc{ComputeSimpleSheet}, Figure~\ref{fig:trace-boundary}). 
After computing its boundary, the simple Reeb sheet $ReebSheet(\alpha, \beta^p)$ is represented by its boundary and the representative arcs $\alpha$ and $\beta^p$. From the stored information, we note that triangulating each simple sheet's interior is straightforward. 
Finally, each \emph{complete} $2$-sheet of a Reeb space $\RS_\f$ is obtained as the union of (adjacent) path-connected simple incomplete $2$-sheets, as described in the procedure \textsc{CompatibleUnion}. 

\begin{algorithm}
\caption{\sc{ComputeReebSpace}}
\label{alg:compute-Reeb-space}
\raggedright\textbf{Input:} $\M, \f$\\
\textbf{Output:} $\RSS_{\f}$
\begin{algorithmic}[1]
\State \% \textit{Computing the Jacobi Structure}
\State $\JSet_{\f} = $\textsc{ComputeJacobiSet}($\M, \f$)
\State $\JStruct_{\f} \gets$ \textsc{ComputeJacobiStructure}($\M, \f, \JSet_{\f}$)
\State \% \textit{Computing the MDRG}
\State $\MDRG_{\f} \gets $ \textsc{ComputeMDRG}($\M, \f, \JSet_{\f}, \JStruct_{\f}$)
\State \% \textit{Computing the Net-Like Structure}
\State $\Net_{\f}\gets$ 
\textsc{ComputeNetLikeStructure}($\JStruct_{\f}, \MDRG_{\f}$)
\State \% \textit{Computing the Simple 2-Sheets}
\State $\RG_{f_1}^{AugIII} \gets $ \textsc{GetFirstDimensionalReebGraph}($\MDRG_{\f}$)
\State Initialize: $simpleSheets \gets \emptyset$
\For{ arc $\alpha \in Arcs(\RG_{f_1}^{AugIII})$}
\State $p \gets $ {\sc GetRepresentativePoint}($\RG_{f_1}^{AugIII}$, $\alpha$)
    \State $\RG_{\widetilde{f_2^p}} \gets \textsc{GetSecondDimensionalReebGraph}(\MDRG_{\f}, p)$
    \For{$\beta^p \in Arcs(\RG_{\widetilde{f_2^p}})$}
        \State $simpleSheet \gets \textsc{ComputeSimpleSheet}(\RG_{f_1}^{AugIII}, \RG_{\widetilde{f_2^p}}, \Net_{\f}, \alpha, \beta^p)$
        \State $simpleSheet.\textsc{SetRepArcs}(\alpha, \beta_p$)
        \State $simpleSheets.\textsc{Add}(simpleSheet$)
    \EndFor
    \EndFor
    \State \% \textit{Computing Complete 2-Sheets}
    \State $completeSheets \gets \textsc{CompatibleUnion}$($simpleSheets$, $\M$, $\Net_\f$)  
        \State $\RSS_{\f} \gets \Net_\f. \textsc{Add}$($completeSheets$)    
\State \Return $\RSS_{\f}$
\end{algorithmic}
\end{algorithm} 
\algoref{alg:compute-Reeb-space} provides the \emph{main algorithm} for computing the correct Reeb space corresponding to $\RS_\f$. In \algoref{alg:compute-Reeb-space}, the procedure \textsc{ComputeJacobiSet} first computes the Jacobi set $\JSet_{\f}$, as described in \secref{subsubsec:jacobiset} (line $1$, \algoref{alg:compute-Reeb-space}). Next, \textsc{ComputeJacobiStructure} computes the Jacobi structure $\JStruct_{\f}$
 by projecting the Jacobi set into the Reeb space as described in \algoref{alg:Jacobi-Structure} (line $2$, \algoref{alg:compute-Reeb-space}). Based on the Jacobi set and Jacobi structure, \textsc{ComputeMDRG} computes  $\MDRG_{\f}$ using \algoref{alg:compute-MDRG} (line $3$, \algoref{alg:compute-Reeb-space}). Then the algorithm computes the net-like structure $\Net_\f$ using \algoref{alg:net-like-structure} (line $4$, \algoref{alg:compute-Reeb-space}).
 The first-dimensional augmented Reeb graph $\RG_{f_1}^{AugIII}$ is retreived from $\MDRG_{\f}$ by the procedure \textsc{GetFirstDimensionalReebGraph} (line $5$, \algoref{alg:compute-Reeb-space}). For an arc $\alpha$ of $\RG_{f_1}^{AugIII}$, \textsc{GetRepresentativePoint} obtains the representative point $p$ on $\alpha$ (line $8$, \algoref{alg:compute-Reeb-space}).  Then  \textsc{GetSecondDimensionalReebGraph} retrieves the corresponding second dimensional Reeb graph $\RG_{\widetilde{f_2^p}}$ from $\MDRG_{\f}$ (line $9$, \algoref{alg:compute-Reeb-space}). Subsequently, for each arc $\beta^p$ in $\RG_{\widetilde{f_2^p}}$, the procedure \textsc{ComputeSimpleSheet} computes the Reeb sheet $ReebSheet(\alpha, \beta^p)$ (line $11$, \algoref{alg:compute-Reeb-space}) which is simple, but may or may not be complete (Figure \ref{fig:trace-boundary}). Each such simple sheet can be uniquely identified by $\alpha$ and $\beta^p$. 
 We set the arcs $\alpha$ and $\beta^p$ as the  representative arcs of $simpleSheet$ (line $12$, \algoref{alg:compute-Reeb-space}).  
 The procedure {\sc CompatibleUnion} computes the union of adjacent incomplete simple  $2$-sheets to obtain the complete $2$-sheets  along with their shared dummy edges (line $16$, \algoref{alg:compute-Reeb-space}). Finally, the computed Reeb space data-structure $\RSS_\f$ corresponding to $\RS_\f$ is obtained as the collection of such complete $2$-sheets along with their connectivity information stored in $\Net_\f$ (line $17$, \algoref{alg:compute-Reeb-space}).  
Next, we discuss the procedure \textsc{ComputeSimpleSheet} in detail.

\paragraph*{Procedure: Computing Simple Reeb Sheets.}
The procedure 
\textsc{ComputeSimpleSheet} first computes 
 the boundary of $ReebSheet(\alpha, \beta^p)$ by tracing the end nodes of the arc $\beta^p\in \RG_{\widetilde{f_2^p}}$ along the Jacobi structure $\JStruct_{\f}$ (in $\Net_\f$), in the monotonically increasing and decreasing directions of $\overline{f_1} \circ \omega_1$ corresponding to the arc $\alpha \in \RG_{f_1}^{Aug III}$. Note that the arc $\alpha$ is directed in the increasing direction with respect to $\overline{f_1}$. First, the start node $p_1$ and end node $p_2$ of the arc $\alpha \in \RG_{f_1}^{Aug III}$ are extracted by the procedures \textsc{GetStartNode} and \textsc{GetEndNode}, respectively (lines $2$-$3$, procedure \textsc{ComputeSimpleSheet}). Let, $minVal$ and $maxVal$ be the values of $f_1$ corresponding to $p_1$ and $p_2$, respectively (lines $4$-$5$, procedure \textsc{ComputeSimpleSheet}). Then, the value of $\overline{f_1} \circ \omega_1$ ranges between $minVal$ and $maxVal$ on $ReebSheet(\alpha, \beta^p)$. Similarly, the start node $p_1'$ and the end node $p_2'$ of the arc $\beta^p\in \RG_{\widetilde{f_2^p}}$, are also extracted by the procedures \textsc{GetStartNode} and \textsc{GetEndNode}, respectively (lines $6$-$7$, procedure \textsc{ComputeSimpleSheet}). Since $p$ is a (regular) point on an arc of $\RG_{f_1}^{Aug III}$, the function $\widetilde{f_2^p}$ is Morse (see \secref{subsec:mdrg-topological-changes} for more details). Therefore, the contour of $\widetilde{f_2^p}$ corresponding to $p_1'$  contains exactly one critical point of $\widetilde{f_2^p}$, say $\x_1$. Moreover, since $\x_1$ lies on the Jacobi set $\JSet_\f$, its projection on $\JStruct_{\f}$ is known from the \algoref{alg:Jacobi-Structure}. \textsc{GetCriticalPoint}
computes the critical point $\x_1=q_{\widetilde{f_2^p}}^{-1}(p_1')\cap \JSet_\f$  (line $8$, procedure \textsc{ComputeSimpleSheet}) and its projection on $\JStruct_{\f}$ is computed as $u_1 = q_{\f}(\x_1)$ (line $14$, procedure \textsc{ComputeSimpleSheet}). Similarly, let $\x_2$ be the unique critical point on the level set of $\widetilde{f_2^p}$ corresponding to $p_2'$, i.e. $\x_2=q_{\widetilde{f_2^p}}^{-1}(p_2')\cap \JSet_\f$,  and $v_1 = q_{\f}(\x_2)$ denote its projection in $\JStruct_{\f}$ (lines $9$ and $15$,  procedure \textsc{ComputeSimpleSheet}).  

\begin{algorithmic}[1]
\Procedure{ComputeSimpleSheet}{$\RG_{f_1}^{Aug III}, \RG_{\widetilde{f_2^p}}, \Net_{\f}, \alpha,\beta^p$}
\label{proc:compute-ReebSpace-Sheets}
    \State $p_1 \gets \textsc{GetStartNode}(\alpha, \RG_{f_1}^{Aug III})$
    \State $p_2 \gets \textsc{GetEndNode}(\alpha, \RG_{f_1}^{Aug III})$
    \State $minVal \gets \overline{f_1}(p_1)$
    \State $maxVal \gets \overline{f_1}(p_2)$
    
    \State $p_1' \gets \textsc{GetStartNode}(\beta^p, \RG_{\widetilde{f_2^p}})$
    \State $p_2' \gets \textsc{GetEndNode}(\beta^p, \RG_{\widetilde{f_2^p}})$ 
    \State $\x_1 \gets  \textsc{GetCriticalPoint}(p_1', \RG_{\widetilde{f_2^p}})$       
    \State $\x_2 \gets  \textsc{GetCriticalPoint}(p_2', \RG_{\widetilde{f_2^p}})$    
    \State\%\textit{First, store the boundary edges of the Reeb sheet}
    \State Initialize: $simpleSheet \gets  \emptyset$
    \State \%Keep a count of dummy edges per $simpleSheet$
    \State $count\gets 0$
    \State $u_1 \gets q_{\f}(\x_1)$
    \State $v_1 \gets q_{\f}(\x_2)$    
    \State \%\textit{Compute the boundary of the simple sheet tracing from $u_1$ and return the endpoint of the path, in the decreasing direction of $\overline{f_1} \circ \omega_1$}
    \State $u_1' \gets  \textsc{ComputeBoundary}(\Net_{\f}, u_1, minVal, simpleSheet,\text{`dec'})$
    \State \%\textit{Compute the boundary of the simple sheet tracing from $v_1$ and return the endpoint of the path, in the decreasing direction of $\overline{f_1} \circ \omega_1$}    
    \State $v_1' \gets  \textsc{ComputeBoundary}(\Net_{\f}, v_1, minVal, simpleSheet, \text{`dec'})$
    \If{$u_1' \neq v_1'$}
        \State $simpleSheet.\textsc{AddDummyEdge}(e(u_1', v_1'))$
        \State $count\gets count+1$
    \EndIf
    \State \%\textit{Compute the boundary of the simple sheet tracing from $u_1$ and return the endpoint of the path, in the increasing direction of $\overline{f_1} \circ \omega_1$}
    \State $u_1'' \gets \textsc{ComputeBoundary}(\Net_{\f}, u_1, maxVal, simpleSheet,\text{`inc'})$
    \State \%\textit{Compute the boundary of the simple sheet tracing from $v_1$ and return the endpoint of the path, in the increasing direction of $\overline{f_1} \circ \omega_1$}
    \State $v_1'' \gets \textsc{ComputeBoundary}(\Net_{\f}, v_1, maxVal, simpleSheet,\text{`inc'})$
    \If{$u_1'' \neq v_1''$}
        \State $simpleSheet.\textsc{AddDummyEdge}(e(u_1'', v_1''))$ 
        \State $count\gets count+1$
    \EndIf
    \State \%\textit{Finally, the simple sheet $ReebSheet(\alpha, \beta^p)$ is set as `complete' or `incomplete' based on the dummy edge count}
\If{$count=0$}
    \State $simpleSheet.\textsc{SetIsComplete}(\mathrm{True})$
\Else
    \State $simpleSheet.\textsc{SetIsComplete}(\mathrm{False})$
\EndIf
\State $simpleSheet.\textsc{SetDummyEdgeCount}(count)$
\State\Return{$simpleSheet$}
\EndProcedure
\end{algorithmic}

Next, starting from $u_1$, the edges of $\JStruct_{\f}$ can be traced along two different directions - in the monotonically decreasing and in the monotonically increasing directions of $\overline{f_1} \circ \omega_1$. First, consider tracing the edges of $\JStruct_{\f}$ along the monotonically decreasing direction of $\overline{f_1} \circ \omega_1$ until encountering a node $u_1'$ such that $\overline{f_1} \circ \omega_1(u_1') = minVal$. Similarly, starting from $v_1$, the Jacobi structure edges are traced until finding a node $v_1'$ such that $\overline{f_1} \circ \omega_1 (v_1') = minVal$. If $u_1' \neq v_1'$, a dummy edge $e(u_1', v_1')$ is added between $u_1'$ and $v_1'$ to the simple sheet boundary (lines $20$-$22$, procedure \textsc{ComputeSimpleSheet}). The dummy edge $e(u_1', v_1')$ corresponds to an arc of a second-dimensional Reeb graph, and along $e(u_1', v_1')$ the $f_2$ value monotonically increases from the start vertex $u_1'$ to the end vertex $v_1'$. We note, the boundary dummy edge $e(u_1', v_1')$ may contain a topological change point of the corresponding second dimensional critical Reeb graph (as discussed in the Note of \lemref{lem:topological-changes}). However, the algorithm does not require explicitly adding this point since these dummy edges of the simple sheets are deleted in the end and the topology of the complete $2$-sheets are computed correctly in the procedure {\sc CompatibleUnion}. On the other hand, if $u_1' = v_1'$, it also signifies a topological change point of the corresponding second dimensional critical Reeb graph, but no dummy edge is required to be added (see \figref{fig:trace-boundary}(b)). We note, the crossing of Jacobi structure edges is not possible in between $u_1$ and $u_1'$ or $v_1$ and $v_1'$.  
Following the same process, starting from $u_1$ (similarly $v_1$), and moving along $\JStruct_{\f}$ in the monotonically increasing direction, the vertices $u_1''$ and $v_1''$ are obtained such that $\overline{f_1} \circ \omega_1 (u_1'') = \overline{f_1} \circ \omega_1(v_1'')=maxVal$.  
If $u_1'' \neq v_1''$, similarly as before  
a dummy edge is added between $u_1''$ and $v_1''$ to the simple sheet boundary (lines 28-30, procedure \textsc{ComputeSimpleSheet}).  
The dummy edge $e(u_1'', v_1'')$ corresponds to an arc of a second-dimensional Reeb graph, and along $e(u_1'', v_1'')$ the $f_2$ value monotonically increases from the start vertex $u_1''$ to the end vertex $v_1''$. The $count$ (lines $22$ and $30$ in \textsc{ComputeSimpleSheet}) counts the number of dummy edges added as the boundary edge to this simple Reeb sheet.
If the dummy edge count is $0$, the simple $2$-sheet is complete, otherwise the simple $2$-sheet is incomplete, this is set in lines $33$-$38$ in \textsc{ComputeSimpleSheet}.

Next, we describe the procedure \textsc{ComputeBoundary} in detail.

\paragraph*{Procedure: Computing the Boundary of a Simple Reeb Sheet.}
As discussed in \textsc{ComputeSimpleSheet}, the procedure \textsc{ComputeBoundary} traces the boundary of a simple $2$-sheet by moving along the Jacobi structure, in the monotonically decreasing or increasing direction of $\overline{f_1} \circ \omega_1$. It starts from a point $u \in \JStruct_{\f}$ and returns the end point $v \in \JStruct_{\f}$, corresponding to a boundary value $boundaryVal$, of the traced boundary along the monotonically decreasing or increasing direction of $\overline{f_1} \circ \omega_1$. In addition, it also associates the traced edges on the Jacobi structure as the boundary edges of the $2$-sheet.
For tracing the boundary along the monotonically decreasing direction (lines $2$-$8$),  starting from $u$, the procedure checks until encountering a vertex $v$ of $\JStruct_{\f}$ such that $\overline{f_1} \circ  \omega_1(v) \leq boundaryVal$. If $\overline{f_1} \circ  \omega_1(v) = boundaryVal$, then no further processing is required and the procedure returns $v$ (line $26$). Otherwise, if $\overline{f_1} \circ  \omega_1(v) < boundaryVal$, then we consider the last processed edge $e(u',v)$ of the while loop.  
We subdivide $e(u',v)$ into two edges by adding a vertex $w$ in the middle such that $\overline{f_1} \circ \omega_1 (w) = bounadaryVal$. Here, the value of $\overline{\f}(w)$ is determined as follows. We note, $\overline{\f}(w)$ is the projection of $w$ onto the range of $\f$, and can be expressed as $\overline{\f}(w) = (\overline{f_1} \circ \omega_1 (w), \overline{f_2} \circ \omega_2(w))$ (see the commutative diagram in \secref{subsec:homeomorphism-proof}). Given that $\overline{f_1} \circ \omega_1 (w) = boundaryVal$, we obtain the value of $\overline{f_2} \circ \omega_2(w)$ by first constructing a parametrization 
$(\delta_{f_1}, \delta_{f_2}): [0,1] \rightarrow \R^2$ of $\overline{\f}(e(u', v))$ so that $\delta_{f_1}(0) = \overline{f_1} \circ \omega_1 (u')$, $\delta_{f_1}(1) = \overline{f_1} \circ \omega_1 (v)$, and $\delta_{f_2}(0) = \overline{f_2} \circ \omega_2 (u')$, and $\delta_{f_2}(1) = \overline{f_2} \circ \omega_2 (v)$. We then determine $t \in [0,1]$ such that $\delta_{f_1}(t) = boundaryVal$, and obtain the value of $\overline{f_2} \circ \omega_2(w)$ as $\delta_{f_2}(t)$.
The new edge $e(u',w)$ is added to the set of boundary edges and the procedure then returns the vertex $w$ as output (lines $17$-$21$).

\begin{algorithmic}[1]
\Procedure{ComputeBoundary}{$\Net_{\f},u, boundaryVal, boundaryEdges, flag$}
\label{proc:trace-boundary}
    \If{$flag =$ `dec'}
        \Do
        \State Get adjacent vertex $v$ of $u$ in $\Net_{\f}$ such that $\overline{f_1} \circ  \omega_1(v) < \overline{f_1} \circ  \omega_1(u)$
        \State Add $e(u,v)$ to $boundaryEdges$
        \State $u'\gets u$
        \State $u \gets v$
        \doWhile{$\overline{f_1} \circ  \omega_1(v) > boundaryVal$}
    \ElsIf{$flag=$`inc'}
        \Do
        \State Get adjacent vertex $v$ of $u$ in $\Net_{\f}$ such that $\overline{f_1} \circ  \omega_1(v) > \overline{f_1} \circ  \omega_1(u)$
        \State Add $e(u,v)$ to $boundaryEdges$
        \State $u'\gets u$
        \State $u\gets v$
        \doWhile{$\overline{f_1} \circ  \omega_1(v) < boundaryVal$}
    \EndIf
    \If{$\overline{f_1} \circ  \omega_1(v) \neq boundaryVal$}
    
        \State Find a parametrization 
        $(\delta_{f_1}, \delta_{f_2}): [0,1] \rightarrow \R^2$ of $\bar{f}(e(u', v))$ satisfying $\delta_{f_1}(0) = \overline{f_1} \circ \omega_1 (u')$, $\delta_{f_1}(1) = \overline{f_1} \circ \omega_1 (v)$, and $\delta_{f_2}(0) = \overline{f_2} \circ \omega_2 (u')$, and $\delta_{f_2}(1) = \overline{f_2} \circ \omega_2 (v)$

        \State Compute: $t\in [0,1]$ such that $\delta_{f_1}(t)=boundaryVal$

        \State $\overline{\f}(w)\gets (\delta_{f_1}(t), \delta_{f_2}(t))$ 
        
        \State Subdivide $e(u', v)$ into edges $e(u', w)$ and $e(w,v)$
        
        \State Delete $e(u',v)$ from $boundaryEdges$
        \State Add $e(u',w)$ to $boundaryEdges$
        \State \Return{$w$}
    \Else
        \State \Return{$v$}
    \EndIf
\EndProcedure
\end{algorithmic}

\begin{figure}[t]
    \centering
    \includegraphics[width=\textwidth]{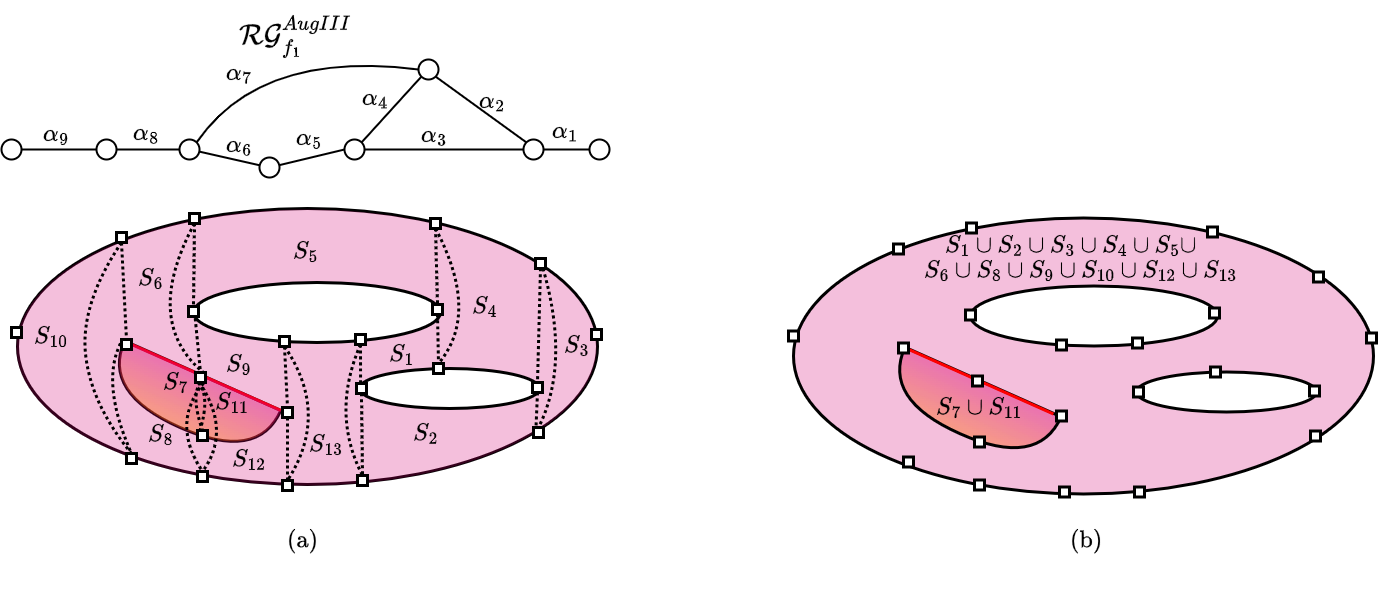}
        \caption{\textbf{Compatible union of simple sheets.} (a) shows the augmented first-dimensional Reeb graph $\RG_{f_1}^{Aug III}$ of $\MDRG_{\f}$ and the boundary of the simple Reeb sheet corresponding to each arc of a second-dimensional Reeb graph of $\MDRG_{\f}$ associated with a representative point of $\RG_{f_1}^{Aug III}$. The sheet boundaries are merged by the \textsc{CompatibleUnion} procedure to obtain two complete sheets, shown in (b). The Jacobi structure edges at the intersection of two sheet boundaries are shown as red solid lines, while the other Jacobi structure edges are depicted as black solid lines. The dummy edges are represented by dotted lines. The two complete Reeb sheets are shaded in different colors, and each simple Reeb sheet in (a) is shaded in the color of the complete Reeb sheet in (b) containing it.}
    \label{fig:compatible-union}
\end{figure}

Next, we discuss the procedure {\sc CompatibleUnion} 
in detail.
\paragraph*{Procedure: Compatible Union.}
We note, not all simple Reeb sheets computed by the procedure {\sc ComputeSimpleSheet} are complete. More specifically, the simple Reeb sheets with `dummy' edges are incomplete and the simple Reeb sheets without any `dummy' edges are complete. In the procedure \textsc{CompatibleUnion}, we compute the union of the adjacent (incomplete) simple Reeb sheets using a Union-Find structure $UF$. The \textsc{Make-Set}$(S)$ procedure in $UF$ creates a new set corresponding to each simple sheet $S$ with representative $S$. $\textsc{Union}(S_1, S_2)$ procedure in $UF$ unites the sets containing $S_1$ and $S_2$, respectively, provided they belong to two different sets and are adjacent in the Reeb sheet. The representative of the resulting set is the representative of either the set containing $S_1$ or $S_2$. \textsc{Find-Set}$(S)$ procedure in $UF$ returns a pointer to the representative of the unique set containing $S$. Two incomplete simple Reeb sheets are adjacent if they have an overlapping dummy edge pair which is checked by the procedure {\sc IsPathConnected} (line 12, procedure {\sc CompatibleUnion}).
We note, if a simple sheet is complete it will be a single component in $UF$.
Each component $C_i$ in $UF$ consists of all the incomplete sheet components that are included in the same complete $2$-sheet. To obtain the complete $2$-sheet from $C_i$ we delete the dummy edges of the incomplete sheets in $C_i$ (line $18$, procedure {\sc CompatibleUnion}). This is illustrated in Figure~\ref{fig:compatible-union}. 

\begin{algorithmic}[1]
 \Procedure{CompatibleUnion}{$simpleSheets$, $\M$, $\Net_\f$}
 \label{proc:compatible-union}
 \State \%Create a Union-Find structure of $simpleSheets$
 \State $UF \gets \emptyset$ 
 
\For{$i\gets 1 \text{ to } simpleSheets.\textsc{length}()$}
\State $S \gets simpleSheets.\textsc{GetSheet}(i)$
\State $UF.$\textsc{Make-Set}$(S)$
\EndFor

\For{$i\gets 1 \text{ to } simpleSheets.\textsc{length}()$}
\State $S_1 \gets simpleSheets.\textsc{GetSheet}(i)$
\For{$j=i+1 \text{ to } simpleSheets.\textsc{length}()$}
\State $S_2 \gets simpleSheets.\textsc{GetSheet}(j)$
        \If{ $UF.$\textsc{Find-Set} $(S_1)$$\neq$ $UF.$\textsc{Find-Set} $(S_2)$ $\And$ \textsc{IsPathConnected}$(S_1, S_2, \M, \Net_\f)$ }
                \State $UF . \textsc{Union}(S_1, S_2)$
            \EndIf
\EndFor
\EndFor
\For{ each component $C_i$ in $UF$}
    \State Delete all Dummy Edges of the Simple Sheets in $C_i$
    \State $completeSheets.\textsc{Add}(C_i)$
\EndFor
    \State \Return $completeSheets$
 \EndProcedure
 \end{algorithmic}

Now one important procedure to decide whether two incomplete sheets belong to the same complete sheet is {\sc IsPathConnected}, which is discussed next.

\begin{figure}[t]
    \centering
    \includegraphics[width=\textwidth]{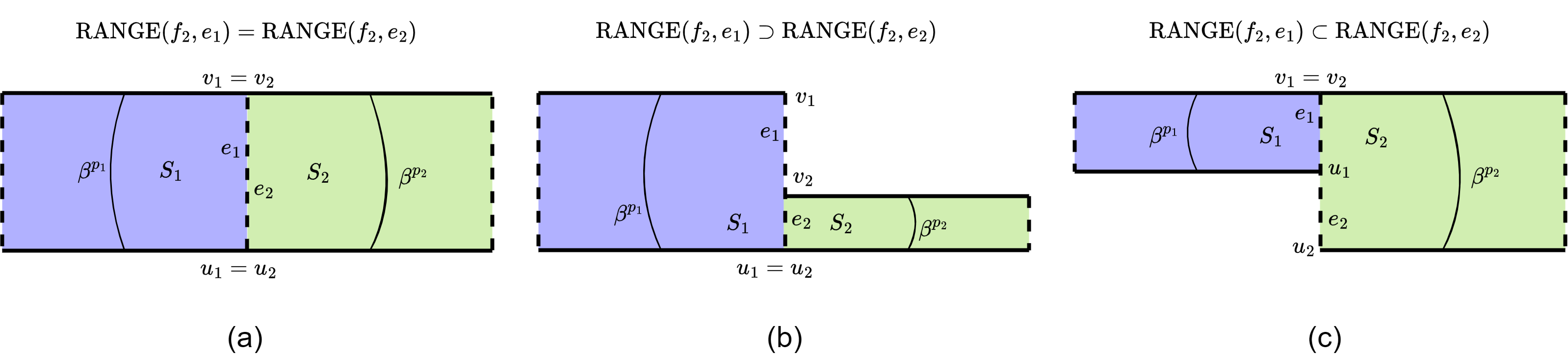}
        \caption{\textbf{Overlapping dummy edges for two adjacent sheets $S_1$ and $S_2$.} $\beta^{p_1}$ and $\beta^{p_2}$ are the representative arcs, and $e_1$ and $e_2$ are the overlapping dummy edges of $S_1$ and $S_2$, respectively. The Jacobi structure edges in a sheet boundary are shown as thick black lines, and the dummy edges as dotted black lines.  Three cases are depicted (from left to right): (a) both the starting and ending vertices coincide, (b) the starting vertices coincide and (c) the ending vertices coincide.}
    \label{fig:overlapping-dummy-edges}
\end{figure}

\begin{figure}[t]
    \centering
    \includegraphics[width=\textwidth]{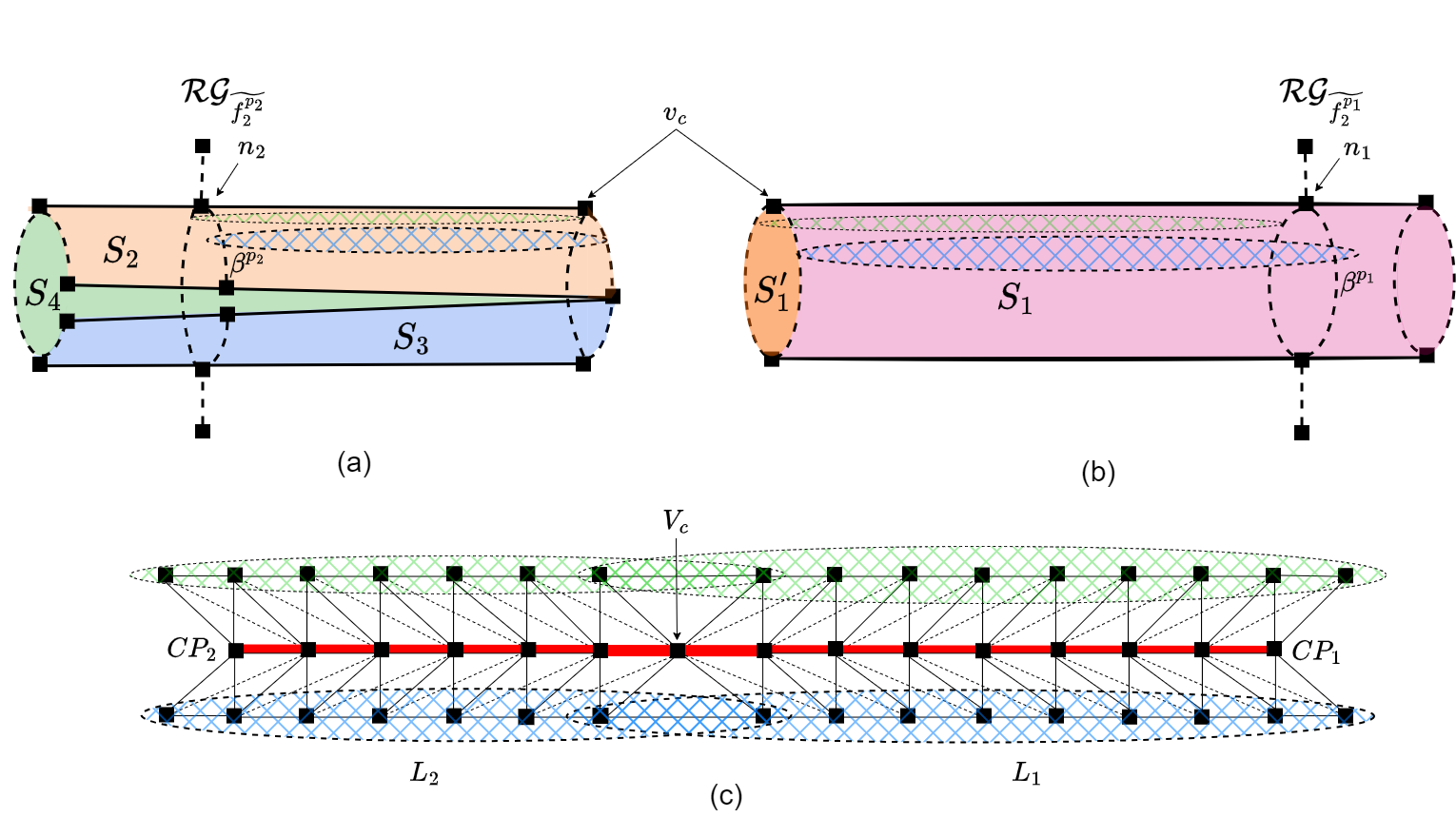}
    \caption{\textbf{IsPath for the union of $S_1$ and $S_2$:} \textbf{Top figures (a) and (b):} Each of the incomplete simple sheets 
    $S_2 \text{ and } S_4$ (in (a)) shares a dummy edge with either of the incomplete simple sheets $S_1 \text{ or } S_1'$ (in (b)) at the common vertex $v_c$.
     The representative arcs of $S_1$ and $S_2$ are denoted by $\beta^{p_1}$ and $\beta^{p_2}$, respectively. Their corresponding end critical nodes are denoted by $n_1$ and $n_2$, respectively. The edges corresponding to the Reeb graphs $\RG_{\widetilde{f_2^{p_1}}}$ and $\RG_{\widetilde{f_2^{p_2}}}$, and the dummy edges, are shown in dotted lines, while the Jacobi structure edges are depicted as solid lines.
    \textbf{Bottom figure (c):} Stars (adjacent tetrahedra) and links of the Jacobi set edges in the domain of the PL-bivariate field. The critical points corresponding to the nodes $n_1, n_2$ and $v_c$ in the second-dimensional Reeb graphs are denoted by $CP_1, CP_2$ and $V_c$, respectively. The link components $L_1$ of the Jacobi set component $\JSet_{\f}(CP_1,V_c)$ associated with $S_1$, and $L_2$ of $\JSet_{\f}(CP_2,V_c)$ associated with $S_2$, are shaded in blue. The other link components of $\JSet_{\f}(CP_1,V_c)$ and $\JSet_{\f}(CP_2,V_c)$ are shaded in green. The projections of these link components in the Reeb space sheets are shown in (b). Since $L_1$ and $L_2$ intersect, $S_1$ and $S_2$ belong to the same complete Reeb space sheet, and the \textsc{IsPath} procedure returns True.}
    \label{fig:is-path}
\end{figure}

\paragraph*{Procedure: {\sc IsPathConnected}.}
Note that two incomplete simple sheets belong to the same (complete) Reeb sheet if a path can connect two interior points of the respective sheets without crossing the boundary Jacobi structure components, or, in other words, if the sheets intersect along a shared dummy edge pair.  The procedure  {\sc IsPathConnected} decides whether two simple sheets $S_1$ and $S_2$ belong to the same Reeb sheet by checking if (i) $S_1$ and $S_2$ have an overlapping dummy edge pair, or, at least one of the vertices of the shared dummy edges are common and (ii) there is a path between an interior point of $p_0\in S_1$ to an interior point $p_n\in S_2$ via. the overlapping part of the dummy edge pair, without crossing the boundary Jacobi structure components of $S_1$ and $S_2$. Condition (i) implies $f_2$-ranges of the corresponding dummy edges overlap. Figure~\ref{fig:overlapping-dummy-edges} illustrates the simple cases of overlapping dummy edges for two incomplete simple sheets. However, if there are more than one simple sheet on both sides of the adjacent dummy edges, then it is challenging to decide which two incomplete sheets belong to the same complete Reeb sheet (as shown in Figure~\ref{fig:is-path}). In this case, in addition to (i), the procedure {\sc IsPathConnected} needs to satisfy condition (ii) which checks if 
there is a path from an interior of the sheet $S_1$ to an interior of $S_2$ via. the overlapping part of the dummy edge pair, without crossing the boundary Jacobi structure components of $S_1$ and $S_2$ (lines $17$, $24$, procedure {\sc IsPathConnected}).

\begin{algorithmic}[1]
 \Procedure{IsPathConnected}{$S_1$, $S_2$, $\M$, $\Net_\f$}
 \label{proc:union}
    \State $n_1\gets S_1.\textsc{GetDummyEdgeCount}()$
    \State $n_2\gets S_2.\textsc{GetDummyEdgeCount}()$

    \If{$n_1=0$ or  $n_2=0$}
        \State \Return{False}
    \EndIf
        
    \For{$i\gets 1 \text{ to } n_1$}
        \State $e_1\gets S_1.\textsc{GetDummyEdge}(i)$
        \For{$j=1 \text{ to } n_2$}
            \State $e_2\gets S_2.\textsc{GetDummyEdge}(j)$
    
            \State $u_1\gets e_1.\textsc{StartVertex}()$
            \State $v_1\gets e_1.\textsc{EndVertex}()$
            \State $u_2\gets e_2.\textsc{StartVertex}()$
            \State $v_2\gets e_2.\textsc{EndVertex}()$
    
            \If{$u_1= u_2$ }
                    \State $flagStartVertex\gets$ True
                    \If{$\textsc{IsPath}(S_1, S_2, \M, \Net_\f, u_1, flagStartVertex)$}    
                        \State \Return{True}
                    \Else
                        \State \Return{False}
                    \EndIf
            \ElsIf{$\neg(u_1= u_2)$ and $(v_1 = v_2)$}  
                    \State $flagStartVertex\gets$ False
                    \If{$\textsc{IsPath}(S_1, S_2, \M, \Net_\f, v_1, flagStartVertex)$}
                        \State \Return{True}
                    \Else
                        \State \Return{False}
                    \EndIf
            \Else
                        \State \Return{False}
            \EndIf
        \EndFor
    \EndFor
\EndProcedure
 \end{algorithmic}

Next, we describe the details of the procedure $\textsc{IsPath}$.

 \paragraph*{Procedure: {\sc IsPath}.}
 The procedure {\sc IsPath} checks if two regular points $\widetilde{p}_1$ and $\widetilde{p}_2$, respectively from two possibly adjacent incomplete simple sheets $S_1$ and $S_2$, can be connected by a path without crossing the Jacobi structure components in the boundaries of $S_1$ and $S_2$ (as in Figure~\ref{fig:is-path}). In other words, it checks if a path exists between two regular points $P_1$ and $P_2$, respectively from two regular fiber components corresponding to $\widetilde{p}_1\in S_1$ and $\widetilde{p}_2\in S_2$, without crossing the Jacobi fiber surface in $\M$. The sheets $S_1$ and $S_2$ are adjacent if the dummy edges of $S_1$ and $S_2$ overlap or have a common point, say $v_c$ (as in Figure~\ref{fig:is-path}). 
We choose the points $\widetilde{p}_1\in S_1$ and $\widetilde{p}_2 \in S_2$ from the second representative arcs $\beta^{p_1}$ and  $\beta^{p_2}$ corresponding to simple sheets $S_1$ and $S_2$, respectively. If a path $\Gamma$ is found between $\widetilde{p}_1$ and $\widetilde{p}_2$ without crossing the Jacobi structure components, then $S_1$ and $S_2$ belong to the same complete Reeb sheet. Such a path $\Gamma$ exists in the Reeb space, if an equivalent path $\gamma$ exists between an interior point $P_1\in q_\f^{-1}(\widetilde{p}_1)$ to an interior point $P_2\in q_\f^{-1}(\widetilde{p}_2)$ without crossing the Jacobi fiber surface, in the domain $\M$. To decide the existence of such a path the procedure \textsc{IsPath} finds the associated upper or lower link components of the corresponding Jacobi set parts of $S_1$ and $S_2$, respectively. If such link components intersect, then a desired path can be found. The details of the procedure \textsc{IsPath} is as follows. 

\begin{algorithmic}[1]
 \Procedure{IsPath}{$S_1, S_2, \M, \Net_\f, v_c, flagStartVertex$}
 \label{alg:Is-Path}
\State \% \textit{To compute  the endpoints of the Jacobi set components to be considered for computing links associated with $S_1$ and $S_2$, respectively.}
\State $\beta^{p_1}\gets S_1.\textsc{GetRepArc2}()$
\State $\beta^{p_2}\gets S_2.\textsc{GetRepArc2}()$
\If{$flagStartVertex$}
    \State $n_1\gets \beta^{p_1}.\textsc{StartVertex}()$
    \State $n_2\gets \beta^{p_2}.\textsc{StartVertex}()$
\Else
    \State $n_1\gets \beta^{p_1}.\textsc{EndVertex}()$
    \State $n_2\gets \beta^{p_2}.\textsc{EndVertex}()$
\EndIf

\State $CP_1 \gets \Net_\f.\textsc{GetCriticalPoint}(n_1)$
\State $CP_2 \gets \Net_\f.\textsc{GetCriticalPoint}(n_2)$

\State $V_c \gets \Net_\f.\textsc{GetCriticalPoint}(v_c)$

\State \% \textit{Computing upper or lower link of the Jacobi set componets}
\If{$flagStartVertex$}
    \State $\ell_1 \gets \Net_\f.\textsc{ComputeJacobiSetLink}(CP_1, V_c, \JSet_\f, `upper')$
    \State $\ell_2 \gets \Net_\f.\textsc{ComputeJacobiSetLink}(CP_2, V_c, \JSet_\f, `upper')$    
\Else
    \State $\ell_1 \gets \Net_\f.\textsc{ComputeJacobiSetLink}(CP_1, V_c, \JSet_\f, `lower')$
    \State $\ell_2 \gets \Net_\f.\textsc{ComputeJacobiSetLink}(CP_2, V_c, \JSet_\f, `lower')$
\EndIf

\State \% \textit{If each computed link has exactly one component, then $S_1$ and $S_2$ must belong to the same sheet.}    

\If{$\textsc{GetNumComponents}(\ell_1)=1$ $\And$ $\textsc{GetNumComponents}(\ell_2)=1$}
    \State \Return True;
\EndIf
\State \% \textit{Else, find the link components associated with $S_1$ and $S_2$, respectively.} 
\State $L_1 \gets \Net_\f.\textsc{FindAssoLinkComp}(\beta^{p_1}, \ell_1)$
\State $L_2 \gets \Net_\f.\textsc{FindAssoLinkComp}(\beta^{p_2}, \ell_2)$

\State \% \textit{ If link components $L_1$ and $L_2$ have non-empty intersection, then a path exists.}

\If{$\textsc{HasIntersection}(L_1, L_2)$}
    \State \Return True;
\Else
    \State \Return False;
\EndIf
 \EndProcedure
 \end{algorithmic}

 The procedure \textsc{IsPath} first computes the endpoints of the Jacobi set components for computing the (upper or lower) link components associated with two adjacent incomplete simple sheets (or two incomplete simple sheets such that dummy edges of the sheets have at least one common intersection point) $S_1$ and $S_2$, respectively. First,  \textsc{GetRepArc2} gets the representative arcs $\beta^{p_1}$ and $\beta^{p_2}$ corresponding to  sheets $S_1$ and $S_2$, respectively (lines 3-4, procedure \textsc{IsPath}). If  $flagStartVertex$ is `True', the start nodes of the dummy edges of  $S_1$ and $S_2$ match. Otherwise, if  $flagStartVertex$ is `False', the end nodes of the dummy edges of  $S_1$ and $S_2$ match. This matched node is the intersection of two Jacobi structure components of $S_1$ and $S_2$, respectively. Let us denote this matched node as $v_c$ (as in \figref{fig:is-path}). If $flagStartVertex$ is `True', the procedure \textsc{IsPath} computes the start critical nodes of $\beta^{p_1}$ and $\beta^{p_2}$, respectively. Otherwise, it computes the end critical nodes of $\beta^{p_1}$ and $\beta^{p_2}$, respectively. The computed critical nodes are denoted as $n_1$ and $n_2$, respectively (lines 5-11, procedure \textsc{IsPath}). The procedure \textsc{GetCriticalPoint} computes the critical points $CP_1$, $CP_2$ and $V_c$ corresponding to $n_1$,  $n_2$ and $v_c$, respectively (lines $12$-$14$, procedure \textsc{IsPath}). Note that $CP_1$, $CP_2$, and $V_c$ are the points on the Jacobi set $\JSet_\f$. If  $flagStartVertex$ is `True', the procedure \textsc{ComputeJacobiSetLink} computes the upper links corresponding to the Jacobi set components $\JSet_\f(CP_1, V_c)$ (between $CP_1$ and $V_c$) and $\JSet_\f(CP_2, V_c)$ (between $CP_2$ and $V_c$). Otherwise, the procedure \textsc{ComputeJacobiSetLink} computes the lower links corresponding to the Jacobi set components $\JSet_\f(CP_1, V_c)$ and $\JSet_\f(CP_2, V_c)$. Computed links are denoted by $\ell_1$ and $\ell_2$ (lines $16$-$22$, procedure \textsc{IsPath}).
 If each of the computed upper or lower links $\ell_1$ and $\ell_2$ has exactly one component, then a desired path exists between $S_1$ and $S_2$ (lines $24$-$26$, procedure \textsc{IsPath}).  
 Else, the procedure \textsc{FindAssoLinkComp} finds the link components $L_1$ and $L_2$ associated with the sheets $S_1$ and $S_2$, respectively (lines $28$-$29$, procedure \textsc{IsPath}).
 Finally, the procedure \textsc{HasIntersection} checks if link components $L_1$ and $L_2$ have a non-empty intersection. In that case, a path exists between $S_1$ and $S_2$. Otherwise, no such path exists (lines $31$-$35$, procedure \textsc{IsPath}).
 Next, we discuss the procedures \textsc{ComputeJacobiSetLink} and 
 \textsc{FindAssoLinkComp} in more details.

\paragraph*{Procedure: {\sc ComputeJacobiSetLink}.}
Each of the Jacobi set components $\JSet(CP_1, V_c)$ and $\JSet(CP_2, V_c)$ is considered as a path consisting of a sequence of edges and their faces (vertices) in $\M$, say $\{\bv_0, e_1, \bv_1, e_2, \bv_2, \ldots, e_n, \bv_{n}\}$, where $e_i=\langle \bv_{i-1}, \bv_{i}\rangle$ for $i=1, 2, \ldots, n$. As described in \secref{subsubsec:jacobiset}, the lower (or upper) link of an edge $e_i$ in the Jacobi set components can be computed by defining a PL height field on $\M$ as $h_{\bu_i}(\x) = \langle \f(\x), \bu_i \rangle$. 
We consider $\f(e_i) \subset \R^2$ and a vector $\bu_i$ normal to $\f(e_i)$ such that the second coordinate of $\bu_i$ should be positive. This is because ``upper'' or ``lower'' corresponds to those of the $f_2$-values. The lower (upper) link of $e_i$ consists of simplices in the link of $e_i$ having $h_{\bu_i}$-values strictly less (greater) than the vertices of $e_i$. Now to find a continuous path via. the lower (upper) link of the Jacobi set components, we also need to compute a restricted lower (upper) links of each vertex on the Jacobi set components. 
For computing the lower (upper) link of a vertex $\bv_i$, we consider a PL height field $h_{\mathbf{n}_{\bv_i}}(\x)$ where unit normal direction $\mathbf{n}_{\bv_i}$ corresponding to $\bv_i$ is chosen by interpolating the normal directions $\bu_{i}$ and $\bu_{i+1}$ of its adjacent edges $e_i$ and $e_{i+1}$. Then the restricted lower (upper) link is computed by deleting the simplices of the lower link intersecting the Jacobi set. Thus for the Jacobi set components $\JSet(CP_1, V_c)$ and $\JSet(CP_2, V_c)$ we obtain restricted lower (or upper) links $\ell_1$ and $\ell_2$, respectively, consisting of one or two components as shown in Figure \ref{fig:is-path}.

\paragraph*{Procedure: {\sc FindAssoLinkComp}.}
Each of the computed lower (or upper) links  
$\ell_1$ and $\ell_2$ corresponding to $\JSet(CP_1, V_c)$ and $\JSet(CP_2, V_c)$, respectively, may have one or two components. {\sc FindAssoLinkComp} associates the component of $\ell_1$ associated with $S_1$ and the component of $\ell_2$ associated with $S_2$.
For that {\sc FindAssoLinkComp} first finds associated link corresponding to the representative arc $\beta^{p_1}$ of $S_1$, say $\ell_1^{p_1}$, and associated link corresponding to the representative arc $\beta^{p_2}$ of $S_2$, say $\ell_2^{p_2}$. 
Next, it finds the component $L_1$ of $\ell_1$ which has a non-empty intersection with $\ell_1^{p_1}$ and the component $L_2$ of $\ell_2$ which has a non-empty intersection with $\ell_2^{p_2}$. We note, only the link component associated with $S_1$ will have a non-empty intersection with the link component associated with $\beta^{p_1}$ and only the link component associated with $S_2$ will have a non-empty intersection with the link component associated with $\beta^{p_2}$.

Next, we provide the proof of the correctness of our algorithm.

\subsection{Proof of Correctness}
\label{subsec:proof-correctness}
In this subsection, we show the Reeb space obtained by  \algoref{alg:compute-Reeb-space} is topologically correct which is followed by - (i) computation of correct MDRG, (ii) computation of a topologically correct embedding of the second dimensional Reeb graphs in the MDRG as a net-like structure corresponding to the Reeb space and (iii)  computation of correct complete $2$-sheets of the Reeb space in the net-like structure. 
The following lemma proves the correctness of our algorithm.

\begin{lemma}
 Let $\f = (f_1, f_2):\M\rightarrow\R^2$ be a generic PL bivariate field defined on a triangulation $\M$ of a compact, orientable  $3$-manifold without boundary. 
 Let $\f$ satisfy the genericity conditions (i)-(iii) in \secref{sec:theory-contributions}.
    Then \algoref{alg:compute-Reeb-space} computes the topologically correct Reeb space corresponding to $\RS_{\f}$.
\label{lemma:proof-of-correctness}
\end{lemma}

\begin{proof}
From \propref{prop:homeo}, we note, the $\MDRG_{\f}$ is homeomorphic to $\RS_{\f}$. Specifically, the second-dimensional Reeb graphs of $\MDRG_{\f}$ have an embedding in $\RS_{\f}$ (see Lemma \ref{lem:homeomorphism-embedding}). Therefore, by examining the variation in the topology of the second-dimensional Reeb graphs $\RG_{\widetilde{f_2^p}}$, as $p$ varies along arcs of $\RG_{f_1}$, the topology of the Reeb space is effectively captured. Let $\alpha$ be an arc in the Reeb graph $\RG_{f_1}^{AugIII}$, which is augmented based on the points of topological change. Then the Reeb graphs in $\{\RG_{\widetilde{f_2^p}} \mid p \in \alpha\}$ are topologically equivalent (see \lemref{lem:topological-changes}). Therefore, for capturing the topology of these Reeb graphs, it is sufficient to choose a representative point $p$ in $\alpha$ for computing the embedding of the Reeb graph $\RG_{\widetilde{f_2^p}}$ into $\RS_{\f}$. 

However, it is essential to capture the topological variations in the second-dimensional Reeb graphs $\RG_{\widetilde{f_2^p}}$ as $p$ varies across different arcs of $\RG_{f_1}^{Aug III}$. We note, the points of topological change on $\RG_{f_1}^{Aug III}$ of $\RG_{\widetilde{f_2^p}}$ correspond to the critical points of $f_1$ or where $\widetilde{f_2^p}$ violates one of the two Morse conditions (Lemma~\ref{lem:topological-changes}). These critical points are on the Jacobi set $\JSet_{\f}$. Since $\JStruct_{\f}$ is the projection of $\JSet_{\f}$ to $\RS_{\f}$, the nodes of $\RG_{\widetilde{f_2^p}}$ embedded in $\RS_{\f}$ are located on $\JStruct_{\f}$. Thus, $\JStruct_{\f}$ tracks the topological changes of the second-dimensional Reeb graphs embedded in $\RS_{\f}$. Therefore, \algoref{alg:net-like-structure} computes a topologically correct embedding of the second-dimensional Reeb graphs in $\MDRG_{\f}$ corresponding to  $\RS_\f$.

Furthermore, the procedures ComputeSimpleSheet and CompatibleUnion 
compute the $2$-sheets of $\W_\f$ in the computed $\Net_\f$, which are 
correct by the following reasons. First, when we vary $p \in \alpha$, 
vertices of the embedded Reeb graphs $\RG_{\widetilde{f_2^p}}$ sweep out the Jacobi 
structure $\JStruct_\f$. Furthermore, by the proof of \propref{prop:homeo} 
we see that the edges of $\RG_{\widetilde{f_2^p}}$
sweep out simple $2$-sheets of $\W_\f$. Thus, $\W_\f$ can be
constructed by attaching these simple $2$-sheets to the Jacobi structure
along their boundaries in a correct manner.
For each arc $\alpha$ of $\RG_{f_1}^{Aug III}$ and each arc
$\beta^p$ of $\RG_{\widetilde{f_2^p}}$ with $\{p\} = \alpha \cap P_R$,
ComputeSimpleSheet provides a correct output of the
corresponding simple $2$-sheet. In order to
attach each simple $2$-sheet correctly to the Jacobi structure to get
the correct Reeb space, it is straightforward to see
the Jacobi set edges on the boundary along which
we attach a given simple $2$-sheet. 
However, if a simple $2$-sheet is not
complete, then it should be attached to another
incomplete $2$-sheet along dummy edges in a correct way.
This is done by our procedure CompatibleUnion,
with the help of the IsPathConnected procedure.

Thus, the output $\RSS_\f$ of Algorithm 4 is topologically equivalent or 
homeomorphic to $\W_\f$.
\end{proof}

Next, we discuss the complexity of the proposed algorithms.

\section{Complexity Analysis}
\label{sec:complexity-analysis}
In this section, we analyze the complexity of the proposed algorithm for computing the Reeb space of a generic PL bivariate field $\f = (f_1,f_2): \M \rightarrow \R^2$,  defined on a triangulation $\M$ of a compact, orientable $3$-manifold without boundary. Let the numbers of vertices, edges, triangles, and tetrahedra in $\M$ be denoted as $n_v$, $n_e$, $n_t$, and $n_T$ respectively, and the total number of simplices is $n = n_v + n_e + n_t + n_T$. Let $j_{v}$ and $j_e$ represent the numbers of vertices and edges of the Jacobi set $\JSet_{\f}$, respectively. Further, when the Jacobi set is projected in the range of $\f$, for a pair of non-adjacent edges $e(\bu,\bv)$ and $e(\bu',\bv')$ of the Jacobi set $\JSet_\f$, their projections $\f(e(\bu,\bv))$ and $\f(e(\bu',\bv'))$  may intersect. Let $c_{int}$ denote the number of such intersections.
Moreover, note that the link $\Link{\tilde{e}}$ of an edge $\tilde{e}$ in $\M$ is a $1$-sphere consisting of vertices and edges in $\M$. To obtain our complexity bound, we also assume the upper bound on the number of simplices in $\Link{\tilde{e}}$ or $|\Link{\tilde{e}}|$ is $c_L$ for any edge $\tilde{e}\in \M$. 

First, we provide the complexity analysis for computing the Jacobi structure $\JStruct_{\f}$ (\algoref{alg:Jacobi-Structure}). Next, we analyze the complexity of computing $\MDRG_{\f}$ (\algoref{alg:compute-MDRG}). Then, we provide the complexity analysis for computing the net-like structure $\Net_{\f}$ corresponding to the Reeb space (\algoref{alg:net-like-structure}). Finally, we determine the complexity for computing the $2$-sheets of  the Reeb space $\RS_{\f}$ (\algoref{alg:compute-Reeb-space}).

\subsection{Complexity of \algoref{alg:Jacobi-Structure}: Computing the Jacobi Structure}
\label{subsec:complexity-jacobi-structure}
First, the Reeb graph $\RG_{f_1}$ is constructed, which takes $\cO(n \log n)$ time (line $3$, \algoref{alg:Jacobi-Structure}). This is the best-known lower bound for computing the Reeb graph \cite{book-tamal-dey}. Line $4$ invokes the procedure \textsc{ComputeJacobiMinima} for computing the minima of $f_1$ restricted to $\JSet_{\f}$. Given that $\JSet_{\f}$ consists of PL $1$-manifold components, each vertex of $\JSet_{\f}$ has at most two neighbours. Thus, determining whether a vertex of $\JSet_{\f}$ is a minimum of $f_1$ requires examining the $f_1$-values of its neighbours, which takes constant time. Consequently, \textsc{ComputeJacobiMinima} requires $\cO(j_v)$ time. The time complexity for computing the maxima is similar (line $5$, \algoref{alg:Jacobi-Structure}). After this step, computing the union of $J_{min}$ and $J_{max}$ takes a time which is linear in the cardinalities of these two sets. Let $j_{min}$ and $j_{max}$ represent the numbers of minima and maxima of $f_1$ restricted to $\JSet_{\f}$, respectively. Then the cardinalities of $J_{min}$ and $J_{max}$ are upper-bounded by $j_{min}$ and $j_{max}$, respectively. Therefore, the time complexity of computing $P'$ is $\cO(j_{min} + j_{max})$ (line $6$, \algoref{alg:Jacobi-Structure}). 

The next step is to augment the Reeb graph $\RG_{f_1}$ based on the points in $P'$ (line $7$, \algoref{alg:Jacobi-Structure}). For each point $\x$ in $P'$, the corresponding arc of $\RG_{f_1}$ is split into two by introducing a node. This operation takes constant time for each point in $P'$. Thus, the complexity of line $7$ is $\cO(|P'|)$, which is upper-bounded by $\cO(j_{min} + j_{max})$. The overall time taken by lines $1$-$7$ is $\cO(n \log(n) + 2j_v + 2(j_{min} + j_{max}))$. 

Next, we assess the complexity of lines $8$-$28$ which compute the Jacobi structure.
The Jacobi structure $\JStruct_{\f}$ is computed by individually processing each edge of the Jacobi set $\JSet_{\f}$ as follows. For an edge $\e(\bu,\bv)$ in $\JSet_{\f}$, the corresponding points $q_{\f}(\bu)$ and $q_{\f}(\bv)$ are taken as $u$ and $v$. Then an edge $e(u,v)$ corresponding to $e(\bu,\bv)$ is added in $\JStruct_{\f}$, which takes constant time (lines $9$-$22$, \algoref{alg:Jacobi-Structure}). After this step, the intersection of the edge $e(u, v)$ is checked with the previously computed edges of $\JStruct_{\f}$ where the corresponding pair of Jacobi edges are non-adjacent (lines $23$-$26$, \algoref{alg:Jacobi-Structure}). To determine the time complexity of these lines, we assess the time complexity for the procedure \textsc{Intersection}, which takes two non-adjacent edges $e(\bu,\bv)$ and $e(\bu',\bv')$ of the Jacobi set as input, and determines the intersection of $q_{\f}(e(\bu, \bv))$ and $q_{\f}(e(\bu',\bv'))$.

The first step in the \textsc{Intersection} procedure is determining the intersection of $\f(e(\bu,\bv))$ and $\f(e(\bu',\bv'))$ which takes constant time (line $3$, procedure \textsc{Intersection}). In the event of an intersection in the range of $\f$, the point of intersection is computed (line $4$, procedure \textsc{Intersection}). Then, the projections of $e(\bu,\bv)$ and $e(\bu',\bv')$ on $\RG_{f_1}^{Aug II}$ (by the quotient map $q_{f_1}$) are examined for intersection, which also takes constant time (line $6$, procedure \textsc{Intersection}). We note that the information of the vertices of $\M$ mapped to an arc of $\RG_{f_1}$ are already stored during the computation of $\RG_{f_1}$. If an intersection is found, then a point $p$ within the intersecting region of $\RG_{f_1}^{Aug II}$ is selected (line $7$, procedure \textsc{Intersection}). Following this, the contour $q_{f_1}^{-1}(p)$ is computed, and the intersection of $q_{f_1}^{-1}(p)$ with the edges $e(\bu,\bv)$ and $e(\bu',\bv')$ are determined, to obtain the points $\x$ and $\y$, respectively (lines $8$-$9$, procedure \textsc{Intersection}). The time complexity of computing $q_{f_1}^{-1}(p)$ and determining the intersections is bounded by $\cO(n_T)$ \cite{1996-Livnar-Isosurface}. The next step is the computation of the Reeb graph $\RG_{\widetilde{f_2^{p}}}$, which takes $\cO(n' \log (n'))$ time, where $n'$ is the number of simplices (vertices, edges, and triangles) of $q_{f_1}^{-1}(p)$ (line $11$, procedure \textsc{Intersection}). The overall complexity of lines $3$-$11$ is $\cO(n' \log (n') + n_T)$. Since $n_T \leq n$ and $n' \leq n$, this bound can be expressed as $\cO(n \log (n) + n)$.

After this step, the adjacency of nodes $q_{\widetilde{f_2^{p}}}(\x)$ and $q_{\widetilde{f_2^{p}}}(\y)$ in $\RG_{\widetilde{f_2^{p}}}$ is examined by checking the presence of $q_{\widetilde{f_2^{p}}}(\x)$ in the adjacency list of $q_{\widetilde{f_2^{p}}}(\y)$ (line $12$, procedure \textsc{Intersection}). We note, the functions $\widetilde{f_2^p}$ are Morse except for a finite set of points in $\RG_{f_1}^{Aug II}$. Therefore, the number of adjacent nodes of $q_{\widetilde{f}_2^{p}}(\x)$ is upper-bounded by $4$ (the bound $4$ is achieved in the case where $q_{\widetilde{f}_2^{p}}(\x)$ is a double fork). Therefore, line $12$ requires constant time. Finally, computing the intersection point of the projections of $e(\bu,\bv)$ and $e(\bu',\bv')$ in $\RS_{\f}$, and then subdividing the edges $e(q_{\f}(\bu),q_{\f}(\bv))$ and $e(q_{\f}(\bu'),q_{\f}(\bv'))$, take constant time (lines $13$-$21$, procedure \textsc{Intersection}). Hence, the total complexity of the procedure \textsc{Intersection} is $\cO(n \log(n) + n)$. However, this bound applies only when the projections of $e(\bu,\bv)$ and $e(\bu',\bv')$ in the range of $\f$ intersect (line $3$, procedure \textsc{Intersection}). Otherwise, the procedure \textsc{Intersection} takes $\cO(1)$ time.

 The for loop in line $23$ of \algoref{alg:Jacobi-Structure} iterates through at most all the edges of $\JSet_{\f}$. Similarly, the for loop in line $8$ iterates over all the edges of $\JSet_{\f}$. Therefore, the complexity for the iterations of both for loops together is $\cO(j_e^2)$. However, the time complexity of the  procedure \textsc{Intersection} is $\cO(n \log(n) + n)$  only for $c_{int}$ pairs of Jacobi set edges. In other instances, it takes $\cO(1)$ time.  Therefore, the time complexity of lines $8$-$28$ of \algoref{alg:Jacobi-Structure} is $\cO(c_{int}(n \log (n) + n) + (j_e^2 - c_{int})) = \cO(j_e^2 + c_{int}(n \log (n) + n)) $. Thus, the total complexity of \algoref{alg:Jacobi-Structure} is $\cO(n \log(n) + 2j_v + 2(j_{min} + j_{max}) + j_e^2 + c_{int}(n \log (n) + n))$. Since $j_v, j_{min}$ and $j_{max}$ are at most $n$, which is in turn upper-bounded by $n \log (n)$, the complexity bound can be simplified as $\cO(j_e^2 + (c_{int}+1)(n \log (n)))$.  In the next subsection, we analyze the time complexity for computing the MDRG (\algoref{alg:compute-MDRG}).

\subsection{Complexity of \algoref{alg:compute-MDRG}: Computing the MDRG}
\label{subsec:complexity-MDRG}
The computation of $\MDRG_{\f}$ begins with the construction of the Reeb graph $\RG_{f_1}$, which takes $\cO(n \log(n))$ time (line $3$, \algoref{alg:compute-MDRG}). We note, this is the best-known lower bound for computing the Reeb graph \cite{book-tamal-dey}. Line $4$ invokes the procedure \textsc{ComputeJacobiMinima} for computing the minima of $f_1$ restricted to $\JSet_{\f}$ which takes $\cO(j_v)$ time (as in \secref{subsec:complexity-jacobi-structure}). The time complexity for computing the maxima is similar (line $5$, \algoref{alg:compute-MDRG}). The procedure $\textsc{DoublePoints}$ identifies the double points of $\JStruct_{\f}$ by examining the degree of every vertex. Therefore, this procedure takes linear time in the number of vertices of $\JStruct_{\f}$. Since $\JStruct_{\f}$ is obtained by the projection of Jacobi edges in  $\JSet_{\f}$ onto the Reeb space where projection of a pair of non-adjacent Jacobi edges may have an intersection, the time complexity of the procedure $\textsc{DoublePoints}$ is $\cO(j_v+c_{int})$ (line $6$, \algoref{alg:compute-MDRG}).

After this step, computing the union of $J_{min},\, J_{max}$, and $DP$ takes a linear time in the cardinalities of these three sets. The cardinalities of $J_{min}$ and $J_{max}$ are upper-bounded by $j_{min}$ and $j_{max}$, respectively (as in \secref{subsec:complexity-jacobi-structure}). Further, the number of double points of $\JStruct_{\f}$ is upper-bounded by $c_{int}$. Therefore, the time complexity of line $7$ is $\cO(j_{min} + j_{max} + c_{int})$. Similar to line $7$ in \algoref{alg:Jacobi-Structure}, line $8$ in \algoref{alg:compute-MDRG} augments the Reeb graph $\RG_{f_1}$ based on the additional points in $P$ is  $\cO(|P|)$ time (see \secref{subsec:complexity-jacobi-structure} for further details). Since $|P|$ is upper-bounded by ($j_{min} + j_{max} + c_{int})$, the overall time taken by lines $1$-$9$ is $\cO(n \log(n) + 3j_v + c_{int} + 2(j_{min} + j_{max} + c_{int}))$. Next, we assess the complexity of lines $10$-$15$.

We note, the nodes in the augmented Reeb graph $\RG_{f_1}^{Aug III}$ constructed at line $8$, correspond to either the critical points (minimum or maximum) of $f_1$ restricted to $\JSet_{f}$, or the double points of $\JStruct_{\f}$. The number of critical points is upper-bounded by $j_{min} + j_{max}$, and the number of double points is at most $c_{int}$. Therefore, the number of nodes of $\RG_{f_1}^{Aug III}$ is at most $(j_{min} + j_{max} + c_{int})$. Given that $f_1$ is a generic PL Morse function, the up-degree (similarly down-degree) of a node of $\RG_{f_1}^{Aug III}$ can be at most $2$ (see \secref{subsec:Reeb-graph} for more details). Thus, the number of arcs of $\RG_{f_1}^{Aug III}$ is at most twice the number of nodes. Let $\mathcal{S}_{f_1}=\{q_{f_1}^{-1}(p_{\alpha}) \mid \alpha \in Arcs(\RG_{f_1}^{AugIII})\}$ represent the set of contours of $f_1$ each corresponding to a representative point $p_{\alpha}$ of arc $\alpha$ in $Arcs(\RG_{f_1}^{AugIII})$. Then, the number of contours in $\mathcal{S}_{f_1}$ is upper-bounded by $2 (j_{min} + j_{max} + c_{int})$. For a representative point $p_{\alpha}$ of an arc $\alpha$ in $Arcs(\RG_{f_1}^{AugIII})$, computing the contour $q_{f_1}^{-1}(p_{\alpha})$ takes $\cO(n_T)$ time \cite{1996-Livnar-Isosurface}. Then, the total time complexity of computing all the contours of $\mathcal{S}_{f_1}$ is $\cO(2 (j_{min} + j_{max} + c_{int}) n_T)$. Next, we analyze the complexity of computing the second-dimensional Reeb graphs of $\MDRG_{\f}$, corresponding to the contours of $\mathcal{S}_{f_1}$.

We assume the mesh $\M$ is sufficiently refined such that each tetrahedron in $\M$ can have intersections with at most $c_{int}+1$ contours in $\mathcal{S}_{f_1}$ ($c_{int}$ is the upper-bound for the number of double points of $\JStruct_{\f}$). Thus, the total number of intersections of all the tetrahedra with all the contours in $\mathcal{S}_{f_1}$ is at most $n_T(c_{int}+1)$. Let $p_{\alpha}$ be the representative point of an arc of $\RG_{f_1}^{Aug III}$. Since $p_{\alpha}$ is a regular point of $\alpha$, $q_{f_1}^{-1}(p_{\alpha})$ is of dimension two and consists of plane sections of tetrahedra of $\M$. For each tetrahedron, this section might be empty, a triangle, or a quadrilateral, and in the last case, it should be further triangulated into triangles. Hence, each tetrahedron of $\M$ has at most four vertices of $q_{f_1}^{-1}(p_{\alpha})$. Similarly, the numbers of edges and triangles of $q_{f_1}^{-1}(p_{\alpha})$ in a tetrahedron of $\M$ are at most five and two, respectively. Thus, the number of simplices of $q_{f_1}^{-1}(p_{\alpha})$ in a tetrahedron is at most $11$. So the total number of simplices of $q_{f_1}^{-1}(p_{\alpha})$ over all tetrahedra, in $\M$, is $11 n_T$. Moreover, the total number of simplices together for all the contours in $\mathcal{S}_{f_1}$ can be given as $\cO(11 n_T(c_{int}+1))$. Hence, the time complexity of computing all the second-dimensional Reeb graphs is $\cO(11 n_T(c_{int}+1) \log (11 n_T)$. 
Therefore, lines $11$-$15$ of \algoref{alg:compute-MDRG} take $\cO(2 (j_{min} + j_{max} + c_{int}) n_T + 11 n_T(c_{int}+1) \log (11 n_T)$ time. Finally, the total time complexity of \algoref{alg:compute-MDRG} is then given by $\cO(n \log(n) + 3 j_v + c_{int}   
+ 2(j_{min} + j_{max} + c_{int}) + 2 (j_{min} + j_{max} + c_{int}) n_T + 11 n_T(c_{int} + 1) \log (n_T))$. Since $n_T, j_v$, and $(j_{min} + j_{max})$ are bounded above by $n$, the complexity bound can be expressed as $\cO(n \log(n) + 5n + 2n^2 + 11n(c_{int} + 1)\log (n) + 3c_{int} + 2c_{int}n) = \cO( n^2 + n\,c_{int}\,\log (n))$.

Next, we analyze the time complexity for computing the net-like structure corresponding to the Reeb space (\algoref{alg:net-like-structure}).

\subsection{Complexity of \algoref{alg:net-like-structure}: Computing the Net-like Structure}
\label{subsec:complexity-net-like-structure}

The lines $1$-$2$ of \algoref{alg:net-like-structure} initialize the net-like structure to the Jacobi structure and retrieve the first-dimensional Reeb graph from the MDRG, both of which take constant time. We analyze the time complexity of lines $3$-$7$ by determining the time complexity of the procedure $\textsc{EmbedReebGraph}$, which embeds the second-dimensional Reeb graphs of $\MDRG_{\f}$.

For a representative point $p$ of an arc in $\RG_{f_1}^{Aug III}$, consider the second-dimensional Reeb graph $\RG_{\widetilde{f_2^p}}$. For an arc $\beta^p$ of  $\RG_{\widetilde{f_2^p}}$, let $p_1$ and $p_2$ denote its start and end nodes. Then, the contour $q_{\widetilde{f_2^p}}^{-1}(p_1)$ (similarly $q_{\widetilde{f_2^p}}^{-1}(p_2)$) contains at least one critical point of $\widetilde{f_2^p}$. From \lemref{lem:topological-changes}, it follows that $\widetilde{f_2^p}$ is a Morse function. Therefore, $q_{\widetilde{f_2^p}}^{-1}(p_1)$ contains exactly one critical point, say $\x_1$, as the presence of more than one would violate the second Morse condition. Since $\x_1$ is a critical point of $f_2$ restricted to a level set of $f_1$, it lies on the Jacobi set. To project $\x_1$ into the Reeb space (by the quotient map $q_{\f}$), we need to determine the edge of $\JSet_{\f}$ containing $\x_1$. This requires examining all edges of $\JSet_{\f}$, and takes $\cO(j_e)$ time. Thus line $4$ (and similarly line $6$) of the procedure \textsc{EmbedReebGraph} takes $\cO(j_e)$ time. After this step, the addition of an edge to $\Net_{\f}$ corresponding to the projection of $\beta^p$ takes constant time (line $7$, procedure \textsc{EmbedReebGraph}). The complexity of the for loop in line $2$ of the procedure \textsc{EmbedReebGraph} is bounded by the number of arcs of $\RG_{\widetilde{f_2^p}}$. Since $\widetilde{f_2^p}$ is Morse, the number of arcs in $\RG_{\widetilde{f_2^p}}$ is at most twice the number of nodes (as discussed in \secref{subsec:complexity-MDRG}). Let $c_{\widetilde{f_2^p}}$ denote the number of critical points of $\widetilde{f_2^p}$. Then, the time complexity of the procedure \textsc{EmbedReebGraph} is $\cO(2 c_{\widetilde{f_2^p}} (2 j_e)) \simeq \cO(4 c_{\widetilde{f_2^p}} j_e)$.

The for loop in line $3$ of \algoref{alg:net-like-structure} takes time linear in the number of arcs of $\RG_{f_1}^{Aug III}$. However, the total number of critical points $c_{\widetilde{f_2^p}}$, over the representative points of all the arcs, is at most the number of edges in the Jacobi set $j_e$. In other words, we have $\sum_{\alpha \in \RG_{f_1}^{Aug III}} c_{\widetilde{f_2^p}} \leq j_e$, where $p$ is the representative point of the arc $\alpha$. Therefore, lines $3$-$7$ of \algoref{alg:net-like-structure} take $\cO(4 j_e^2)$ time. The total time complexity of \algoref{alg:net-like-structure} is then $\cO(4 j_e^2)$.

Next, we analyze the total time complexity of the algorithm for computing the Reeb space (\algoref{alg:compute-Reeb-space}). 

\subsection{Complexity of \algoref{alg:compute-Reeb-space}: Computing the Reeb Space}
\label{subsec:complexity-Reeb-Space}
The computation of the Reeb space starts with the construction of the Jacobi set $\JSet_{\f}$, which takes $\cO(n_e)$ time (line $2$, \algoref{alg:compute-Reeb-space}) \cite{2016-Julien-ReebSpace}. Next, the computation of the Jacobi structure $\JStruct_{\f}$ takes $\cO(j_e^2 + (c_{int}+1)(n \log (n)))$ time (line $3$, \algoref{alg:compute-Reeb-space}). Then, the MDRG of $\f$ is computed, which takes $\cO(n^2 + n\,c_{int}\,\log (n))$ time (line $5$, \algoref{alg:compute-Reeb-space}). Next, the computation of the net-like structure takes $\cO(4 j_e^2)$ time (line $7$, \algoref{alg:compute-Reeb-space}). After this step, the first-dimensional Reeb graph $\RG_{f_1}^{Aug III}$ is retrieved from the MDRG, which takes constant time (line $9$, \algoref{alg:compute-Reeb-space}). For each arc $\alpha$ of  $\RG_{f_1}^{Aug III}$, we first obtain its representative point $p$, and retrieve the second-dimensional Reeb graph $\RG_{\widetilde{f_2^p}}$ from $\MDRG_{\f}$ (lines $12$-$13$, \algoref{alg:compute-Reeb-space}). These steps also take constant time. Then, for each arc of $\RG_{\widetilde{f_2^p}}$, we compute the simple Reeb sheet $ReebSheet(\alpha,\beta_p)$ by the procedure \textsc{ComputeSimpleSheet} (line $15$, \algoref{alg:compute-Reeb-space}). Next, we analyze the time taken by this procedure for an arc $\beta^p$ of $\RG_{\widetilde{f_2^p}}$, where $p$ is the representative point of an arc $\alpha$ of $\RG_{f_1}^{Aug III}$.

The procedure \textsc{ComputeSimpleSheet} begins by retrieving the start and end nodes ($p_1$ and $p_2$) of $\alpha$, and their corresponding $\overline{f_1}$ values (lines $2$-$5$, procedure \textsc{ComputeSimpleSheet}). Similarly, for $\beta^p$, the start and end nodes ($p_1'$ and $p_2'$) are retrieved (lines $6$-$7$, procedure \textsc{ComputeSimpleSheet}). We note, the contour $q_{\widetilde{f_2^p}}^{-1}(p_1')$ contains at least one critical point of $\widetilde{f_2^p}$. From \lemref{lem:topological-changes}, it follows that $\widetilde{f_2^p}$ is a Morse function. Therefore, $q_{\widetilde{f_2^p}}^{-1}(p_1')$ contains exactly one critical point, say $\x_1$, as the presence of more than one would violate the second Morse condition (line $8$, procedure \textsc{ComputeSimpleSheet}). Since $\x_1$ is a critical point of $f_2$ restricted to a level set of $f_1$, it lies on the Jacobi set. To project $\x_1$ onto $\RS_{\f}$ (by the quotient map $q_{\f}$), we need to determine the edge of $\JSet_{\f}$ containing $\x_1$. This process involves examining all edges of $\JSet_{\f}$, and takes $\cO(j_e)$ time. Once the edge containing $x_1$ is identified, the projection of $\x_1$ onto $\RS_{\f}$ is determined by projecting the endpoints of the identified edge of $\JSet_{\f}$ containing $\x_1$, a step that takes constant time. Thus lines $8$ and $14$ of the procedure \textsc{ComputeSimpleSheet} together take $\cO(j_e)$ time. Similarly, $q_{\widetilde{f_2^p}}^{-1}(p_2')$ contains exactly one critical point $\x_2$ of  $\widetilde{f_2^p}$, and projecting $\x_2$ onto $\RS_{\f}$ takes $\cO(j_e)$ time (lines $9$ and $15$, procedure \textsc{ComputeSimpleSheet}). Lines $11$-$13$ initialize the sheet boundary and the dummy edge count, which take constant time. Thus, lines $1$-$15$ of the \textsc{ComputeSimpleSheet} procedure takes $\cO(2j_e)$ time.

Next, the procedure \textsc{ComputeBoundary} computes the boundary of a simple sheet $ReebSheet(\alpha, \beta^p)$ corresponding to $\alpha\in \RG_{f_1}^{Aug III} $ and $\beta^p\in \RG_{\widetilde{f_2^p}}$, by moving along the Jacobi structure in the monotonically increasing and decreasing directions of $\bar{f_1} \circ \omega_1$, (lines $16$-$19$ and $24$-$27$, procedure \textsc{ComputeSimpleSheet}). 
Since no double point can occur in the interior of the traced path on the Jacobi structure, the time complexity of tracing the boundary of $ReebSheet(\alpha, \beta^p)$ is bounded by the number of edges in the Jacobi set, i.e.  $\cO(j_e)$. After tracing the boundaries, at most two additional edges are added, and the dummy edge counts are updated (lines $20$-$23$ and $28$-$31$,  procedure \textsc{ComputeSimpleSheet}). These steps take constant time. After this step, $simpleSheet$ consists of edges forming the boundary of the simple Reeb sheet. Finally, updating the status of $simpleSheet$ as complete or incomplete based on the dummy edge count, and setting the dummy edge count, take constant time (lines $33$-$38$, procedure \textsc{ComputeSimpleSheet}). Thus, lines $16$-$38$ take $\cO(j_e)$ time. Therefore, the overall time complexity of the procedure \textsc{ComputeSimpleSheet} is then $\cO(3 j_e)$.

The bound for the number of iterations of the for loop in line $14$ of \algoref{alg:compute-Reeb-space} is similar to that of the for loop in the procedure \textsc{EmbedReebGraph} (see \secref{subsec:complexity-net-like-structure} for more details). Thus, the time complexity of lines $14$-$18$ is $\cO(2 c_{\widetilde{f_2^p}} (3 j_e)) \simeq \cO(6 c_{\widetilde{f_2^p}} j_e)$. The for loop in line $11$ takes time which is linear in the number of arcs of $\RG_{f_1}^{Aug III}$. However, the total number of critical points $\displaystyle\sum_{\alpha \in \RG_{f_1}^{Aug III}}c_{\widetilde{f_2^{p_{\alpha}}}}$, where $p_{\alpha}$ is the representative point of the arc $\alpha$, is at most the number of edges in the Jacobi set ($j_e$). In other words, we have $\displaystyle\sum_{\alpha \in \RG_{f_1}^{Aug III}} c_{\widetilde{f_2^{p_{\alpha}}}} \leq j_e$. Therefore, lines $11$-$19$ of \algoref{alg:compute-Reeb-space} take $\cO(6 j_e^2)$ time. Next, we examine the time complexity of the procedure \textsc{CompatibleUnion}.

The procedure \textsc{CompatibleUnion} begins by initializing the union-find data structure by creating a set for each simple Reeb sheet, which takes a time linear in the number of simple Reeb sheets (lines $2$-$7$). We note, for each arc $\alpha$ of $\RG_{f_1}^{Aug III}$, a simple Reeb sheet is computed corresponding to each arc in the second-dimensional Reeb graph $\RG_{\widetilde{f_2^{p_{\alpha}}}}$ (lines $11$-$19$, procedure \textsc{ComputeReebSpace}). As discussed in Sections \ref{subsec:complexity-jacobi-structure} and \ref{subsec:complexity-MDRG}, the number of arcs in $\RG_{\widetilde{f_2^p}}$ is at most twice the number of nodes. Therefore, the total number of simple Reeb sheets in $simpleSheets$ is at most $2 \sum_{\alpha}c_{\widetilde{f_2^{p_{\alpha}}}}$, Since $\sum_{\alpha}c_{\widetilde{f_2^{p_{\alpha}}}}$ is bounded above by $j_e$, the lines $2$-$7$ of the procedure \textsc{CompatibleUnion} take $\cO(2j_e)$ time. Next, we analyze the time complexity of the procedure \textsc{IsPathConnected} for two simple Reeb sheets $S_1$ and $S_2$ (line $12$, procedure \textsc{CompatibleUnion}).

The for loops in lines $7$ and $9$ of this procedure iterate through the dummy edges of $S_1$ and $S_2$. Since a simple Reeb sheet can have at most two dummy edges, each for loop iterates at most twice. The lines $8$, $10$-$14$ obtain the dummy edges and their start and end vertices. Therefore, these lines take constant time. The lines $15$ and $22$ check for the equivalence of vertices, which also takes constant time. Thus, the time complexity of the procedure \textsc{IsPathConnected} is determined by the complexity of the \textsc{IsPath} procedure (lines $17$ and $24$, procedure \textsc{IsPathConnected}).

Lines $1$-$11$ of the \textsc{IsPath} procedure retrieve the representative arcs in the second-dimensional Reeb graphs (in the MDRG) corresponding to the sheets $S_1$ and $S_2$, and the corresponding start or end nodes, $n_1$ and $n_2$. Lines $12$-$14$ retrieve the critical points $CP_1, CP_2$, and $V_c$ corresponding to the Reeb graph nodes $n_1, n_2$ and $v_c$, respectively. All of these steps take constant time. Next, the links of the Jacobi set edges along the path between the points $CP_1$ and the $V_c$ ($\JSet_{\f}(CP_1, V_c)$), and the path between $CP_2$ and $V_c$ ($\JSet_{\f}(CP_2, V_c)$) are computed by the procedure \textsc{ComputeJacobiSetLink} (lines $16$-$22$, procedure \textsc{IsPath}). The time complexity for computing the links depends on the number of simplices in the star of each of the edges in $\JSet_{\f}(CP_1, V_c)$ and $\JSet_{\f}(CP_2, V_c)$. Thus, the time taken by lines $16$-$22$ is at most $\cO(\sum_{e(\bu,\bv) \in j_{S_1,S_2}} |\Star{e(\bu,\bv)}|)$, where $j_{S_1,S_2} = \JSet(CP_1, V_c) \cup \JSet(CP_2, V_c)$, and  $|\Star{e(\bu,\bv)}|$ is the number of simplices in the star of $e(\bu,\bv)$. After this step, the procedure \textsc{GetNumComponents} computes the number of components in a link, which takes time linear in the number of simplices in the link. Thus, line $24$ takes $\cO(2 \sum_{e(\bu,\bv) \in j_{S_1,S_2}} |\Star{e(\bu,\bv)}|)$ time. The time complexity of lines $1$-$26$ of the \textsc{IsPath} procedure is $\cO(3 \sum_{e(\bu,\bv) \in j_{S_1,S_2}} |\Star{e(\bu,\bv)}|)$. 

The procedure \textsc{FindAssoLinkComp} finds the link component of $\ell_1$ associated with the representative arc $\beta^{p_1}$ (line $28$, procedure \textsc{FindAssoLinkComp}). This procedure first computes the link of $CP_1$ in $q_{f_1}^{-1}(p_1)$, which takes a time linear in the number of simplices in the star of $CP_1$ in $q_{f_1}^{-1}(p_1)$. If $\Star{CP_1}$ denotes this star, then the link of $CP_1$ is computed in $\cO(|\Star{CP_1} |)$ time. Since $CP_1$ is a critical point of $f_2$ restricted to a level set of $f_1$, it lies on an edge $e(\bu', \bv')$ of the Jacobi set $\JSet_{\f}$. The number of simplices in $\Star{CP_1}$ (denoted by, $|\Star{CP_1}|$) depends on the number of simplices in $\Star{e(\bu',\bv')}$ (denoted by, $|\Star{e(\bu',\bv')}|$). Thus, the time complexity for computing the link of $CP_1$ is $\cO(|\Star{e(\bu',\bv')}|)$.

After computing the (upper or lower) link of $CP_1$ in $q_{f_1}^{-1}(p_1)$, the component of this link associated with the arc $\beta^{p_1}$  (i.e. $\ell_1^{p_1}$) is determined. This step takes time linear in the number of simplices in the link, which is given by $\cO(|\Link{CP_1}|) = \cO(|\Link{e(\bu',\bv')}|)$. Next, the intersection of $\ell_1^{p_1}$ with the link of $\JSet_{\f}(CP_1,V_c)$ is computed. We note, $\ell_1$ is the (upper or lower) link of all the edges of $\JSet_{\f}(CP_1,V_c)$. However, to determine its intersection with $\ell_1^{p_1}$, it is sufficient to compute the intersection between $\ell_1^{p_1}$ and the link of the edge $e(\bu',\bv')$ of $\JSet_{\f}(CP_1, V_c)$ which contains $CP_1$  (and not the links of all the edges in $\JSet_{\f}(CP_1, V_c)$). The time complexity of determining this intersection is linear on the product of the number of simplices in the two links, which is given by $\cO(|\Link{e(\bu',\bv')}| |\Link{CP_1}|) = \cO(|\Link{e(\bu',\bv')}|^2)$. Thus, the time taken by the procedure \textsc{FindAssoLinkComp} is $\cO( |\Star{e(\bu',\bv')}| + |\Link{e(\bu',\bv')}| + |\Link{e(\bu',\bv')}|^2) = \cO(|\Star{e(\bu',\bv')}| + |\Link{e(\bu',\bv')}|^2)$. Since $|\Link{e(\bu',\bv')}| \leq c_L$, we have $|\Link{e(\bu',\bv')}|^2 \leq c_L^2$. Since $|\Star{e(\bu',\bv')}| \leq n$, the time complexity of the procedure \textsc{FindAssoLinkComp} is $\cO(n + c_L^2)$. Thus, lines $28$-$29$ of the \textsc{IsPath} procedure take $\cO(2 (n + c_L^2))$ time.

Line $31$ of the \textsc{IsPath} procedure computes the intersection of the associated link components, $L_1$ and $L_2$, which takes time linear in the number of simplices in $L_1$ and $L_2$. This in turn depends on the total number of simplices in the stars of the edges in $\JSet_{\f}(CP_1, V_c)$ and $\JSet_{\f}(CP_2, V_c)$. Thus, line $31$ takes $\cO(\sum_{e(\bu,\bv) \in j_{S_1,S_2}} |\Star{e(\bu,\bv)}|)$ time. Since $\sum_{e(\bu,\bv) \in j_{S_1,S_2}} |\Star{e(\bu,\bv)}| \leq n$, the total time taken by the \textsc{IsPath} procedure is $\cO(4 \sum_{e(\bu,\bv) \in j_{S_1,S_2}} |\Star{e(\bu,\bv)}| + 2(n + c_L^2)) = \cO(6n + 2c_L^2)$. This is also the complexity of the \textsc{IsPathConnected} procedure, when the conditions in lines $15$ or $22$ of the procedure are satisfied. Otherwise, the \textsc{IsPathConnected} procedure takes constant time. Next, we obtain the complexity bound for lines $8$-$16$ of the $\textsc{CompatibleUnion}$ procedure.

Since the number of simple Reeb sheets is at most $2j_e$, the total number of iterations of the two for loops in lines $8$ and $10$ is $\cO(4j_e^2)$. However, each sheet $S_1$ has at most two dummy edges. Based on our assumptions, a dummy edge can overlap with at most two other dummy edges (see \figref{fig:is-path}). Hence, for every iteration of the loop in line $10$, the \textsc{IsPathConnected} procedure in line $12$ takes $\cO(6n + 2c_L^2)$ time for at most $4$ iterations (because of the call to the \textsc{IsPath} procedure), and takes constant time during the remaining iterations. This is because the \textsc{IsPathConnected} procedure calls the \textsc{IsPath} procedure when only when $S_1$ and $S_2$ have overlapping edges. The check for overlapping edges is performed at lines $15$ and $22$ of the \textsc{IsPathConnected} procedure. We note, the $\textsc{Find-Set}$ operation in line $12$ of the procedure \textsc{CompatibleUnion} takes constant time \cite{Book-Algorithms-CLRS}. Therefore, the complexity of line $12$ over all iterations of the for loop in line $10$ is $\cO(4 (6n + 2c_L^2)) = \cO(24n + 8c_L^2)$. Since the for loop in line $8$ iterates $\cO(2j_e)$ times, the complexity of line $12$ over all iterations of the for loop in line $8$ is $\cO(2 j_e (24n + 8c_L^2)) = \cO(48 n j_e + 16 j_e c_L^2)$. Next, we analyze the complexity of line $13$ of the procedure \textsc{CompatibleUnion}.

Line $13$ performs the union of two simple Reeb sheets. Since the number of times two simple Reeb sheets from different sets of $UF$ are merged is at most the number of simple Reeb sheets, which is upper-bounded by $2j_e$, the total time complexity of line $13$ over all iterations of the for loops in lines $8$ and $10$ is $\cO(2j_e \log(2j_e))$ \cite{Book-Algorithms-CLRS}. Therefore, the time complexity of lines $8$-$16$ of procedure \textsc{CompatibleUnion} is $\cO(4j_e^2 + 48 n j_e + 16 j_e c_L^2 + 2j_e \log(2j_e))$, where $4j_e^2$ is the number of iterations of the for loops in lines $8$ and $10$, and the terms $48 n j_e + 16 j_e c_L^2$ and $2j_e \log (2j_e)$ are the complexity bounds for the lines $12$ and $13$, respectively, over all the iterations of the for loops. Since the number of simple Reeb sheets is at most $2j_e$, the number of components in $UF$ is upper-bounded by $2j_e$. Further, every simple Reeb sheet has at most $2$ dummy edges. Thus, lines $17$-$20$ take $\cO(2j_e)$ time. The total complexity of the procedure \textsc{CompatibleUnion} is $\cO(4j_e^2 + 48 n j_e + 16 j_e c_L^2 + 2j_e \log(2j_e) + 2j_e)$.

The time complexity of \algoref{alg:compute-Reeb-space} is then $\cO(n_e +j_e^2 + (c_{int}+1)(n \log (n)) +  n^2 + n\,c_{int}\,\log (n) + 4 j_e^2 + 6 j_e^2 + 4j_e^2 + 48 n j_e + 16 j_e c_L^2 + 2j_e \log(2j_e) + 2j_e)$. Since the terms $n_e$ and $j_e$ are upper-bounded by $n$, the complexity bound can be simplified as $\cO(n^2 + n\,c_{int}\,\log \,n + nc_L^2)$.

\section{Conclusion and Future Work}
\label{sec:conclusion} 
In the current paper, we introduce the first algorithm for computing a topologically correct Reeb space of a generic PL bivariate field without relying on range-quantization.
The time complexity of our algorithm is $\cO(n^2 + n\,c_{int}\,\log\, n + nc_L^2)$, where $n$ is the total number of simplices in $\M$, $c_{int}$ is the number of intersections of the projections of the non-adjacent Jacobi set edges on the range of the bivariate field and $c_L$ is the upper bound on the number of simplices in the link of an edge of $\M$. The proposed algorithm is comparable with the fastest algorithm available in the literature. Furthermore, existing algorithms in the literature suffer from the correctness issue, whereas we provide proof of topological correctness of the computed Reeb space using our algorithm. Our algorithm of computing correct Reeb space is based on computing a correct MDRG
which is first proven to be homeomorphic with the Reeb space. To build our main algorithm, we introduce four novel algorithms for (1) computing the Jacobi structure, (2) computing the MDRG, (3) computing a net-like structure embedded in the Reeb space and (4) computing the complete $2$-sheets of the Reeb space.

However, the theory and algorithms introduced in the current paper are specifically designed for bivariate fields on combinatorial $3$-manifolds without boundary. Future work will focus on extending the results for generic PL multi-fields. Moreover, our algorithm admits a potential generalization to piecewise-linear fields defined over arbitrary finite simplicial complexes, thereby extending beyond the more restrictive framework of combinatorial 3-manifolds without boundary.
It is important to highlight that the net-like structure of the Reeb space for a bivariate field encapsulates the joint topological features of both fields in a concise $1$-dimensional structure and is the topologically correct version of the joint contour net \cite{2014-Carr-JCN}. Therefore, this work harbors potential for applications across diverse computational domains, requiring exploration in future studies.

\backmatter

\bmhead{Acknowledgements}
The first two authors would like to thank the Science and Engineering Research Board (SERB), India,  Grant Number: SERB/CRG/2018/000702 and MINRO Center (Machine Intelligence and Robotics Center) at  International Institute of Information Technology-Bangalore (IIITB), for funding this project. The third author has been supported in part by JSPS KAKENHI Grant Numbers JP22K18267, JP23H05437, and by Institute of Mathematics for Industry,
Joint Usage/Research Center in Kyushu University (FY2024 Workshop(II) ``Integration of machine learning and mathematical modeling, and deepening of its theory II'' (2024a005)).

\bibliography{references}

@ARTICLE{1991-Shinagawa-Reeb-Graph,
  author={Shinagawa, Y. and Kunii, T.L.},
  journal={IEEE Computer Graphics and Applications}, 
  title={{C}onstructing a {R}eeb {G}raph {A}utomatically from {C}ross {S}ections}, 
  year={1991},
  volume={11},
  number={6},
  pages={44-51},
  doi={10.1109/38.103393}}

@article{2003-Cole-McLaughlin-Reeb-Graph,
author = {Cole-McLaughlin, Kree and Edelsbrunner, Herbert and Harer, John and Natarajan, Vijay},
year = {2003},
month = {05},
pages = {},
title = {{L}oops in {R}eeb {G}raphs of 2-{M}anifolds},
volume = {32},
journal = {Discrete \& Computational Geometry},
doi = {10.1145/777792.777844}
}

@ARTICLE{2009-Tierny-Reeb-Graph,
  author={Tierny, Julien and Gyulassy, Attila and Simon, Eddie and Pascucci, Valerio},
  journal={IEEE Transactions on Visualization and Computer Graphics}, 
  title={{L}oop {S}urgery for {V}olumetric {M}eshes: {R}eeb graphs {R}educed to {C}ontour {T}rees}, 
  year={2009},
  volume={15},
  number={6},
  pages={1177-1184},
  doi={10.1109/TVCG.2009.163}}

@ARTICLE{1996-Livnar-Isosurface,
  author={Livnat, Y. and Han-Wei Shen and Johnson, C.R.},
  journal={IEEE Transactions on Visualization and Computer Graphics}, 
  title={{A} {N}ear {O}ptimal {I}sosurface {E}xtraction {A}lgorithm {U}sing the {S}pan {S}pace}, 
  year={1996},
  volume={2},
  number={1},
  pages={73-84},
  doi={10.1109/2945.489388}}

@book{book-herbert-computational-topology,
  author    = {Herbert Edelsbrunner and
               John Harer},
  title     = {{C}omputational {T}opology - an {I}ntroduction},
  publisher = {American Mathematical Society},
  address = {Providence, RI},
  year      = {2010},
  isbn      = {978-0-8218-4925-5},
  bibsource = {dblp computer science bibliography, https://dblp.org}
}

@ARTICLE{2016-Klacansky,
  author={Klacansky, Pavol and Tierny, Julien and Carr, Hamish and Geng, Zhao},
  journal={IEEE Transactions on Visualization and Computer Graphics}, 
  title={{F}ast and {E}xact {F}iber {S}urfaces for {T}etrahedral {M}eshes}, 
  year={2017},
  volume={23},
  number={7},
  pages={1782-1795},
  doi={10.1109/TVCG.2016.2570215}}

@inproceedings{1973-Golubitsky-Stable-Maps,
  title={{S}table {M}appings and {T}heir {S}ingularities},
  author={Martin Golubitsky and Victor W. Guillemin},
  year={1973},
  url={https://api.semanticscholar.org/CorpusID:119015259}
}

@book{book-tamal-dey, place={Cambridge}, title={Computational Topology for Data Analysis}, DOI={10.1017/9781009099950}, publisher={Cambridge University Press}, address="Cambridge, UK", author={Dey, Tamal Krishna and Wang, Yusu}, year={2022}}

@article{2012-Parsa-Reeb-Graph,
author = {Parsa, Salman},
year = {2012},
month = {06},
pages = {},
title = {A {D}eterministic O(m log m) {T}ime {A}lgorithm for the {R}eeb {G}raph},
volume = {49},
journal = {Discrete \& Computational Geometry},
doi = {10.1145/2261250.2261289}
}

@inproceedings{2010-Harvey-Reeb-Graph,
author = {Harvey, William and Wang, Yusu and Wenger, Rephael},
title = {{A} {R}andomized O(m Log m) {T}ime {A}lgorithm for {C}omputing {R}eeb {G}raphs of {A}rbitrary {S}implicial {C}omplexes},
year = {2010},
isbn = {9781450300162},
publisher = {Association for Computing Machinery},
address = {New York, NY, USA},
url = {https://doi.org/10.1145/1810959.1811005},
doi = {10.1145/1810959.1811005},
booktitle = {Proceedings of the Twenty-Sixth Annual Symposium on Computational Geometry},
pages = {267–276},
numpages = {10},
location = {Snowbird, Utah, USA},
series = {SoCG '10}
}

@article{2007-Mapper,
  title={{T}opological {M}ethods for the {A}nalysis of {H}igh {D}imensional {D}ata sets and 3{D} {O}bject {R}ecognition.},
  author={Singh, Gurjeet and M{\'e}moli, Facundo and Carlsson, Gunnar E and others},
  journal={PBG@ Eurographics},
  volume={2},
  number={091-100},
  pages={90},
  year={2007},
DOI = {10.2312/SPBG/SPBG07/091-100}
}

@ARTICLE{2014-Carr-JCN, 
author={H. {Carr} and D. {Duke}}, 
journal={IEEE Transactions on Visualization and Computer Graphics}, 
title={{J}oint {C}ontour {N}ets}, 
year={2014}, 
volume={20}, 
number={8}, 
pages={1100-1113}, 
doi={10.1109/TVCG.2013.269}, 
ISSN={1077-2626}, 
month={Aug},}

@inproceedings{2008-edels-reebspace,
author = {Edelsbrunner, Herbert and Harer, John and Patel, Amit K.},
title = {{R}eeb Spaces of {P}iecewise {L}inear {M}appings},
year = {2008},
isbn = {9781605580715},
publisher = {Association for Computing Machinery},
address = {New York, NY, USA},
doi = {10.1145/1377676.1377720},
booktitle = {Proceedings of the Twenty-Fourth Annual Symposium on Computational Geometry},
pages = {242–250},
numpages = {9},
location = {College Park, MD, USA},
series = {SCG '08}
}

@ARTICLE{2012-Duke-Nuclear-Scission,
  author={Duke, David and Carr, Hamish and Knoll, Aaron and Schunck, Nicolas and Nam, Hai Ah and Staszczak, Andrzej},
  journal={IEEE Transactions on Visualization and Computer Graphics}, 
  title={{V}isualizing {N}uclear {S}cission through a {M}ultifield {E}xtension of {T}opological {A}nalysis}, 
  year={2012},
  volume={18},
  number={12},
  pages={2033-2040},
  doi={10.1109/TVCG.2012.287}}

@article {2015-Carr-Fiber,
author = {Carr, Hamish and Geng, Zhao and Tierny, Julien and Chattopadhyay, Amit and Knoll, Aaron},
title = {{F}iber {S}urfaces: {G}eneralizing {I}sosurfaces to {B}ivariate {D}ata},
journal = {Computer Graphics Forum},
volume = {34},
number = {3},
issn = {1467-8659},
doi = {10.1111/cgf.12636},
pages = {241--250},
year = {2015},
}

@ARTICLE{2021-Ramamurthi-MRS,  author={Ramamurthi, Yashwanth and Agarwal, Tripti and Chattopadhyay, Amit},  journal={IEEE Transactions on Visualization and Computer Graphics},   title={A {T}opological {S}imilarity {M}easure between {M}ulti-resolution {R}eeb {S}paces},   year={2021},  volume={},  number={},  pages={1-1},  doi={10.1109/TVCG.2021.3087273}}

@book{Book-Algorithms-CLRS,
author = {Cormen, Thomas H. and Leiserson, Charles E. and Rivest, Ronald L. and Stein, Clifford},
title = {Introduction to Algorithms, Third Edition},
year = {2009},
isbn = {0262033844},
publisher = {The MIT Press},
address ={Cambridge, MA, U.S.A.}, 
edition = {3rd}
}

@article{2003-Chiang-Simplification-Tetrahedral-Meshes,
author = {Chiang, Yi-Jen and Lu, Xiang},
title = {{P}rogressive {S}implification of {T}etrahedral {M}eshes {P}reserving {A}ll {I}sosurface {T}opologies},
journal = {Computer Graphics Forum},
volume = {22},
number = {3},
pages = {493-504},
doi = {https://doi.org/10.1111/1467-8659.00697},
eprint = {https://onlinelibrary.wiley.com/doi/pdf/10.1111/1467-8659.00697},
year = {2003}
}

@article{2012-Doraiswamy-Reeb-Graph,
author = {Doraiswamy, Harish and Natarajan, Vijay},
year = {2012},
month = {04},
pages = {},
title = {{C}omputing {R}eeb {G}raphs as a {U}nion of {C}ontour {T}rees},
volume = {19},
journal = {IEEE transactions on visualization and computer graphics},
doi = {10.1109/TVCG.2012.115}
}

@article{2004-edelsbrunner-jacobi-set,
	Author = {Herbert Edelsbrunner and John Harer},
	Journal = {In Foundations of Computational Matematics, Minneapolis, 2002},
	Note = {Cambridge Univ. Press, 2004},
	Pages = {37-57},
	Title = {Jacobi {S}ets of {M}ultiple {M}orse {F}unctions},
	Year = {2004}}

@ARTICLE{2016-Julien-ReebSpace,
author={Tierny, Julien and Carr, Hamish},
journal={IEEE Transactions on Visualization and Computer Graphics}, 
title={{J}acobi {F}iber {S}urfaces for {B}ivariate {R}eeb {S}pace {C}omputation}, 
year={2017},
volume={23},
number={1},
pages={960-969},
doi={10.1109/TVCG.2016.2599017}}

@article{edelsbrunner-simulation-of-simplicity,
author = {Edelsbrunner, Herbert and M\"{u}cke, Ernst Peter},
title = {{S}imulation of {S}implicity: {A} {T}echnique to {C}ope with {D}egenerate {C}ases in {G}eometric {A}lgorithms},
year = {1990},
issue_date = {Jan. 1990},
publisher = {Association for Computing Machinery},
address = {New York, NY, USA},
volume = {9},
number = {1},
issn = {0730-0301},
url = {https://doi.org/10.1145/77635.77639},
doi = {10.1145/77635.77639},
journal = {ACM Trans. Graph.},
month = {jan},
pages = {66–104},
numpages = {39}
}

@book{cerf-1968-diffeomorphismes,
  title={Sur les diffeomorphismes de la sphere de dimensions trois ({G}amma 4=0)},
  author={Cerf, J.},
  isbn={9783540042235},
  lccn={68024615},
  series={Lecture Notes in Mathematics},
  url={https://books.google.co.in/books?id=smYvvQEACAAJ},
  year={1968},
address={Berlin-New York},
  publisher={Springer Berlin Heidelberg}
}

@inproceedings{2014-EuroVis-short,
  author       = {Amit Chattopadhyay and
                  Hamish A. Carr and
                  David J. Duke and
                  Zhao Geng},
  editor       = {Niklas Elmqvist and
                  Mario Hlawitschka and
                  Jessie Kennedy},
  title        = {{E}xtracting {J}acobi {S}tructures in {R}eeb {S}paces},
  booktitle    = {16th Eurographics Conference on Visualization, EuroVis 2014 - Short
                  Papers, June 9-13, 2014},
address = {Swansea, UK},
  publisher    = {Eurographics Association},
  year         = {2014},
  url          = {https://doi.org/10.2312/eurovisshort.20141156},
  doi          = {10.2312/EUROVISSHORT.20141156},
  bibsource    = {dblp computer science bibliography, https://dblp.org}
}

@article{2016-Chattopadhyay-CGTA-simplification,
	Author = {A. Chattopadhyay and H. Carr and D. Duke and Z. Geng and O. Saeki},
	journal = {Computational Geometry: Theory and Application},
	publisher = {Elsevier},
	Pages = {1-24},
	Title = {{M}ultivariate {T}opology {S}implification},
	Volume = {58},
	Year = {2016}}

@article{2015-Strodthoff,
title = {{L}ayered {R}eeb {G}raphs for {T}hree-{D}imensional {M}anifolds in {B}oundary {R}epresentation},
journal = {Computers \& Graphics},
volume = {46},
pages = {186-197},
year = {2015},
note = {Shape Modeling International 2014},
issn = {0097-8493},
doi = {https://doi.org/10.1016/j.cag.2014.09.026},
url = {https://www.sciencedirect.com/science/article/pii/S0097849314001149},
author = {B. Strodthoff and B. Jüttler}
}

@article{2008-Edelsbrunner-Time-varying-Reeb-graphs,
title = {{T}ime-varying {R}eeb {G}raphs for {C}ontinuous {S}pace–{T}ime {D}ata},
journal = {Computational Geometry},
volume = {41},
number = {3},
pages = {149-166},
year = {2008},
issn = {0925-7721},
doi = {https://doi.org/10.1016/j.comgeo.2007.11.001},
url = {https://www.sciencedirect.com/science/article/pii/S0925772107001071},
author = {Herbert Edelsbrunner and John Harer and Ajith Mascarenhas and Valerio Pascucci and Jack Snoeyink}
}

@article{2004-Saeki-topoology-of-singulat-fibers,
  title={{T}opology of {S}ingular {F}ibers of {D}ifferentiable {M}aps},
  author={Osamu Saeki},
  journal={SUGAKU},
  volume={60},
  number={1},
  pages={46-67},
  year={2008},
  doi={10.11429/sugaku.0601046}
}

@article{Osamu-2013-Triangulating,
title = "Triangulating {S}tein factorizations of generic maps and Euler characteristic formulas",
author = "Hiratuka, J. T. and Osamu Saeki",
year = "2013",
month = apr,
language = "English",
volume = "B38",
pages = "61--89",
journal = "RIMS K\^oky\^uroku Bessatsu",
issn = "1881-6193",
publisher = "Kyoto University",
}

@InProceedings{2014-saeki-visualizing-multivariate-data,
author="Saeki, Osamu
and Takahashi, Shigeo
and Sakurai, Daisuke
and Wu, Hsiang-Yun
and Kikuchi, Keisuke
and Carr, Hamish
and Duke, David
and Yamamoto, Takahiro",
editor="Wakayama, Masato
and Anderssen, Robert S.
and Cheng, Jin
and Fukumoto, Yasuhide
and McKibbin, Robert
and Polthier, Konrad
and Takagi, Tsuyoshi
and Toh, Kim-Chuan",
title="{V}isualizing {M}ultivariate {D}ata {U}sing {S}ingularity {T}heory",
booktitle="The Impact of Applications on Mathematics",
year="2014",
publisher="Springer Japan",
address="Tokyo",
pages="51--65",
isbn="978-4-431-54907-9"
}

@inproceedings{Kushner-1984,
  author    = {Kushner, León and Levine, Harold and Porto, Paulo},
  title     = {{M}apping {T}hree-{M}anifolds into the {P}lane. I.},
  booktitle = {Bol. Soc. Mat. Mexicana},
  pages     = {11--33},
  year      = {1984},
  volume    = {29},
  number    = {1}
}

@book{Levine-2006,
  author    = {Levine, Harold},
  title     = {{C}lassifying {I}mmersions into {R}4 over {S}table {M}aps of 3-{M}anifolds into {R}2},
  publisher = {Lecture Notes in Math., Springer},
  address="Berlin, Heidelberg",
  year      = {2006},
  isbn      = {978-3-540-39700-7}
}

@String{Computing = "Computing" }

@String{Computer = "{IEEE} Computer" }

@String{Springer = "Springer-Verlag" }

@ArtifactSoftware{R,
    title = {R: A Language and Environment for Statistical Computing},
    author = {{R Core Team}},
    organization = {R Foundation for Statistical Computing},
    address = {Vienna, Austria},
    year = {2019},
    url = {https://www.R-project.org/},
}
\end{document}